# Strategies to improve the energy storage properties of perovskite lead-free relaxor ferroelectrics: a review

**Vignaswaran Veerapandiyan [1], Federica Benes [1], Theresa Gindel [1] and Marco Deluca [1,*]**

[1] Materials Center Leoben Forschung GmbH, Roseggerstrasse 12 and A-8700 Leoben (Austria); vignaswaran.veerapandiyan@mcl.at (V.V.); federica.benes@mcl.at (F.B.); theresa.gindel@mcl.at (T.G.)

* Correspondence: marco.deluca@mcl.at (M.D.)

**Electrical** energy storage systems (EESSs) with high energy density and power density are essential for the effective miniaturization of future electronic devices. Among different EESSs available in the market, dielectric capacitors relying on swift electronic and ionic polarization-based mechanisms to store and deliver energy already demonstrate high power densities. However, different intrinsic and extrinsic contributions to energy dissipations prevent ceramic-based dielectric capacitors from reaching high recoverable energy density levels. Interestingly, relaxor ferroelectric-based dielectric capacitors, because of their low remnant polarization, show relatively high energy density and thus display great potential for applications requiring high energy density properties. Here, some of the main strategies to improve the energy density properties of perovskite lead-free relaxor systems are reviewed. This includes (i) chemical modification at different crystallographic sites, (ii) chemical additives that do not target lattice sites and (iii) novel processing approaches dedicated to bulk ceramics, thick and thin films, respectively. Recent advancements are summarized concerning the search for relaxor materials with superior energy density properties and the appropriate choice of both composition and processing route to match various needs in the application. Finally, future trends in computationally-aided materials design are presented.

## 1. Introduction:

The challenges associated with growing world population and the increased degree of interconnection of electronic devices worldwide bring about an increase in energy consumption, which needs to be tackled off-grid by a new generation of stand-alone Electrical Energy Storage Systems (EESSs) compensating the discontinuity of renewable energy sources [1]. Renewable energies, in fact, are unavailable for long periods of time (e.g. solar energy is predominantly available in the daytime and wind energy in the early mornings). Hence, converting harvested renewable energy to electrical energy and storing it to be readily available anytime for the needs of electronic devices is the primary solution. To achieve this, efficient EESSs tuned to specific applications are needed. EESSs can be broadly classified into four main classes such as 1. Solid oxide fuel cells; 2. Traditional batteries (Li-ion batteries); 3. Electrochemical capacitors and 4. Dielectric capacitors [2]. The appropriateness of any of these EESSs classes for a specific application is generally decided by two important parameters, namely the energy density (ED) and the power density. The ED is the energy stored in a given amount of substance, which can be expressed in volume (volumetric ED: Wh/L or J/cm$^3$) or mass (Specific ED: Wh/kg). Power density is the measure of power output from a particular amount of substance, and is often expressed in W/kg. A Ragone plot, named after David. V. Ragone [3], is often used to show the energy and power available for a certain load, i.e., energy density vs. power density. Here it is important to note that the Ragone plot depicts the maximum energy in a finite power region that is based on the type of EESSs. The loss mechanisms such as leakage currents, internal heating etc. are not included in a Ragone plot, although they are crucial for end applications.

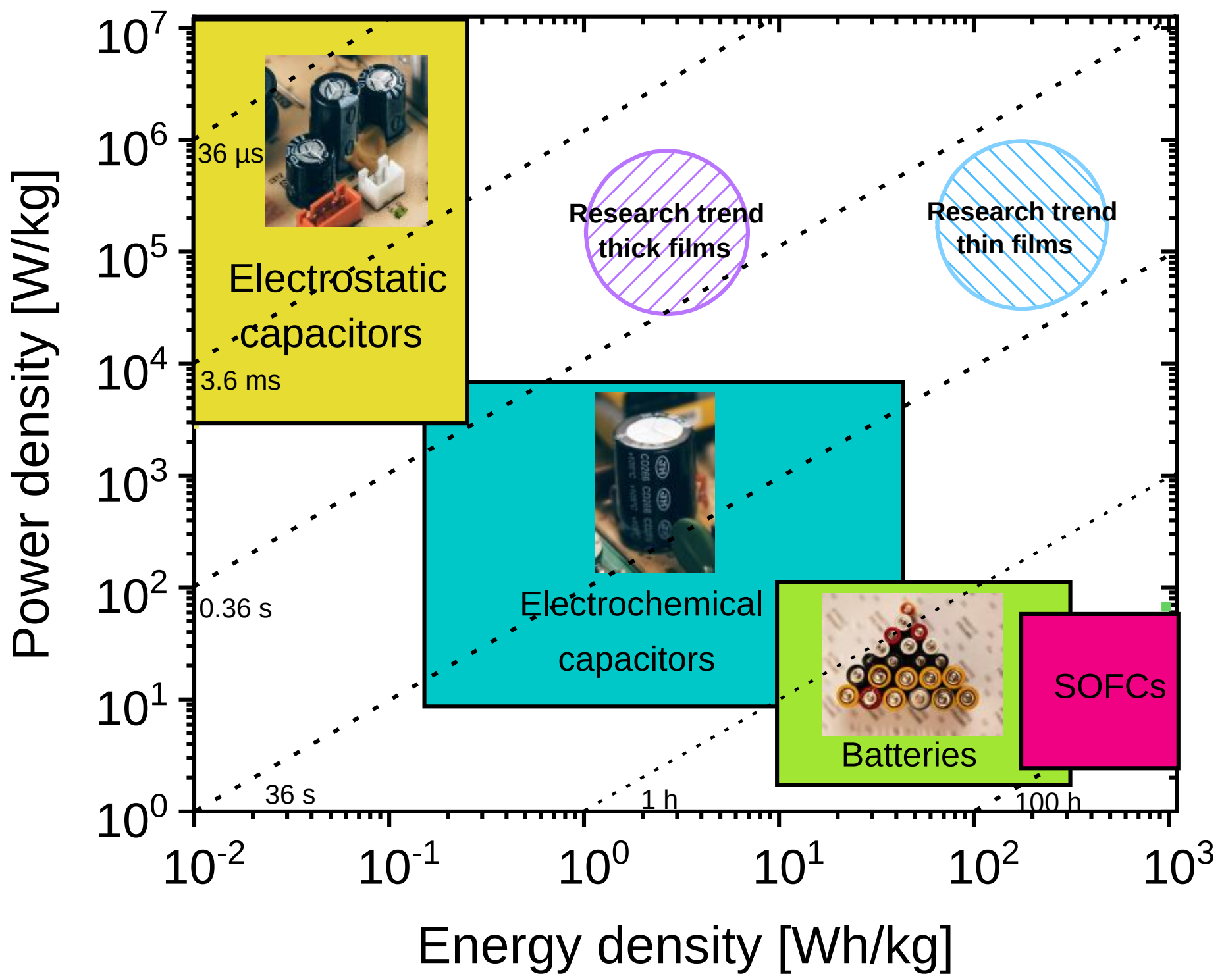

**Figure 1.** Ragone plot comparing energy density against power density for different EESSs.

From the Ragone plot shown in Figure 1, it is clear that EESSs have to be chosen depending on the needs, because of the unavailability of any EESS that combines high power as well as ED. In much simpler terms, this plot shows why traditional batteries can supply energy for a longer time (> 100 s) but need more time to replenish compared to a dielectric capacitor (< 0.01 s). In spite of the low ED of dielectric capacitors (cf. Figure 1), higher operating voltages, lower cost, size flexibility, thermal and cyclic stability and range of possibilities to tune the leakage currents are some of the major advantages. Realizing high ED in a dielectric capacitor while retaining its high power density would thus set up new possibilities towards versatility, cost-effectiveness, miniaturization, etc [4].

Dielectrics are materials with high electrical resistivity, typically greater than $10^8$ Ω.m and can store electrical energy through lattice polarization resulting from the formation or reorientation of electric dipoles. When a dielectric is placed in an electric field, there is no long-range flow of charge; however, atoms or ions locally react to oppose the electric field by polarizing or setting up a dipole moment that opposes the external applied electric field [5]. Hence, dielectric capacitors can deliver charges very quickly while traditional batteries rely on chemical reactions making them less time-efficient. Dielectric capacitors can also have a longer lifetime for the very reason in contrary to batteries in which the chemical reactions are not always completely reversible.

For a ceramic dielectric, the stored ED, $J_s$, is given by,

$$J_s = \frac{1}{2}\varepsilon_0\varepsilon_r E^2 \quad (1)$$

Where $\varepsilon_0$ is the permittivity of the free space, $\varepsilon_r$ is the dielectric permittivity of the ceramic material and $E$ is the applied electric field. $J_s$ can be represented as an integral function of polarization ($P$) since P=$\varepsilon_0\varepsilon_r E$,

$$J_s = \int_0^{P_s} E dP \quad (2)$$

The above equations represent the amount of energy that can be stored in a ceramic dielectric when the polarization is increased from 0 to polarization saturation ($P_s$) under the applied field increasing from 0 to $E_{max}$, respectively.

Whereas, the recoverable energy density will be,

$$J_r = \int_{P_r}^{P_s} E dP \quad (3)$$

Where $P_r$ is the remnant polarization. Based on these equations, for superior ED properties, a ceramic dielectric should have high $\varepsilon_r$, large $P_s$, low $P_r$, and high dielectric breakdown strength (BDS). The BDS is one of the primary deciding factors of ED properties of EESSs [6]. Dielectrics with all the stated properties originate from the broad class of ferroelectric materials.

The aim of this review is to introduce perovskite based relaxor ferroelectrics and then summarize some of the common strategies that are used to tune and promote the ED properties in chemically modified high-permittivity perovskite-based dielectric systems that are often, but not always, relaxor ferroelectrics. In addition to the chemical modification that is essential to attain a relaxor state, novel fabrication methods are discussed that are crucial to control the microstructure and thereby the BDS, which is mandatory to achieve high ED properties in any FE systems. Given the amount of works available in the literature on these topics, and the complexity of the problem, this review cannot be comprehensive of all published material. Nevertheless, it seeks to provide researchers with clear guidelines on how lead-free relaxor-based systems could be modified to enhance ED properties.

## 2. Perovskite based relaxor ferroelectrics

Ferroelectrics (FE) are polar materials with spontaneous polarization that can be reoriented by changing the direction of the external applied electric field. In general, the overall polarization of the ferroelectric crystal is zero because of the equal number of domains oriented in random directions. As $E$ increases, the cations obtain sufficient energy to overcome the local electrical potential barrier and will be able to jump from one random potential well position to another permissible well position most closely aligned with the field, which results in switching of domains. At strong enough $E$ ($E_{max}$), switching will result in a domain saturation state (i.e. at the field above which no further domain reorientation in field direction is possible) at which the exhibited polarization is the $P_s$. Upon reducing and reversing $E$, the converse process takes place, but traces along a new path consistent with the creation of new domains in the opposite direction. The polarization exhibited at zero field after field reduction is $P_r$, which is not equal to zero in a FE material. The required $E$ that can switch the domains of the ferroelectric material back and forth is the coercive field ($E_c$). Once poled, the material continues to follow the hysteresis loop and will return to zero net polarization at -$E_c$ or if the material is raised above $T_c$, but not at $E$ = 0 [7]. This phenomenon is called polarization-electric field (PE) hysteresis, also shown in Figure 2 for a ferroelectric material. Also shown in Figure 2 is the P-E for relaxors and linear dielectrics.

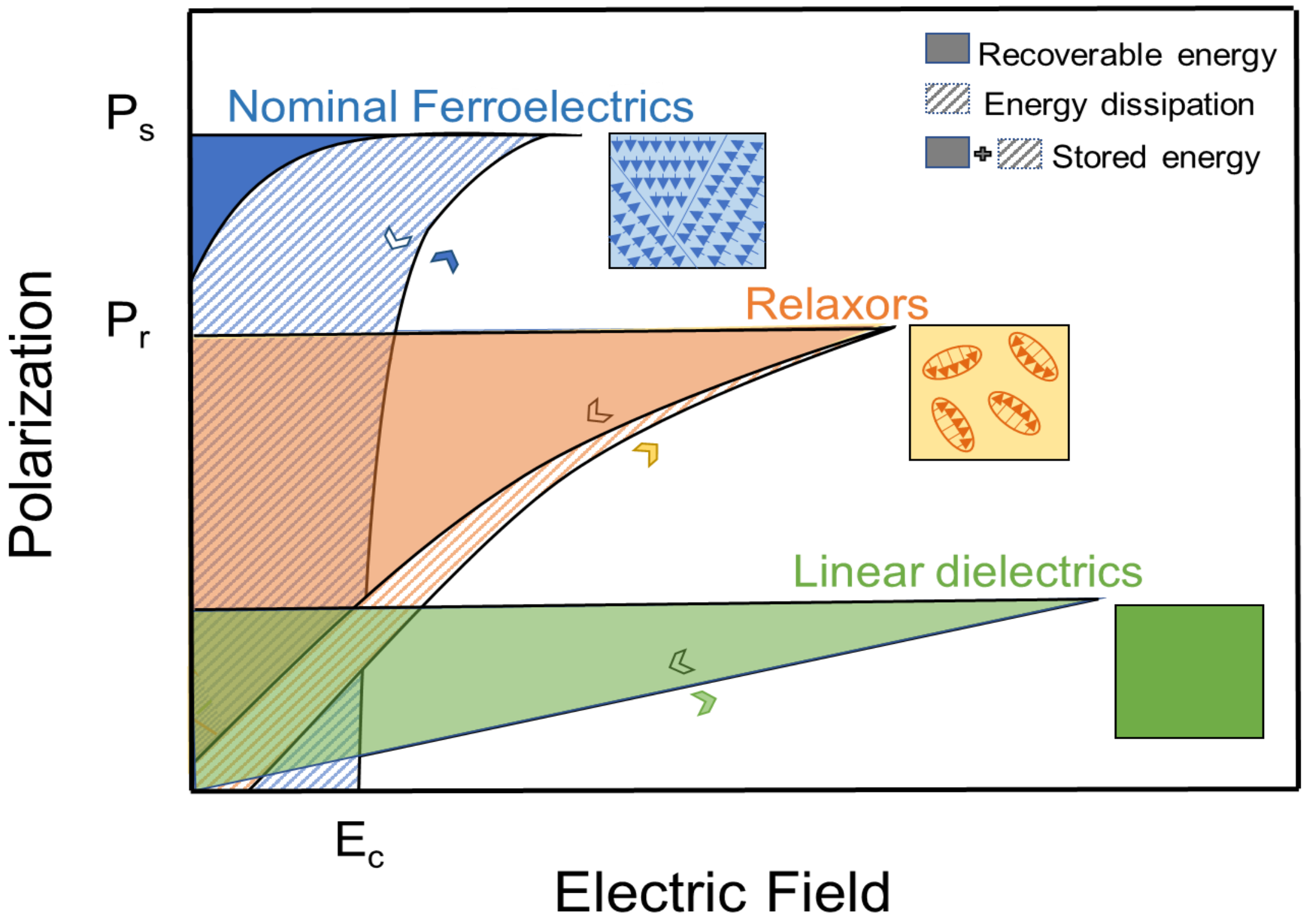

**Figure 2.** Polarization versus Electric field (P-E) for typical ferroelectrics, relaxors and linear dielectrics.

Because of PE hysteresis, the recoverable ED, $J_r$, is usually smaller than $J_s$ in ferroelectric ceramics, as shown in Figure 2. The figure clearly marks the difference in stored and recovered energy in ferroelectric materials by stripes and fillings. The ratio of $J_r$ and $J_s$ is the energy-storage efficiency $\eta$.

$$\eta = \frac{J_r}{J_s} \quad (4)$$

The difference in $J_s$ and $J_r$ is a direct consequence of non-zero $P_r$ and $J_r$ can be drastically different with different $P_r$ values, also shown in Figure 2 (nominal ferroelectrics vs. relaxors).

Ferroelectricity is reported in four material classes: 1. Oxygen octahedral group (i.e. perovskite) 2. Pyrochlore group 3. Tungsten-bronze group and 4. Bismuth layer-structure group [8]. From a structural point of view, FE materials belong to non-centrosymmetric point groups with orientable spontaneous polarization. From an electrical point of view, a FE material, in addition to the defined P-E loops, will exhibit a sharp rise in the temperature dependent $\varepsilon_r$ response when the material undergoes a transition from non-centrosymmetric FE state (where the spontaneous polarization exists) to a centrosymmetric paraelectric (PE) state. This can be seen in Figure 4, that shows the $\varepsilon_r$ response of ferroelectric barium titanate ceramics. This transition temperature is called the Curie temperature ($T_c$). The most studied FE material class is the oxygen octahedral group, also categorized as perovskite [9], and we will refer further only to this material class in this review. In addition to the recent interests on perovskite based relaxors for EESSs, which is addressed in this review, this material class has gained interest for various applications such as photovoltaics, catalysis, smart windows, etc., because of their versatile structure and the possibility to achieve a wide range of electrical, magnetic, optical and mechanical properties. We also caution the reader that also non-perovskite systems may possess high ED properties (for instance, tetragonal tungsten bronzes [10, 11], and that the discussion about using additives and novel processing methods included in this review (Chapters 4. and 5.) may apply to those systems as well.

Perovskite is the classification name given to materials based on the general crystal structure and bonding arrangement of the mineral calcium titanate ($CaTiO_3$) [12]. At room temperature, $CaTiO_3$ has the orthorhombic *Pbnm* crystal structure and undergoes reversible phase transformation to tetragonal $I4/mcm$ at ~1240 °C. It transforms to ideal cubic $Pm\bar{3}m$ at a temperature of ~1360 °C and remains $Pm\bar{3}m$ cubic until its melting temperature of ~1975 °C. According to displacive model, in the ideal cubic perovskite structure ($ABO_3$), $Pm\bar{3}m$, also as shown in Figure 3, atoms have a face centred arrangement and the structure is cubic close-packed with larger A-site (A) cations and C-site (C) anion forming an FCC lattice with the smaller B-site (B) cation possessing octahedral coordination with anions. This octahedral coordination of the B-site cations classifies perovskites under *oxygen octahedral group*. In the perovskite structure, the co-ordination number of A cation is twelve and the B cation and C anion coordination numbers are six each [7].

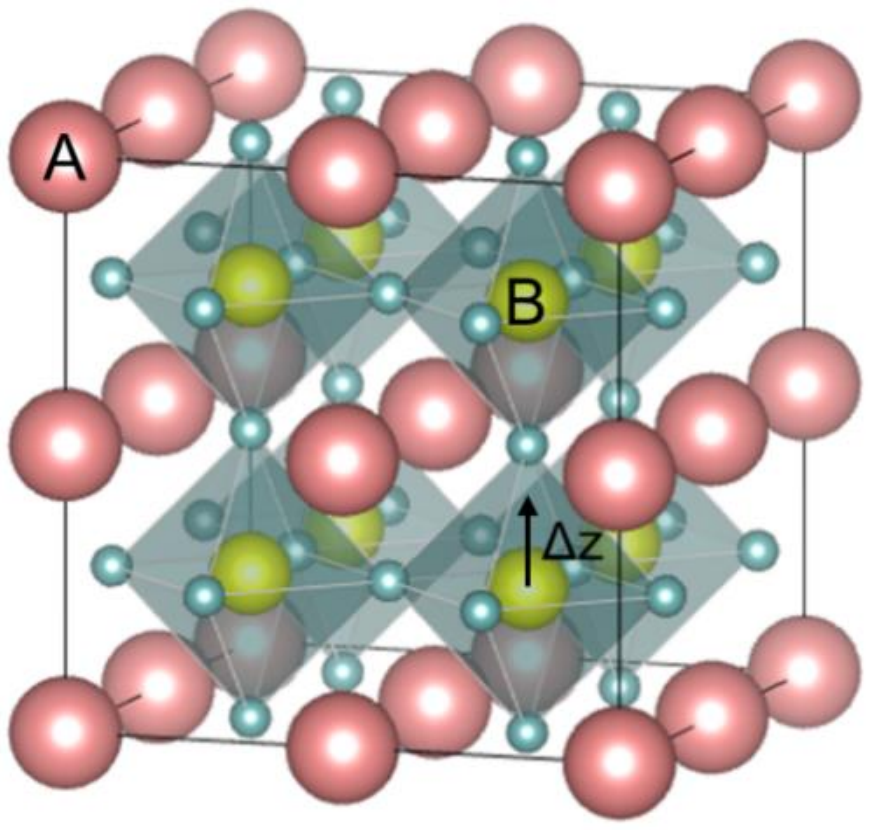

**Figure 3.** Perovskite $ABO_3$ structure with face centred arrangement. The B cation sits with VI-fold coordination at the centre of the oxygen octahedral.

For ferroelectrics and related material systems, the perovskite structure can tolerate a wide range of substitutions in the A and B site, which can result in significant variations of material properties because of the substitutions changing the polarization energy per unit volume, band structure, etc [13, 14]. The theoretical packing density of the close packed perovskite structure can range from 0.52 to 0.76 and can be increased further by selective elemental substitution. Each lattice site may incorporate multiple ions of unique ionic radii and valence states that can lead to complex perovskites like $PbMg_{1/3}Nb_{2/3}O_3$ (PMN), $Na_{1/2}Bi_{1/2}TiO_3$ etc. As a result, a perovskite can take on a wide range of crystal structures depending upon the nature of the incorporated atoms and thus the material rarely forms the ideal cubic perovskite structure. The non-cubic or non-ideal perovskite structure typically transforms into the ideal cubic perovskite structure at elevated temperatures.

Perovskite materials are often structurally understood by applying a semi-empirical relationship known as the Goldschmidt tolerance factor (GTF) [15], which is expressed by the following equation,

$$t = \frac{R_A + R_C}{\sqrt{2}(R_B + R_C)} \quad (1)$$

where $R_A$, $R_B$ and $R_C$ are the ionic radii of the A, B, and C-site atom(s), respectively. In Goldschmidt's formalism, $T$ ranges from about 0.77 to about 1.05 with the "ideal" cubic perovskite forming when $T$ is about 1.00. For $T > 1$, the material is often associated with high permittivity material properties, which include ferroelectric materials. For $T < 1$ is often associated with low symmetry materials. The scientific community has utilized the GTF, as a relatively simple tool for nearly a century to guide discovery and development of new perovskite materials, although it does not consider effects deviating from pure ionic bonding behaviour and thus might not be applicable to all perovskite systems [16].

Most of the technologically relevant perovskite materials are based on $PbTiO_3$, where the A-site of the lattice is occupied by $Pb^{2+}$. The lone electron pair of $Pb^{2+}$ induces a hybridization with the neighbouring oxygen anions, thereby shifting the bonding character to covalent. As a result, the $Pb^{2+}$ cation goes off-centre, which has important implications in the giant electromechanical properties of $PbZr_{1-x}Ti_xO_3$ (PZT) and $PbMg_{1/3}Nb_{2/3}O_3$-$PbTiO_3$ (PMN-PT) solid solutions [17]. Lead-based FE materials, however, are subject of restrictions due to the toxicity of lead-containing compounds especially during processing steps, and because of the risk of Pb leaking to the environment after end-of-use of electronic components [18]. The study of lead-free FE materials is however far from being concluded and it is yet unclear how lead-free materials have to be designed to attain desired properties. It is thus the scope of the present work to review the state-of-the-art of lead-free perovskites especially for EESSs.

One of the most widely studied lead-free perovskite-based FE materials is barium titanate ($BaTiO_3$, BTO). Historically, BTO was discovered simultaneously in the United States by Wainer and Salomon in 1942, in Russia by Vul in 1944 and in Japan by Ogawa in 1944. The crystal structure of BTO was first reported by Megaw [19] and Von Hippel [20]. BTO is an ideal cubic structure above 120 °C - 128 °C (Curie temperature-$T_c$) and follows a Curie-Weiss law:

$$\frac{1}{\varepsilon_r} = \frac{T - T_c}{C} \quad (6)$$

Where, C is Curie constant.

Below $T_c$, BTO undergoes two ferroelectric-ferroelectric phase transitions: a structural phase transformation from tetragonal (space group: P4mm) to orthorhombic (space group: Amm2) at 6 - 12 °C, followed by a transition from orthorhombic to rhombohedral (space group: R3m) at -77 – (-92) °C [21]. Figure 4. shows the dielectric and structural properties of BTO ceramics.

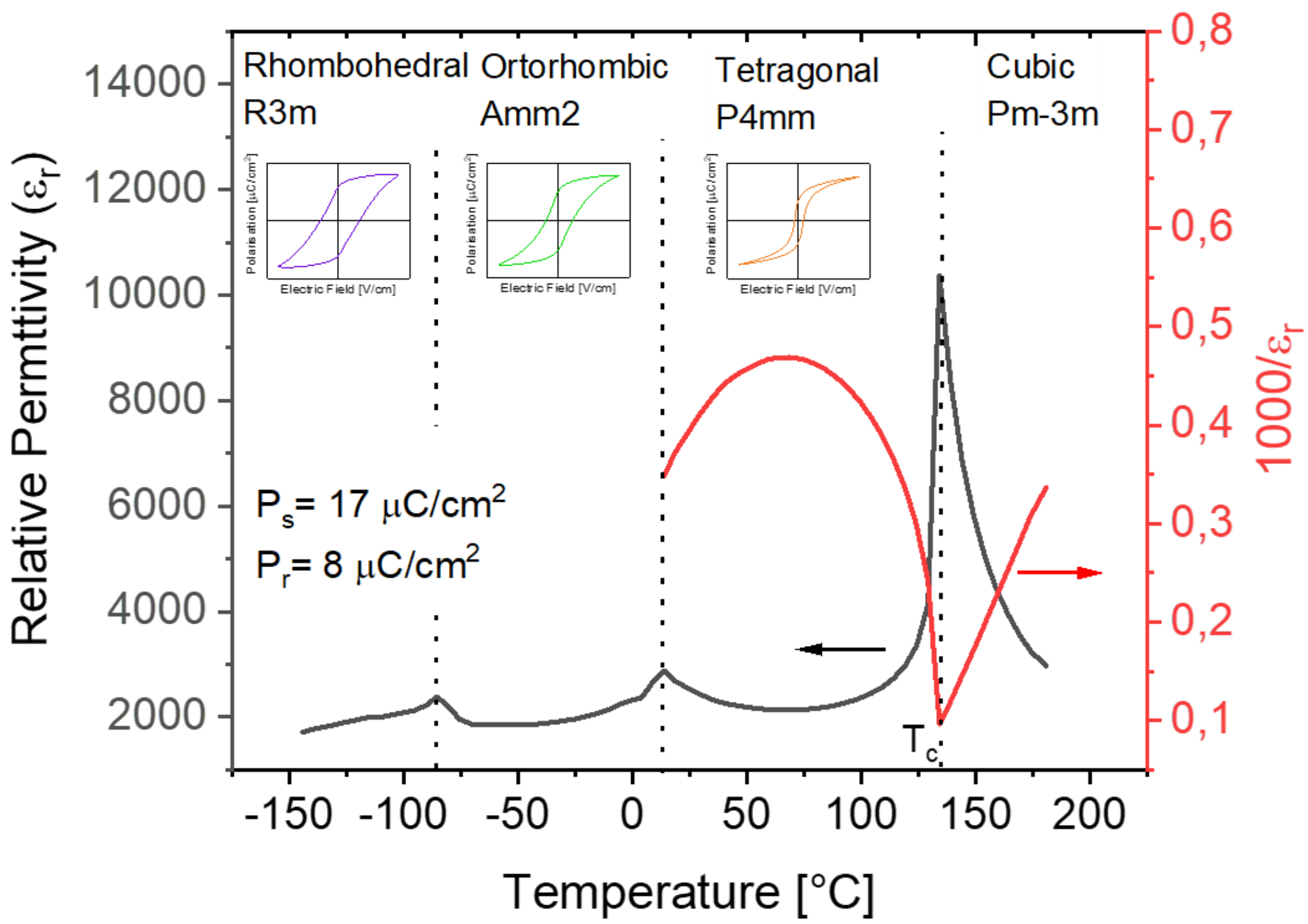

**Figure 4.** Dielectric and structural properties of BTO ceramics. Values of saturation polarization ($P_s$) and remnant polarization ($P_r$) are at 30 °C.

Chemical modification that includes both chemical substitution and additives is a very common approach in BTO to enhance or tune the material properties. This is later combined with novel fabrication methods for various technological needs [22]. In addition to macroscopic material properties, chemical alterations have profound effects on the fundamental FE nature of the parent material. Here parent material can be referred to any FE perovskite system that is constituted by one variation of atomic species at every crystallographic site in a perovskite structure without affecting the translational symmetry. Interestingly, chemical alterations even on an atomic level in a FE parent material can have notable impact on the macroscopic properties. From now on, considering the scope of this review, discussions will be based primarily on changes observed in macroscopic properties (dielectric properties, PE loops, microstructure etc.) as a consequence of chemical alterations, although changes at atomic scale cannot be completely ignored. Upon chemical alteration, mainly substitutions at a lower concentration level, the lattice continuum of the FE matrix is disrupted resulting in increased diffusivity of the temperature dependent $\varepsilon_r$ in addition to changes in the well-defined features of the P-E loops that are typical for a FE material. This type of FE material, although chemically modified, retains the long-range FE order and so exhibits a diffuse FE-PE phase transition at the $T_c$ and follows the Curie-Weiss law above $T_c$. Such FE systems with a broad $\varepsilon_r$ response, are categorized as FE with 'diffuse phase transition-DPT' [23, 24]. The nature of FE phase transitions in DPT is controversial and it is outside of the scope of this review. Upon further chemical modification, a very peculiar material state called 'relaxor-state' can be achieved. There are three main features that characterize relaxors: 1. A diffuse temperature dependent relative permittivity response, 2. The dispersion of the relative permittivity maximum as a function of frequency and 3. The absence of macroscopic symmetry breaking as a function of temperature (the permittivity maximum is hence denoted as $T_m$ instead of $T_c$). The transition from FE state to a DPT state and finally to a relaxor state is reported in many FE parent materials and this transition sequence is especially valid in BTO based systems [25–28]. The compositions in between a DPT state and a full relaxor state are generally called "crossover compositions". The exact concentration for this series of composition-driven transitions until relaxor state varies with different substituting ions and will be discussed more in detail in the subsequent sections. This article will primarily focus on BTO-based systems showing relaxor characteristics at high substituent content.

Relaxors are attractive for EESSs because of their relatively high BDS, high $\varepsilon_r$ and slim P-E loops (i.e. low remnant polarization-$P_r$). Figure 2 depicts the drastic difference in the appearance of the PE loops of relaxors compared to a FE system [29]. The slim PE loops are a direct consequence of the chemical heterogeneity that results from chemical modifications, disrupting the long-range polar FE order into a fragmented short-range polar-state [30]. The disruptions are of different origin and there are numerous theories available in the literature that differentiate 'relaxor' states based on the nature of the substituents [31–34]. The most recent theory suggests the occurrence of slush-like polar structures, which primarily elucidates the dynamics of electric dipoles as a function of temperature [35]. Electric dipoles can originate both from static (i.e. defect-induced) and dynamic (cation hopping-induced) lattice disorder. These dipoles are the source of random electric fields (RF) that play a crucial role in inducing relaxor behaviour independent of the substitution types. Dynamic disorder is temperature-driven and is not only specific to relaxors but is present also in the cubic phase of BTO, resulting in broad Raman spectra well above the $T_c$ [36, 37]. Intrinsic static disorder is related to defects and is present even in single crystal BTO – for example due to oxygen vacancies [38]. It is understood that relaxor behaviour might have a different origin in Pb-based or Ba-based systems, since cation off-centring has a role in local lattice polarization.

## 3. Tuning Energy density by chemical substitution

A relaxor state can only be attained upon chemical modification and there is a certain 'compositional window' inside the *solubility limit* or *'x' mole percentage* (for a solid solution) within which the relaxor properties can further be tuned. Here, the term *chemical modification* for a compound refers to modification of its chemical composition by addition or removal of an atom or molecule. In the field of materials science, this practice is often called as *doping* or *substitution*. The term 'doping' is widely used in semiconductor science, and refers to the introduction of a foreign ion in the material predominantly to modify the electrical properties [39]. This foreign ion can replace an existing atom in the equivalent crystallographic site, otherwise called as 'substitutional doping' or can take an interstitial lattice site without being incorporated in the lattice, otherwise called as 'interstitial doping'. Please note that doping is carried out in very small quantities (in the range of parts per million-ppm). On the other hand, substitution in materials science refers to replacement of an atom in the crystallographic site with a suitable atom of same or different oxidation state compared to the atom that is replaced, as shown in Figure 5a. Substituents go on a definite crystallographic site and are usually added in higher percentages (up to 40 at% in barium zirconium titanate to attain relaxor state). When it comes to perovskite relaxors, using the term substitution for chemical modification is appropriate since replacement of atoms are preferred over interstitial impurities to tuning the material properties. Also, unlike doping, substitution percentages inducing relaxor behaviour can vary from as low as 10 at% ($Nb^{5+}$ modified BTO-BNbT) [40] to as high as 40 at% ($Zr^{4+}$ modified BTO-BZT) [41], depending on the type of chemical substitution.

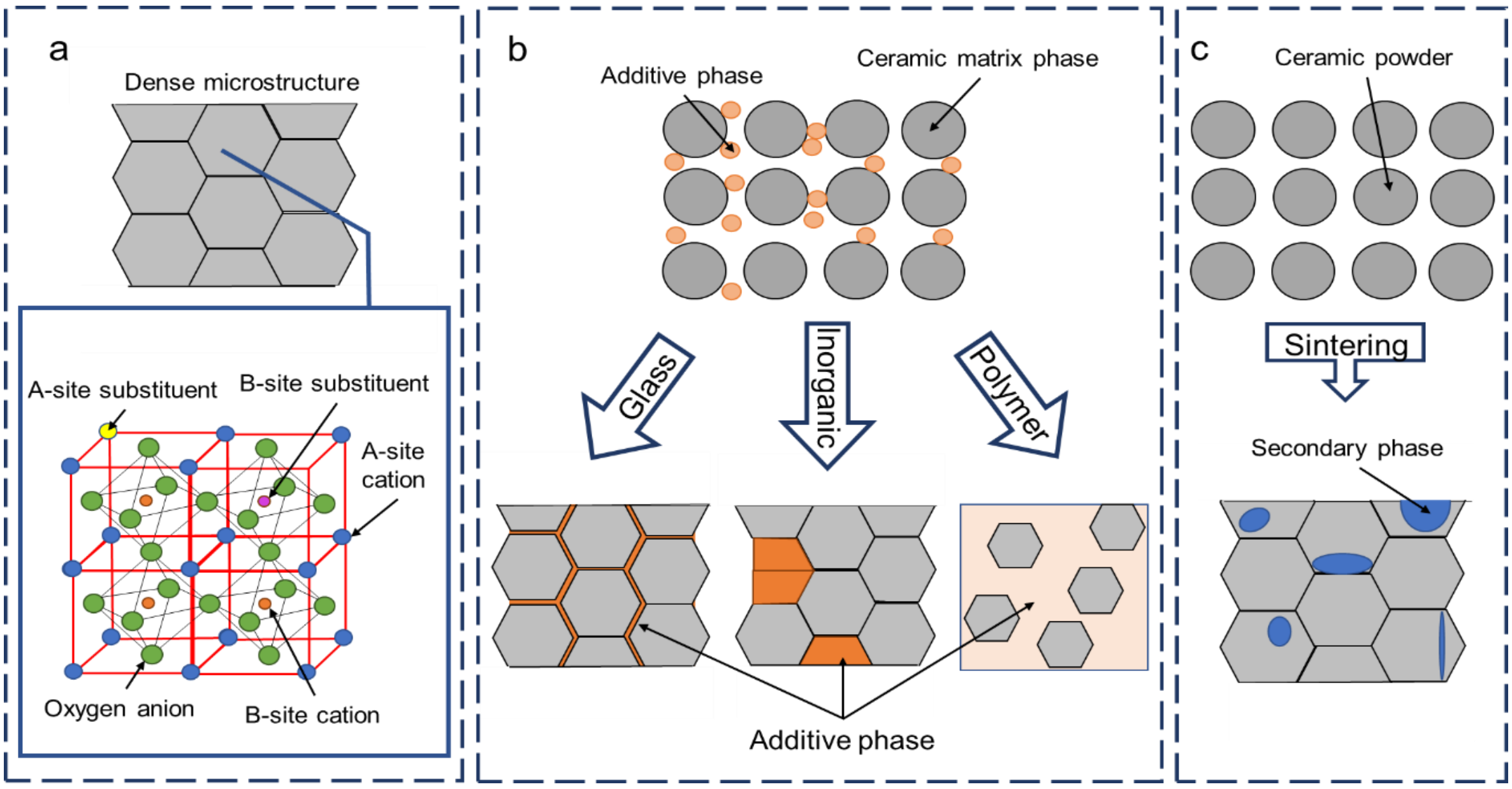

**Figure 5.** Graphical representation of chemical substituents, additive phases and secondary phases.

Substitution can be either be homovalent (HoV) or heterovalent (HeV) in nature. In most cases, substitution in perovskite structure is carried out at the A and/or B site of the $ABO_3$ structure (c.f. Figure 5a). Here, homovalent substitution refers to replacement of a cation in the lattice site with an ionic species of same oxidation state, whereas in case of heterovalent substitution the substituent ion has a different oxidation state to that of replaced cation. In both cases, it should have a permissible ionic radius as stated by the GTF (c.f. Eqn. 5). Relaxors, as defined previously, are chemically heterogenous and disrupted FE systems where the long-range FE order is disrupted by a mechanism that has different origin based on the substitution type. In a few words, for HoV substitution, the disruption is considered to be a simple 'bond-breaking methodology' (Polar cluster + matrix), when the atoms do not go off-centre and weak random fields (RF) originate from steric effect because of the difference in the ionic radii. This kind of relaxor systems can therefore be called as 'fragmented ferroelectrics' and a typical example is BZT [42, 43]. On the other hand, in HeV substitution, the origin of FE disruption is more complex depending on the donor (lower oxidation state compared to substituted cation) or acceptor (higher oxidation state compared to substituted cation) state of the substituent. The charge imbalances as a result of difference in the oxidation states must be compensated by free electrons or cationic vacancies to retain the electrical neutrality [44]. Hence, the long-range FE order is disrupted in the HeV case by strong random electric fields that emerge from off-centring of substituted cations in addition to the defect complexes arising from charge compensation schemes. This kind of relaxor systems can therefore be called as 'disordered ferroelectrics' and a typical example is BNbT [38]. Detailed reports on origin of relaxor behaviour upon chemical modification can be found elsewhere [45].

Our previous [38, 46] and other reported works have overwhelming evidence that HeV substitutions are more effective in inducing the 'relaxor state' and have profound effect on the dielectric properties for comparatively small substituent concentrations compared to HoV substitution [47, 48]. Often multiple HeV substituents are used simultaneously at the A and/or B site to preserve an overall charge neutrality and limit the formation of cationic vacancies that could deteriorate the dielectric performance of the material at the same time introducing strong RFs for effective disruption of long-range FE order. In this chapter, HoV and HeV substitutions will be treated individually to limit the complexity and to give a general outline on the effect of different chemical modification on the dielectric properties relevant to enhancing and stabilising the ED.

*3.1. Effect on Curie temperature:*

One important effect of chemical modification is the shift in the $T_c/T_m$. This effect is technologically relevant since the $\varepsilon_r$ value (and hence the ED) is usually greater around $T_c/T_m$ (c.f. Eqn. 1). It is thus important to shift the $T_c/T_m$ in a temperature range where high ED is desired. Figure 6 shows the evolution of $T_c/T_m$ with the substitution concentration in BTO. Except for $Ca^{2+}$ in small percentages at the A-site, $T_c/T_m$ decreases with the increase in substitution concentration. The crossover to relaxor compositions is marked as asterisks for every substituent. As a general guideline, the effect of a substituent on the $T_c/T_m$ is given by the shift of the B atom (c.f. Figure 3-marked by an arrow) from the ideal cubic (centrosymmetric) position in oxygen octahedra, denoted by $\Delta z$. If a substituent limits $\Delta z$ as a result of ionic radii or bonding environment, thereby limiting also the distortion of the octahedra, $T_c$ drops as a consequence. Ravez et al. thoroughly investigated the influence of substituent cations on the octahedral distortion and hence on the $T_c/T_m$ [47]. Considering BTO as a reference material, some of the most important factors that control the shift in $T_c/T_m$ are the ionic radii of substituent (as seen for $Zr^{4+}$) at the B-site, the coordination number, the presence of a lone electron pair (as it occurs for cations $Bi^{3+}$ or similar) at the A-site, the Jahn-Teller effect induced by substituents like $Mn^{3+}$or $Cr^{2+}$, and a cationic order within the lattice that can favour a cooperative effect and distorts the crystalline structure [49]. HoV substituents at the B-site ($Zr^{4+}$, $Sn^{4+}$, $Hf^{4+}$, $Ce^{4+}$) influence the $T_c$ more when compared to A-site ($Ca^{2+}$, $Sr^{2+}$) cations that are outside of the octahedra. The reason for this behaviour is that B-site cations have a direct effect on $\Delta z$ and so on the octahedral distortion, whereas A-site cations have an effect only on the strength of the Ti-O bond. A weaker $\pi_{Ti\text{-}O}$ bond is reflected by a smaller octahedral distortion and therefore a more subtle effect on the $T_c/T_m$ [49]. This difference is also evident from the slope of $T_c/T_m$ evolution for HoV at both A (filled circles) and B-sites (empty circles). HeV substituents both at the A-site ($La^{3+}$, $Pr^{3+}$) and B-site ($Nb^{5+}$) show a rapid decrease in the $T_c/T_m$, not only for their effect on the $\Delta z$, but also for the creation of cationic vacancies and the presence strong RFs originating from defect complexes [40, 50, 51]. Figure 6 clearly depicts the rapid decrease in the $T_c/T_m$ for HeV at the A site (filled squares) and B-site (empty squares) respectively.

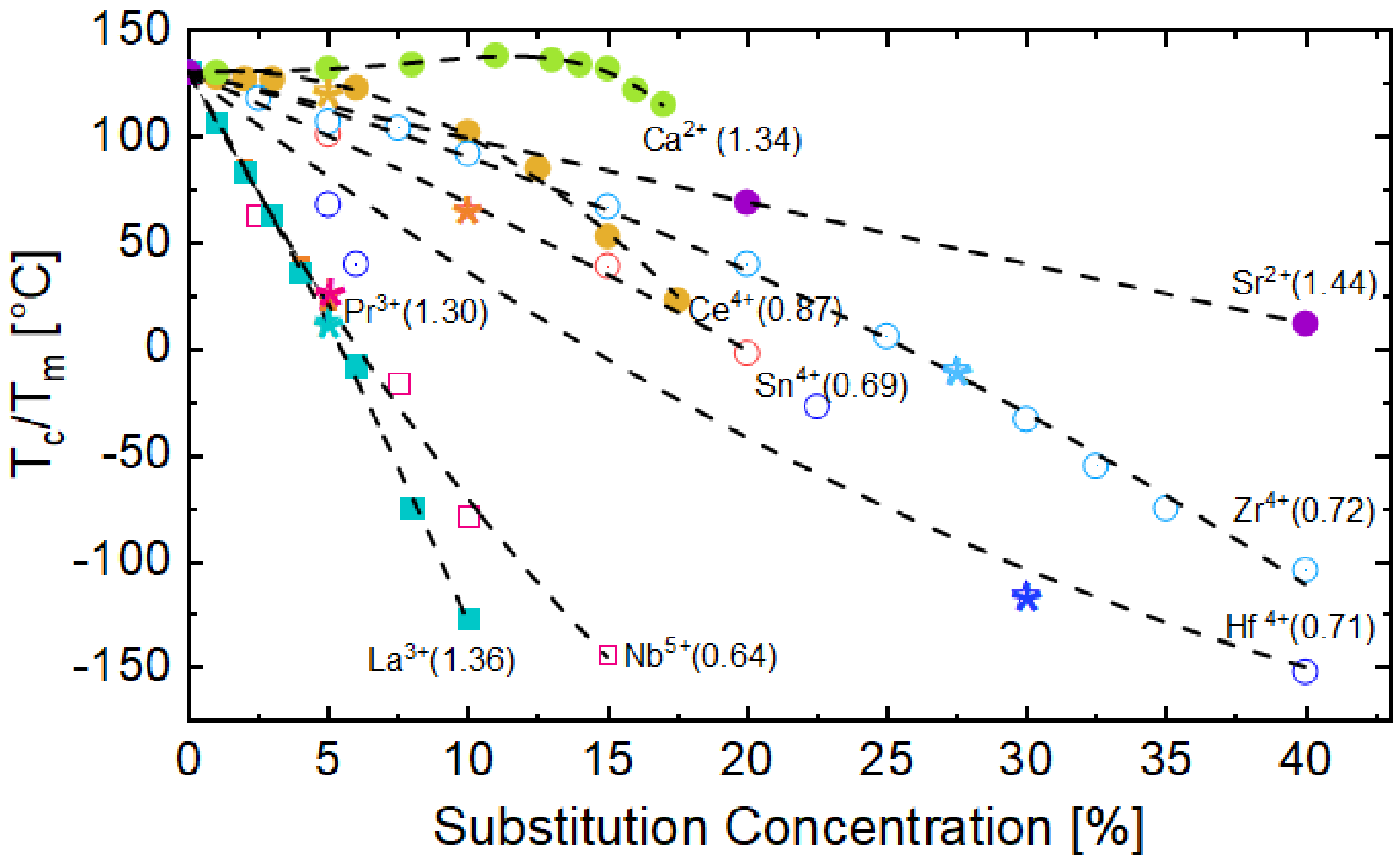

**Figure 6.** Dependence of $T_c/T_m$ on substituent percentage for A-site HoV (filled circles) and B-site HoV (empty circles), A-site HeV (filled squares) and B-site HeV (empty squares) substituted BT [40, 50–57]. In brackets, the ionic radii in Å obtained from Shannon [58]. The crossover to relaxor compositions is marked as asterisks for every substituent.

*3.2. Effect on relative permittivity:*

Another important parameter that influences the ED properties of relaxor systems is the absolute value of the relative permittivity ($\varepsilon_r$) as shown in Eqn. 1. In most BTO based systems, with the increase in substituent concentration, the maximum relative permittivity ($\varepsilon_{r\ max}$) value increases along with the increase in diffusivity of the temperature dependent dielectric response until it becomes a relaxor, at which the $\varepsilon_{r\ max}$ drops. This initial increase in the permittivity until the material becomes a relaxor is attributed to the increased number of polarization states available due to the co-existence of different phases or polar lattice entities (e.g. broken-bond fluctuations in HoV or charged defect complexes in HeV); more details on the enhancement of different dielectric properties at the crossover compositions can be found elsewhere [59]. In any case, for chemically modified FE systems, the diffusivity in the dielectric response is characterized by the *diffuseness parameter,* $\gamma$ [60]. This parameter, derived from the modified Curie-Weiss law, is used to characterize the type of phase transitions with a value ranging from 1 for a typical FE material to a maximum of 2 for a complete diffuse phase transition [61],

$$\frac{1}{\varepsilon} - \frac{1}{\varepsilon_{max}} = \frac{(T - T_m)^{\gamma}}{C'} \qquad (7)$$

Where, $C'$ is the Curie-Weiss constant and other terms as explained previously in the text. This model does not invoke the frequency dependence of relaxor systems and hence cannot be taken as a measure of relaxor behaviour.

Importantly, microstructural properties such as grain size and porosity have shown to influence the dielectric permittivity of FE systems including relaxors. It has been reported that the $\varepsilon_r$ decreases with the decrease in grain size (increasing the fraction of grain boundaries per unit volume) primarily because of the non-ferroelectric nature of the grain boundaries. Also, the diffusivity of $\varepsilon_r$ response is increased with the decrease in grain size, which is related to the suppression of latent heat of different subsequent phase transitions [21]. The porosity has also an effect on the permittivity. Pores in a ceramic material are filled with air, that possess a much lower $\varepsilon_r$, thus the permittivity decreases with the increasing of the porosity [62, 63]. These microstructural properties are mainly controlled with the choice of processing conditions and/or use of additives that will be discussed in chapter 4 and 5, although they can be modified with substituents as well. Substituents are commonly used to suppress abnormal grain growth and obtain a fine-grained microstructure. When a foreign atom is homogeneously dispersed in the material matrix, the grain boundary mobility can be hindered by the 'solute drag effect' [63, 64]. For instance: (1) $Hf^{4+}$ substituted BTO presents a smaller grain size [65], but also high porosity [52], leading to an overall reduction in $\varepsilon_r$; (2) $Nb^{5+}$ substituted BTO ceramics showed a small and uniform microstructure due to better diffusion of $Nb^{5+}$ into the BTO grains inhibiting the abnormal grain growth. Also, microstructural density was improved by introduction of $Nb^{5+}$ that ultimately led to enhanced $\varepsilon_r$ [66].

In addition, different HoV and HeV substitutions results in heterogeneities of several forms (defect complexes, random dipoles, non-polar entities etc.) in the FE lattice, which induce different types of dielectric relaxations. Understanding the defect chemistry of the material system thus becomes very critical. Being most relaxors heavily substituted systems (10 % or more of substituting atoms), a transition from electronic to ionic compensation is proven to be preferred in several BTO based systems with few exceptions. These cases typically comprise – in some BTO systems – a transition from semiconductor behaviour at low substitution concentration to insulator at higher substitution concentration. The nature of ionic vacancies (A-, B-site or oxygen vacancies) impacts the dielectric response of the material differently, thus affecting the ED properties. In fact, recent reports suggest that A- or B-site vacancies can be an effective disruptor of FE order (thereby inducing relaxor state) as well as promote ED properties if carefully designed.

One more interesting possibility with chemical modification is to produce a core-shell microstructure that can have remarkable impact on the ED properties. A core-shell by definition is when a grain has inhomogeneous chemical distribution with selective enrichment of certain elements in the core compared to the shell. This method is normally used to produce a broad permittivity response by superposition of multiple permittivity peaks because of the core-shell structure. Interestingly, Wang et al. [67] demonstrated electrical homogeneity in samples with such structures in spite of the above mentioned chemical heterogeneity that resulted in enhanced electrical resistivity in (1-x)$BiFeO_3$–0.3$BaTiO_3$-xNd($Zr_{0.5}Zn_{0.5}$)$O_3$ (BF-BT-NZZ) based relaxor systems. Most of the grains had no macro domains and there were nanodomains in the core regions. This clever engineering of domain distribution might have helped to suppress several grain boundary related energy dissipation mechanisms that ultimately led to enhanced ED properties. The same approach can be used to add an additive phase in the matrix and will be discussed in Chapter 4.

### *3.3. Effect on polarization*

In polycrystalline ceramics, the material is composed of numerous individual grains, which individually can be treated as single crystals. Each grain is constituted of multiple ferroelectric domains in which the polarization is randomly oriented [68]. The boundary between two domains is called a domain wall. When an electric field is applied to the material, the electric polarization within each domain will align parallel to that of the applied field and so the domain walls will move to a new position. This continues to happen until polarization saturates ($P_s$) at a sufficiently high electric field ($E$). In single crystals, after the removal of the electric field the polarization value decreases only slightly, that is, the remanent polarization $P_r$ has a value close to $P_s$. On the other hand, in relaxors the polarization decreases until $P_r$ approaches values close to zero at E = 0 [69]. Hence, to achieve high ED properties, the difference between $P_s$ and $P_r$ ($P_s$-$P_r$) has to be maximized [70] by minimizing the energy dissipations (dielectric losses) in material (c.f. Figure 2) – in other words obtaining a slim hysteresis loop. The main contributors to dielectric losses are extrinsic: polarization rotation and domain wall movements. These effects are extensively explained in a review by Liu et al. [68]. Among the methods to maximize the $P_s$-$P_r$ are controlling the grain size, the addition of secondary phases and most commonly, as mentioned previously, by disrupting FE order using chemical substitution. A nominal FE material like BTO (c.f. Figure 2) possesses a small $P_s$-$P_r$ value, making it not suitable for energy storage applications because of the resulting low $J_r$. Chemical substitutions in the lattice introduce local disorder and random fields that disrupt the long-range polar order, so that no well-developed FE domains exist but only short-range localized polar entities [70]. Moreover, HeV substitutions can also introduce defects such as vacancies as charge compensation scheme that restrict the domain wall movement by "pinning effect" resulting in lower mechanical losses upon field reversal. All these produce a lower $P_r$ and a slimmer P-E loop (c.f. Figure 2). The same "pinning" effect can be obtained with smaller grains due to the stabilization of domain walls by grain boundaries. To summarize, any chemical modification that influences the degree of lattice disorder, will influence $P_r$ as well and so the ED properties.

### *3.4. Effect on breakdown strength*

In addition to the requirement of a large $P_s$-$P_r$ value, a high BDS is essential to realize high performance dielectric based EESSs. The BDS of a material is the maximum electric field that can be applied on a sample of given thickness before catastrophic electrical breakdown (i.e. a disruptive discharge) occurs. Since for improving the ED properties a large $P_s$ has to be achieved, the relaxor materials will in general have to withstand high electric fields to reach the $P_s$ value. The BDS is directly linked to the ED properties by Eqn. 2 and is strongly influenced by both microstructure and the band gap of the material. In general, smaller grains, high density, phase purity and wide band gap are critical in determining the BDS. It has been already explained how substituents can restrict abnormal grain growth, by reducing the boundary migration, therefore ensuring a uniform dense microstructure [64]. This enhances ED properties in many relaxor systems [71]. More details on the dependence of BDS on the microstructure will be covered in the subsequent chapters. The electrical breakdown of a material can also occur via avalanche breakdown and electronic breakdown as explained by Seitz [72] and Fröhlich [73] in two comprehensive works. Electronic breakdown occurs when sufficiently high electric field is applied to a ferroelectric material for electrons in the valence band to cross the energy gap and accumulate in the conduction band, ultimately leading to breakdown by continued field application [63]. This is connected to the band gap and therefore can be tuned by substituents. For instance, Zhao et al. showed that Ta incorporation for Nb in $BaTiO_3$-$Bi(Zn_{2/3}Nb_{1/3})O_3$ improves the BDS due to the wider band of $Ta_2O_5$ compared to $Nb_2O_5$ [74].

## 4. Tuning Energy density by chemical additives

Chemical additives are intended as chemical modifications that do not modify the crystalline lattice (i.e. differently from the case treated in Chapter 3), but are willingly dispersed as a second phase in the ceramic matrix along the grain boundaries in the sample microstructure (c.f. Figure 5b). This should not be confused with the often-used terminology, '*secondary phase*', that is neither foreseen in material fabrication nor favourable to material properties (c.f. Figure 5c). Usually, additives are mixed with the starting ceramic powders and are distributed uniformly within the microstructure of bulk ceramics and thick film multilayer architectures after sintering [75]. Figure 5 schematically differentiates substitutions, additives and secondary phases that are commonly found in the field of ceramic science. Additives mainly facilitate lowering of the sintering temperature, increase density ($\rho$), increase the BDS, etc. The use of additives such as glasses, metal oxides and polymers have shown to achieve homogenous microstructure by reducing the grain growth. The additive particles, when mixed in the ceramic matrix phase, due to the aforementioned solute drag effect get distributed preferably along the material grains during the sintering process and limit the grain boundary mobility and grain growth. This effect is common when additives or chemical modifiers are involved [64]. In addition to the refinement of the microstructure, additives can overall improve BDS in relaxor systems, while retaining the intrinsic properties of the matrix phase such as relatively large $\varepsilon_r$ and $P_s$ [76–78].

### *4.1. Glass additives*

Although glass shows lower $\varepsilon_r$ [76] compared to ceramics, ceramic-glass composites have been studied since the 1950s to benefit from their superior BDS properties [79]. During ceramic fabrication, glass powders are mixed with the ceramic starting powders until they are uniformly dispersed and then formulated to various needs. During sintering, the glass phase additive, due to its lower melting temperature, forms a liquid phase in between the solid particles, promoting the dissolution-precipitation process for enhanced sintering activity (c.f. Figure 5b). This contributes to the reduction of the sintering temperature [80], an increase in $\rho$ [81] and a refined grain size [82]. An ideal glass phase additive should possess the following: (1) low melting temperature to reduce the sintering temperature and to limit grain growth, (2) low reactivity with the solid phase to avoid the formation of secondary phases, (3) low viscosity to promote mobility for easy redistribution around the matrix phase grains and (4) relatively high $\varepsilon_r$ [81].

The borosilicate based glasses such as $B_2O_3$-$SiO_2$ are commonly used in connection with BTO-based relaxors due to their high BDS and good wettability with BTO-based powders [78, 83]. For example, Wang et al. studied the effect of BaO-$B_2O_3$-$SiO_2$-$Na_2CO_3$-$K_2CO_3$ glass content on the dielectric properties of $Ba_{0.4}Sr_{0.6}TiO_3$ (BST). The chemistry of glass content strongly influenced the dielectric properties of BST ceramics with a maximum achievable $J_r$ of 0.72 J/cm$^3$ and a BDS of 280.5 kV/cm that is substantially better than its ceramic counterparts [84]. Yang et al. showed that the $J_s$ of $Ba_{0.85}Ca_{0.15}Zr_{0.1}Ti_{0.9}O_3$ ceramic can be tuned from 0.205 J/cm$^3$ to 1.15 J/cm$^3$ when an optimized 5 wt % of $B_2O_3$-$Al_2O_3$-$SiO_2$ glass additive is used [85]. Recently, Yang et al. incorporated $Bi_2O_3$, which is not a glass former, in a conventional $B_2O_3$-$SiO_2$ glass composition, and then added the mixture as an additive to enhance ED properties of BST ceramics because of the high polarizability of the $Bi^{3+}$ ions. At an ideal glass additive concentration, a maximum BDS of 279 kV/cm and $J_r$ of 1.98 J/cm$^3$ was realized [83]. Only very few examples of the vastly available literature are highlighted here and most glass additives are based on $B_2O_3$–$SiO_2$, BaO–$B_2O_3$–$SiO_2$, $Bi_2O_3$–$SiO_2$–$B_2O_3$–ZnO and BaO–$Bi_2O_3$–$B_2O_3$ systems [86]. With glass additives, the ratio of ceramic powder to glass concentration has to be carefully evaluated for the following reasons:

- Although the BDS is improved with glass addition, the $\varepsilon_r$ of the composite is reduced with the increase in percentage of glass additive due to its lower $\varepsilon_r$ [87, 88].
- During sintering, excessive liquid phase formation with higher glass content could promote grain growth, and increase microstructural defects such as porosity, causing the so-called "de-sintering" phenomenon, ultimately degrading the electrical properties of the system [81].

### *4.2. Inorganic additives*

Metal oxides such as $SiO_2$, MgO and ZnO and are often employed as sintering additives to enhance densification and refine the microstructure of ceramics. Just like glass additives, inorganic additives with a high BDS and low dielectric loss are mixed with the ceramic powders and sintered to investigate its effects on the overall ED properties of the system. It is important to note that in contrast to glass additives, ionic species from metal oxides can often diffuse into the lattice causing the formation of secondary phases by chemical reactions [89] or substitute a cation in the lattice resulting in unwanted chemical modification [90].

*Silica ($SiO_2$)* is a common additive employed in relaxor ceramics [89]. In addition to limiting the grain growth, $SiO_2$ particles with small $\varepsilon_r$ were found to experience a high local electric field compared to the total applied field. These field localizations combined with excellent insulating properties of $SiO_2$ facilitate the enhancement of ED properties of ceramics [91]. Diao et al. showed that the $SiO_2$ additive can be a cost-effective way to tune ED properties by demonstrating a $J_r$ and BDS of 0.86 J/cm$^3$ and 134 kV/cm in BST ceramics, which is slightly better than conventional counterparts [92]. On the other hand, Zhang et al. demonstrated that a core-shell approach can be more beneficial to the ED properties in BTO systems. The core-shell structure is generally achieved by chemical modification, which was already discussed previously, but here it was realized by coating the matrix phase particles with an additive phase. This is different from traditional additives approach where additives are dispersed uniformly within the starting powders and sintered. In the case of coating, a chemical route is adopted where tetraethoxysilane is added to BTO nanoparticles, followed by ammonia assisted hydrolysis in alcoholic media. The technique offers the possibility to even control the thickness of the coating layer and the effective dispersion of additive phase. BTO with 2 wt % $SiO_2$ reported a $J_r$ of 1.2 J/cm$^3$ and a BDS of 201.8 kV/cm that is substantially superior to that of pure BTO ceramics [89]. A similar approach was used on BST powders synthesized by the sol-gel method and a mere 8 wt% of $SiO_2$ coating resulted in a $J_r$ of 1.6 J/cm$^3$, BDS of 400 kV/cm, and a higher $\eta$ compared to pure BST ceramics [93].

*Magnesia (MgO)* is another important inorganic additive in BTO based composites especially for its high BDS (~1000 kV/cm) in spite of its low $\varepsilon_r$ (~10). Zhang et al. showed that MgO as an additive is also effective in achieving a uniform and small grained microstructure, thereby ensuring a superior BDS. A maximum BDS and $J_s$ of 330 kV/cm and 1.14 J/cm$^3$, respectively, was achieved on BST-MgO composites [94]. One of the main drawbacks of MgO based composites is the reactivity of $Mg^{2+}$ ions that often produce chemical substitution or trigger chemical reaction to form a secondary phase. This is indeed a major problem for all inorganic additives, if not chosen wisely. This reactivity of MgO was limited using a rapid spark plasma sintering approach by Huang et al., resulting in synthetized BST-MgO composites with a significantly enhanced BDS and $J_r$ of 330 kV/cm and 1.49 J/cm$^3$, respectively. Here it is interesting to note that although the $P_s$ of BST-MgO composites decreased from 14.20 μC/cm$^2$ to 11.50 μC/cm$^2$ with the increase in MgO concentration from 0 to 10 wt %, the overall $J_r$ still increased from 1.20 J/cm$^3$ to 1.49 J/cm$^3$ [95]. This again suggests that optimizing the concentration of additives can be very critical. Similar work using SPS to limit the $Mg^{2+}$ reactivity and fabricate $BaTi_{0.85}Sn_{0.15}O_3$/MgO composites was done by Ren et al. Just 10 wt % of MgO resulted in a $J_s$, BDS and $\eta$ of 190 kV/cm, 0.51 J/cm$^3$ and 93%, respectively [96].

*Zinc Oxide (ZnO)* can also be employed as an additive phase to enhance ED properties in BTO based systems for its role in enhancing dielectric properties and also as a sintering additive [97]. Dong et al. showed that $Ba_{.3}Sr_{0.7}TiO_3$ ceramics with 1.6 wt% of ZnO additive resulted in a $J_s$ of 3.9 J/cm$^3$ at an applied electric field of 400 kV/cm. The ED properties improved with ZnO concentration, peaking at 1.6 wt% of ZnO, followed by gradual degradation again when compared to $J_s$ and BDS of 2.2 J/cm$^3$ and 340 kV/cm, respectively, for pure BST [90]. Similarly, Yao et al. showed that 1.0% of ZnO additives in $Na_{0.5}Bi_{0.5}TiO_3$-$BaTiO_3$-$NaNbO_3$ relaxor ceramics can achieve a $J_r$ of 1.27 J/cm$^3$ and an electric field endurance as high as 100 kV/cm, in addition to great temperature stability in the ED properties [98]. Tao et al. later showed that in $Bi_{0.5}Na_{0.5}TiO_3$-$BaTiO_3$-$K_{0.5}Na_{0.5}NbO_3$ (BNT-BT-KNN) the $\varepsilon_r$ response can be effectively tuned by utilising competing local electric field induced by matrix phase and ZnO based polar entities. BDS increased from 750 kV/cm for pure BNT-BT-KNN to 1230 kV/cm for 40% ZnO addition [99].

In summary, inorganic additives ensure a fine and dense microstructure and decrease the sintering temperature, ensuring a uniform and dense microstructure. In contrary to glass additives, inorganic additives can also act as a substituent by diffusing into the material lattice making it difficult to control the chemistry of the matrix phase. An appropriate amount and type of inorganic additive has shown to be beneficial to the ED properties of relaxors, above which undesirable secondary phases are inevitable.

### *4.3. Polymer additives*

Polymers as a chemical additive have attracted a lot of attention due to their high BDS, flexibility and superior mechanical properties. In spite of their very low $\varepsilon_r$, giant BDS has made polymers a suitable chemical additive to improve ED properties in relaxor ceramics [100]. The integration of a high $\varepsilon_r$ ferroelectric ceramic matrix phase and polymer additive is beneficial to achieve high BDS while retaining the advantages of ferroelectric matrix such as high $\varepsilon_r$ and $P_s$. Among different polymers, ferroelectric polymers possess a high $\varepsilon_r$ and therefore are the most suitable for energy storage applications. One such polymer is Poly(vinylidene fluoride) (PVDF) and its co-polymers that are widely used in capacitor applications due to the large electronegativity. This is mainly created by the presence of fluorine and hydrogen atoms in the polymeric chain leading to the formation of numerous dipoles that contribute to a high $\varepsilon_r$ [101]. PVDF is a nontoxic material and it exists in four crystalline forms: $\alpha$, $\beta$, $\gamma$ and $\delta$. In the $\beta$ form, the chain formation presents the highest dipole moment resulting in high piezo and ferroelectric properties and therefore is the most suitable for energy storage applications [102].

$Ba_{0.95}Ca_{0.05}Zr_{0.15}Ti_{0.85}O_3$ (BCZT)-PVDF flexible composites were successfully fabricated by Luo et al. with BCZT percentages ranging from 6 vol% to 61 vol%. The BDS of the composites decreased drastically with the decrease in PVDF concentration since the ceramic matrix phase is increasingly deciding the overall performance. At 61 vol% of BCZT, a BDS of ~680 kV/cm and a $J_s$ of 2.0 J/cm$^3$ was achieved [100]. In a following study by Luo et al., BTO-PVDF based composites were produced using a more environmentally friendly procedure for treating the BTO surface. This surface treatment is essential to achieve homogenous dispersion of ceramic and polymer phase. The BDS of the samples showed a decrease with increasing amount of BTO powder from 3300 kV/cm for 20 vol % BTO to 1870 kV/cm for 50 vol% BTO concentration. The maximum $J_s$ of 8.13 J/cm$^3$ was achieved at an optimum BTO vol% of 20 while the efficiency remained relatively low like any other polymer based ceramic composite [103]. The use of conventional sintering methods to fabricate polymer-ceramic composites is not feasible due to the low melting temperature of the polymeric matrix. Alternatively, the polymer solution and ceramic powders are often mixed and cast.

Generally, the BDS decreases with increase in ceramic phase concentration and so the vol% of polymer phase is relatively high (as high as 90 vol%) unlike other additive phases. Achieving a homogeneous dispersion of the ceramic powder in the polymeric matrix is vital as agglomeration resulting from difference in the surface energy of ceramic and polymer phase can lead to accumulation of defects and inevitably lower the BDS and ED performance of the composite. Therefore, surface compatibility between ceramic powder and polymeric additives has been extensively studied and various modification to ceramic phase morphology [104] and surface activation [105] are employed to ensure sufficient chemical homogeneity in the polymer/ceramic composite.

## 5. Tuning energy density by processing methods

Superior ED performance can be effectively obtained also by controlling the microstructural properties of materials, including grain size, microstructural defects (i.e., porosity) and density. Although tuning $\varepsilon_r$, $T_c$ and $P$ ($P_s$ and $P_r$) by chemical modification has profound effects on the material properties to promote ED properties, novel processing methods ensure that a material system with (theoretically) high ED will have a favourable microstructure to demonstrate its high ED properties in reality. This is undoubtedly true when it comes to tuning BDS, which is one of the most vital parameters (c.f. Eqn. 1) influencing the ED. Importantly, advancements in processing routes are essential in upscaling and commercialization of certain technologies to make them available in the commercial market. This chapter is only dedicated to novel processing methods of different relaxor systems in different available forms (bulk ceramics, multilayer thick and thin films) that mainly ensure tight control on microstructural properties to produce reliable high ED materials. Please note that this section is only an introduction to some of the most recent and impactful processing methods to fabricate high ED relaxor systems and is definitely not a complete guide.

### *5.1. Bulk ceramics*

Historically, some of the first relaxor systems were fabricated by the conventional solid-state sintering route. This fabrication procedure, from now on abbreviated as 'CS', includes powder compaction usually by uniaxial (or isostatic) pressure followed by sintering at temperatures below the melting point of the material, which has the ultimate goal of achieving a dense material with a favourable microstructure. During sintering, several changes such as grain growth, change in pore sizes, pore density etc. occur as a consequence of simultaneous densifying and non-densifying diffusion mechanisms [106]. After sintering, microstructural defects ascribed to sintering, organics evaporation, or poor compaction might still be present in the material. For this reason, although ceramic dielectrics including relaxors are expected to have excellent ED properties because of their large BDS, those defects largely undermine their performance in reality. Some of the defects in bulk ceramics can include pores, impurities (i.e., conducting particles), agglomerates, cracks, secondary phases etc. These defects in ceramics under a large applied electric field can act as 'field intensification regions', where the applied field can vary largely compared to rest of the sample leading to accelerated local degradation followed by catastrophic electrical breakdown [107]. In later stages, advancements in multilayer ceramic fabrication technology helped to overcome some of these problems shifting the research and development activities to thick films with controlled microstructure, as will be discussed in detail in the subsequent section.

Recently, there are renewed interests in studying the ED properties of bulk ceramics especially due to innovations in fabrication routes, like sintering approaches that made high density ceramics (with less microstructural defects) more realizable than ever, to benefit from the true BDS of material systems. Also, recent advancements in rapid sintering approaches [108], with external stimuli such as pressure, electric field, plasma etc. are efficient in retaining small grain size and ensure high density to achieve high BDS. In spite of advancements in both thick and thin film technology, bulk ceramics with high ED properties are still valuable in certain applications for the ease in industrial upscaling and cost-effectiveness. In addition, demonstrating high ED properties on bulk ceramics has become critical in the advent of searching for new material compositions with superior properties that can later be adopted to thick or thin film technologies. In the subsequent discussion, some of the most successful fabrication approaches of bulk relaxor systems for EESSs will be summarized without considering the downsides of bulk ceramics in comparison with thick or thin films. When stating some novel fabrication approaches, process parameters such as the sintering temperature, pressure and time will be reported in addition to the relevant properties such as relative density, grain size (if required) and the corresponding ED for the discussed relaxor composition. The compositions included in this section are not based on any particular interest but just on the availability of studies in their respective fabrication methods as well as by CS method for comparison. Relatively simpler compositions are preferred if available.

a) Role of temperature

In general, sintering is an essential step in any ceramics to develop a favourable microstructure. As a consequence of different diffusion and mass transport, densification and coarsening occur during sintering. Grain coarsening occurs normally in the final stage of the process to retain energetic equilibrium between grain boundary and the rest of the bulk material. Grain growth hinders the densification and can be represented by the following equation [109],

$$\frac{d\rho}{\rho\, dt} \propto \left(\frac{\delta\ D}{G^3}\right) \quad (8)$$

Here, the term on the left side is the densification rate where $\rho$ is density, $t$ is time, $\delta$ is the grain boundary thickness, $D$ is grain boundary diffusivity and $G$ is grain size. This equation clearly depicts the inverse relationship of densification rate and $G$. A common approach to attain a dense microstructure is to follow a longer sintering time making the abnormal grain growth inevitable, which is also explained in the above equation [64]. In such a scenario, a very interesting rate controlled sintering approach was proposed in 2000 by Chen et al., where grain growth is limited by taking advantage of the difference in kinetics between the densification and grain growth effectively by freezing the grain network [110, 111]. In this approach, a critical density is achieved by rapidly heating the material to a high temperature (first stage-$T_1$) and then cooling to relatively lower temperatures (second stage-$T_2$) where it is held for a certain time like in the CS. This novel approach of achieving fine microstructure just by tweaking the sintering temperature is called two-step sintering (TSS). In this method, densification occurs during the second stage without grain growth as long as the critical density is achieved in the first stage of the sintering. This approach is the simplest method that can be adopted by industries in fabricating high ED bulk ceramics as well as thick film Multilayer Ceramic Capacitors, MLCCs (which will be discussed in the subsequent section). Wang et al. fabricated $Ba_{0.94}(Bi_{0.5}K_{0.5})_{0.06}Ti_{0.85}Zr_{0.15}O_3$ based relaxor systems that are rapidly heated to 1500°C ($T_1$) and held at a different $T_2$ to study the effect of TSS in the ED properties [112]. The relative density increased with decrease in $T_2$ up to a certain cut-off $T_2$ below which the temperature might be too low to activate any densifying mechanisms. Obviously, this observation was in direct correspondence with the electric field endurance of the samples and the ED properties. With the decrease in $T_2$ from 1400°C to 1250°C, $J_r$ increased from 0.33 to 0.95 J/cm$^3$. This wide range of achievable $J_r$ just by changing the $T_2$ shows how grain growth mechanisms can be tightly manipulated by TSS to have positive impact on the ED properties of relaxor systems. A similar work was done on $0.89Bi_{0.5}Na_{0.5}TiO_3$–$0.06BaTiO_3$–$0.05K_{0.5}Na_{0.5}NbO_3$ FE system to show that the $J_r$ was substantially improved by TSS [113] compared to CS [114]. This technique has demonstrated the crucial role of temperature in achieving favourable material properties in any ceramic material.

b) Role of Pressure

From this point onwards, sintering variants that use an external stimulus in addition to heat to modify ED properties in relaxor systems will be discussed. One possibility is to use pressure-assisted sintering techniques such as pressure sintering or hot pressing (HP). Hot pressing is an effective way to limit grain growth while promoting densification as a result of high uniaxial pressure in a constrained geometry [115]. From a microstructural viewpoint, applied pressure promotes densification through mechanisms such as lattice diffusion, grain boundary diffusion and grain sliding, and can be described by the following power-law creep ($\varepsilon$) equation,

$$\varepsilon \propto \left(\frac{b}{G}\right)^{m} \left(\frac{\sigma}{g}\right)^{n} \quad (9)$$

Where $b$ is the Burgers vector, $\sigma$ is the applied stress, $g$ the shear modulus, and $G$ again is the grain size. It is important to note that the exponent to stress intensification factor ($n$) is close to unity while the grain size exponent ($m$) dominates the densification process in a nominal hot-pressing method in BTO based ceramics that excludes plastic deformation of grain boundary as possible promotor of densification [116]. In any case, it is important to note that pressure will be an added but benefit-bringing parameter on the top of an already complex sintering process with parameters such as sintering temperature, heating rate and time in all variants of pressure assisted sintering. Again, this technique satisfies the ultimate objective to attain high microstructural density with smaller grains, which is a requirement for superior ED properties. HP strategy was used to demonstrate improved ED properties of $(Bi_{0.5}K_{0.5})TiO_3$-$0.06La(Mg_{0.5}Ti_{0.5})O_3$ ceramics [117]. The $J_r$ values obtained on ceramics fabricated by HP is 2.08 J/cm$^3$ compared to 0.96 J/cm$^3$ by CS. Interestingly, HP is also reported to be advantageous to consolidate volatile compounds without the problem of secondary phase resulting from volatilization and stochiometric imbalance that can deteriorate material properties. Since we are discussing the importance of smaller grain size to achieve high density, please note that ferroelectricity is progressively diluted with increasing grain boundary density (as a result of decreasing G), because of the non-ferroelectric nature of the grain boundaries [21]. For the same reason, the $\varepsilon_r$ values of hot-pressed ceramics can be lower than that of ceramics sintered by CS, but by Eqn. 1 this is counteracted by the increased BDS as a result of higher density with uniform grain size distribution, which ultimately improves the ED properties [117, 118]. A variation of HP is a 'rapid hot-pressing', in which powder compacts are sintered at a very fast heating rate (hundreds of °C/min) at a high pressure (several hundreds of MPa). To summarize, these pressure-assisted sintering techniques are excellent to promote ED properties, are time efficient and can have better control over the stoichiometry of volatile compositions [119].

c) Role of electric field

Field-assisted sintering refers to densification occurring also as a result of applied small electric pulses leading to internal heating (Joule heating). When used in combination with high uniaxial pressure, further reduction of the temperature requirements and the obtainment of a high-density nano-structured ceramics in a very short time was demonstrated. This technique is effectively a combination of HP and field-assisted sintering, and is called 'Spark Plasma Sintering-SPS', although no evidence of presence of plasma was reported so far. In this technique, the ceramic powders are loaded in a conductive (usually graphite) die under vacuum conditions and simultaneous pulsed DC current and pressure are applied to consolidate the powders. A conductive die is essential for non-conductive samples like ceramics, where the electric pulses first flow through the die resulting in its Joule heating and subsequently heating the sample itself. By controlling the electric pulses, extreme heating rates of up to 1000 °C/min can be realized, allowing rapid densification of ceramics [120]. All the above stated novelties of SPS facilitate possibilities to retain the nanocrystallinity of grains in fully dense relaxor systems, which is profitable to ED properties. The densification mechanism of 'SPS' for non-conducting ceramic samples are reported to be same as that of pressure assisted sintering except the source of heat during sintering [121]. But, due to rapid sintering nature of SPS, accumulation of charged defects along the grain boundaries limits the grain growth by restricting grain boundary mobility. This defect induced 'pinning mechanism' was already discussed in the earlier section about chemical modification.

Comprehensive reviews on field assisted sintering techniques and their advancements can be found elsewhere in the literature [108, 120, 122]. To utilize the advantages of SPS and to retain the grain size in nanometre range, SPS was performed on $Ba_{0.4}Sr_{0.6}TiO_3$ powders synthesized by sol-gel method. A high-density ceramic was fabricated at 1000 °C in just 5 minutes under vacuum conditions. The samples showed relatively high $J_r$ of 1.23 J/cm$^3$ with a very high efficiency compared to just 0.37 J/cm$^3$ for CS counterparts [123]. In another work, barium zirconate titanate ($BaTi_{0.7}Zr_{0.3}O_3$) ceramics fabricated by SPS and CS are compared for their ED properties. SPS samples showed a very high BDS and $J_r$ of 170 kV/cm and 0.51 J/cm$^3$ compared to just 40 kV/cm and 0.12 J/cm$^3$ for CS, respectively. From finite element analysis investigation, it is shown that a high BDS in SPS samples is a consequence of a uniform electric field distribution resulting from small grain sizes and narrow grain size distribution. On the other hand, the CS sample showed great inhomogeneity in the electric field distribution resulting in 'local field intensification' and leading to an early 'electrical breakdown' [71].

Figure 7 shows the superior energy density performance of bulk ceramics produced by novel sintering approaches compared to conventional sintering.

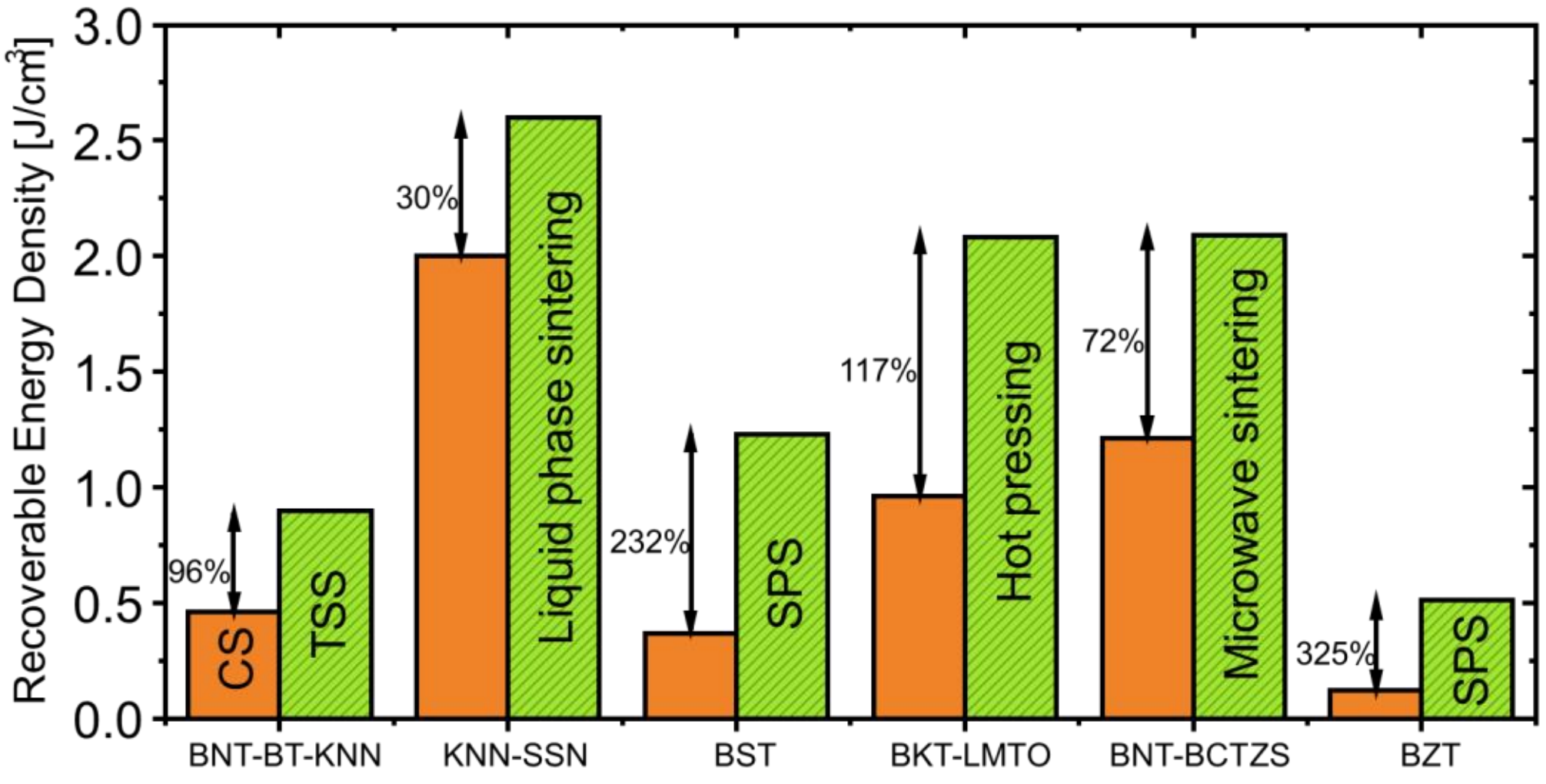

**Figure 7.** Superior energy density performance of bulk ceramics produced by novel sintering approaches compared to conventional sintering (CS) [71, 113, 117, 123–125].

### *5.2. Multilayer thick films*

In general, research efforts on designing EESSs using guidelines from Chapter 3 are mainly demonstrated on bulk ceramics, and designing multilayer architectures is an extension to further enhance the functionality of material systems. However, due to a very different fabrication approach, not all the concepts on enhancement of ED properties by novel processing routes on bulk ceramics can be translated into multilayer technology except some of the guiding principles of effects of $\rho$, grain size ($G$) etc on ED properties (there are exceptions such as two-step sintering which will be discussed later in this section [110]). Multilayer architectures are attractive for designing EESSs for the following fundamental reasons:

- Capacitance and available area ($A$) of the dielectric material

$$C = \frac{\varepsilon_o \varepsilon_r \mathrm{A}}{d} \tag{10}$$

Here, $d$ is the thickness of the dielectric layer. From the above equation, it is clear that capacitance is impacted positively by increasing $A$ and decreasing $d$. The dielectric thickness, $d$, cannot be lower than the grain size ($G$) and, with reducing G, progressive dilution of $\varepsilon_r$ response happens [126], also discussed in the previous chapter.

- Reduction in voltage requirements

$$V = E * d \tag{11}$$

With the reducing $d$, the voltage requirements are kept low, which is advantageous for relaxor systems where a very high E is required to induce high ED properties. This is also a requirement to make EESSs safe for consumer technologies.

- BDS and sample thickness

$$BDS \propto d^{-n} \tag{12}$$

The BDS is primarily controlled by presence of defects and Weibull statistics states that the probability of occurrence of defects increases with increase in material volume [127]. Wang et al. experimentally presented this dependence by fabricating a relaxor based multilayer architecture, where the BDS increased from 511 to 1047 kV/cm with decrease in $d$ from 26 to 5 μm [128]. In general, a material system that demonstrates high ED properties in the bulk form fabricated by CS can then be adopted to fabricate multilayer architectures to make use of all the above stated advantages to fabricate high performance EESSs, as shown in Figure 8. It is clear that $J_r$ of the same material can be substantially amplified using multilayer technology compared to bulk ceramics fabricated by CS.

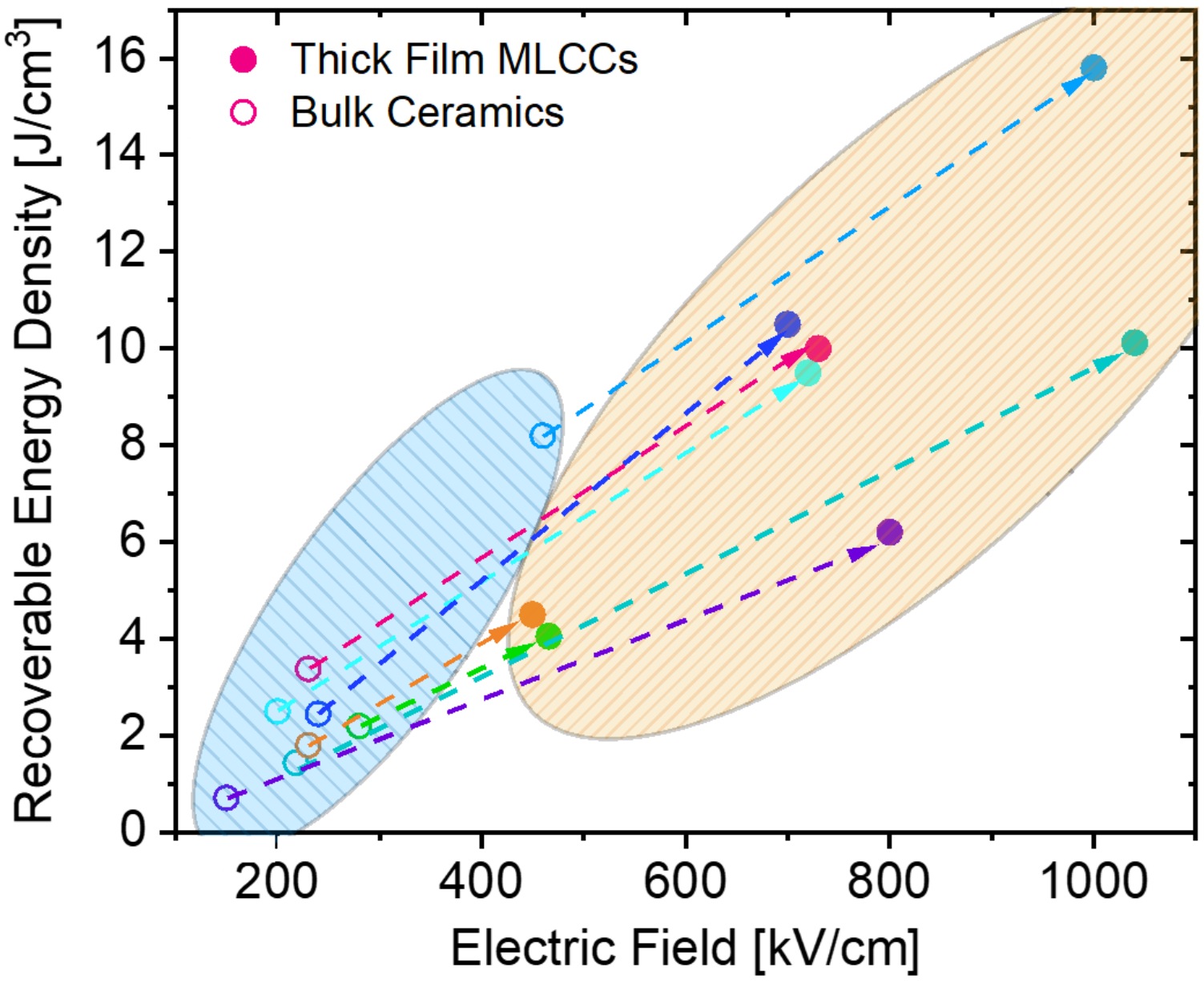

**Figure 8.** Recoverable energy density of several relaxor based systems in bulk form compared against multilayer architectures [67, 74, 129–135]. Dashed arrows are merely guides to the eye.

In spite of a dilution in the $\varepsilon_r$ response with reducing $d$, the volumetric dielectric efficiency is uplifted to certain limit because of increasing $A$ and BDS. Below this limit, special considerations to powder synthesis and thick film forming procedures have to be taken and also tackle ways to overcome handling difficulties of fine powders and ultra-thin sheets. Please note that this section is dedicated only to multilayer 'thick-films', with film thickness of more than 1 μm. 'Thin film' based multilayer ceramic capacitors (MLCCs) and strategies associated with it to enhance ED properties will be discussed in the next chapter. The most successful forming procedure of relaxor based MLCCs for EESSs is tape casting in addition to other less commonly studied techniques such as ink-jet printing, screen printing, electrophoretic deposition, dip coating etc. A detailed report on each of this technique can be found elsewhere [136]. A detailed description of each steps followed in fabrication of MLCCs is outside the scope of this review and we refer to Pan et. al. for details [137]. Figure 9 shows a standard multilayer thick film architecture for capacitor applications (MLCC).

In addition to the above discussions that primarily explained the need for multilayer architectures, this chapter will continue briefing some unique strategies that are successfully used to enhance ED properties by thick film multilayer processing.

a) Bulk ceramics to MLCCs:

The most common strategy to tune ED properties is by compositional tuning as discussed in Chapter 3 and this is applicable to MLCCs as well. In general, when a material with particular chemistry demonstrates superior ED properties in bulk form, the MLCC fabrication procedure can be adopted to further amplify the ED properties because of the previously discussed advantages of multilayer architectures. One such example is the bismuth titanate-strontium titanate solid solutions (BT-ST) which was previously investigated as bulk ceramics for their attractive electrical and magnetic properties [138, 139]. Lu et. al. carefully adopted the composition and tuned it to reduce the charge carriers and demonstrated its effect on the ED properties [135]. The loss of volatile $Bi_2O_3$ causes oxygen vacancies which is suppressed by introducing $Nb^{5+}$ substitution at the B-site. A mere 3% Nb substitution showed a dramatic increase in resistivity and electrical homogeneity of the ceramics. They went further to optimize $Nb^{5+}$ concentration in a BT-ST based composition for maximum electrical resistivity and electrical homogeneity to ensure high BDS and recorded a maximum $J_r$ of 8.2 J/cm$^3$ in bulk ceramics. The optimized composition was then adopted to MLCCs fabrication procedure to record a $J_r$ of 15.8 J/cm$^3$ and BDS of 1000 kVcm$^{-1}$ for a $d$ of 8 μm. This work is one of the classic examples that show a strategic use of guidelines from Chapter 3 producing bulk ceramics to demonstrate the high performance of the composition and then fabricate MLCCs to show the real potential of the material system. Similarly, chemical modification was carried out to break long-range order in BTO by $Zr^{4+}$ substitution at the B-site to attain a relaxor system (more details in chapter 3) and then the MLCCs of the same composition was fabricated with a dielectric layer thickness of 20 μm and recorded a $J_r$ value of 6.2 J/cm$^3$ which is at least three times larger than that of its bulk counterpart. In this case, considering the previous discussions on impact of $d$ on BDS, the ED properties can further be amplified by decreasing the layer thickness.

There are very few overlaps in the fabrication procedures of bulk ceramics and MLCCs as mentioned previously at various instances. One is the two-step sintering (TSS) that was discussed in chapter 5.1 and is easily adopted to MLCC fabrication. Advantages of using TSS for MLCCs are:

- The sintering temperature in TSS can be substantially lower than the conventional sintering which makes it beneficial to co-fire electrodes that are cheap and have lower melting temperatures.
- The G is low in TSS, making it feasible to reduce $d$ of dielectric layers that can significantly reduce the defect concentration and increase BDS.

Zhao et al. shown that $0.87BaTiO_3$-$0.13Bi(Zn_{2/3}(Nb_{0.85}Ta_{0.15})_{1/3})O_3$ based MLCCs fabricated by TSS have a $J_r$ of 10.12 J/cm$^3$ at 1047 kV/cm [74]. An extension of this work by Cai et al. demonstrated possibilities to tune the ED properties by controlling the heating rate of TSS. Increasing the heating rate from 4 °C/min to 40 °C/min for $T_1$ in TSS substantially improved the quality of the interface by reducing the occurrence of defects and ensured superior bonding between electrodes and the dielectric layer [140]. Further, a finite element method was employed to calculate the electric field distribution in the microstructures with and without pores at the interfaces. It was clear that defects such as pores act as 'field intensification regions' and trigger an electrical breakdown at substantially low applied electric field.

b) Composite multilayer architectures

One of the main drawbacks of polycrystalline relaxor systems for ED applications is the large electric field requirements that gives rise to large electrostrictive strains. These strains can cause micro-cracks, which is one of the primary reasons for a lower BDS in spite of reducing the defect concentration in MLCCs. Li et al. proposed and demonstrated an approach to reduce such electric-field induced strains by grain orientation engineering.

A $(Sr_{0.7}Bi_{0.2})TiO_3$-$(Na_{0.5}Bi_{0.5})TiO_3$ (SBT-NBT) composition that showed a $J_r$ value of 10 J/cm$^3$ in bulk form was selected to engineer the grain orientation in multilayer form to validate this approach. $SrTiO_3$ templates with high aspect ratio were synthesized and mixed with SBT-NBT to form films with 20 μm thickness by tape casting method. The textured SBT-NBT thick-film MLCCs showed an impressive 1030 kV/cm BDS and a $J_r$ value of 21.5 J/cm$^3$. This is one of the latest developments in search for new strategies to enhance ED properties of thick-film relaxor based MLCCs [141].

Compositionally gradient MLCCs are multilayer architectures with layers of different materials systems alternated with one another to control the electric field distribution. Here, high permittivity FE/relaxor based layers are arranged in different periodicity along with linear dielectric materials that show high BDS. The basic idea is to get a superimposed effect that combines advantages of different material systems that are integrated in one composite structure. This design strategy can be a breeding ground to innovative ideas since the periodicity can be varied widely (for instance periodic and non-periodic connectivity) to realize superior material properties [142]. Yan et. al. demonstrated such $SrTiO_3$-$0.94Bi_{0.54}Na_{0.46}TiO_3$-$0.06BaTiO_3$ (ST-BNBT) based multilayer structures showing a $J_r$ of 2.41 J/cm$^3$ at 237 kV/cm. The BDS of constituent systems individually are 300 kV/cm and 128 kV/cm for ST and BNBT, respectively. ST is a linear dielectric and BNBT is a FE material and the multilayer architecture showed a relaxor like PE loops. This shows that a relaxor like high $Ps$-$Pr$ value can be attained not only by chemical modification but also by designing such multilayer structures. Please note that the dielectric thickness of BNBT layer was as thick as 50 μm and the ED properties are expected to be further improved by reducing the $d$. In this work it was shown by simulation that ST based dielectric layers experienced higher electric field compared to the FE layers and helped in weakening the pace and stopping the electrical surge fronts and thereby ultimately improving the BDS of the multilayer structure.

c) Strategies related to multilayer design

One of the aspects of MLCCs that has not changed much over the years is the electrode design, even though there is a lot of ongoing research on alternative electrode materials that are cheap and can withstand high temperatures, but not on the higher voltages that are required when used on relaxors for EESSs. The internal electrodes are applied on the stacked dielectric layer with a small margin at the alternating ends to connect it parallelly using terminal electrodes [143] (c.f. Figure 9). One important design aspect of electrodes is to decrease the margin, to increase the $A$ and hence the $\varepsilon_r$ and $J$ (c.f. Eqn. 10). Also, the tip of the electrode experiences the maximum electric field concentration and an electrical breakdown usually initiates around that region. In principle, BDS can be tuned by changing the electrode design or the margin length (the distance between the tip of the internal electrode to the terminal electrode). For instance: Yoon et al. could tune the BDS by designing the electrode patterns from 1450 V to 1650 V on BTO MLCCs [144]. In another instance, Cai et al. used a phase-field model to study the initiation of electrical breakdown in MLCC designs. A larger margin length is seen to be necessary to reduce the inhomogeneous distribution of electric field strength and to enhance the BDS [145]. Based on the insights from this study, they went on to design $0.87BaTiO_3$-$0.13Bi(Zn_{2/3}(Nb_{0.85}Ta_{0.15})_{1/3})O_3$ (BT-BZNT) based MLCCs [128] with different margin lengths. It was shown that the BDS can be tuned from 783 kV/cm to 895 kV/cm just by changing the margin length from 100 μm to 400 μm [146].

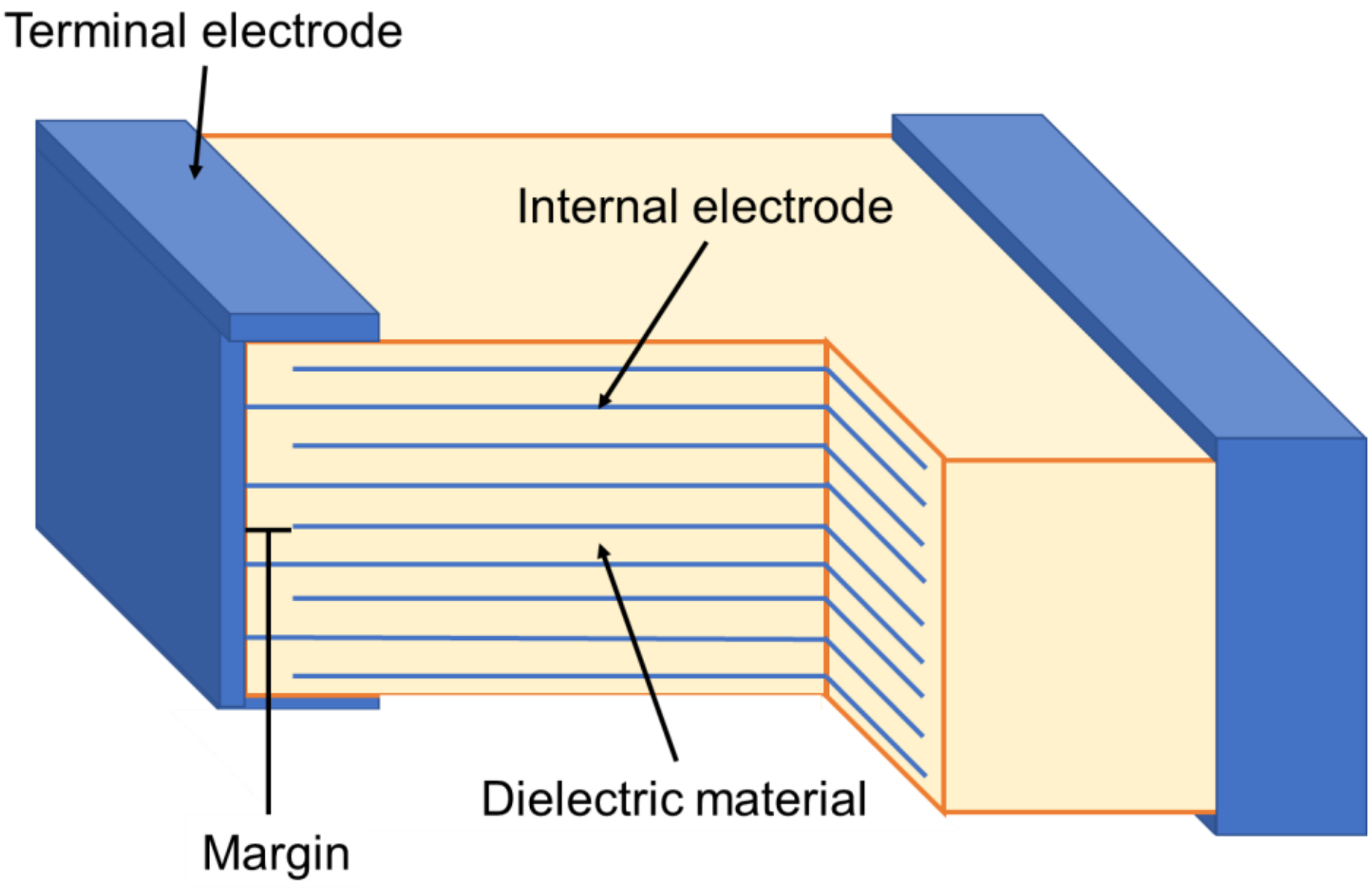

**Figure 9.** Multilayer thick film architectures describing individual components in it.

d) Strategies related to tape casting parameters

The most commonly used tape casting technology for MLCC fabrication is a complex process that needs understanding of the powder characteristics, slip rheology, co-firing and sintering behaviour. Traditional 'one variable at a time' experimental approaches can be very tedious because of the large number of variables in the process. From a processing point of view, for high ED properties, a uniform and dense microstructure with a small G is beneficial. Considering the complexity of tape casting, Yoon et al. carried out a systematic variation of processing parameters using design of experiments tools to study the influence of individual parameters in the final microstructural properties of the fabricated MLCCs. Different process variables such as the choice of starting powders, dispersant, binder, solvents and binder were studied. Although all the process variables had a notable impact on the slip properties and green density, not all parameters had significant effects on the microstructure except for different starting powders, which is related to the presence of different amounts of low-melting impurities. Also, to attain a fine microstructure, solvent-based systems are favourable than water-based systems. This is mainly because the water-based systems leach out $Ba^{2+}$, which was proven to promote abnormal grain growth in the system [147, 148].

*5.3. Thin films*

After the discovery of high permittivity FE materials such as BTO, initial focus resided mostly on fabricating bulk ceramics by conventional methods. It was not until 1970's, the focus moved from innovating new FE material compositions to translating bulk ceramic form to thin film form (100-1000 nm thickness) to benefit mainly from the reduced $d$ [149]. Research on FE thin films was triggered by reports on possibility to use FE materials for non-volatile memories [150]. From there, sophisticated techniques such as chemical vapour deposition (CVD), metal-organic deposition (MOD) and chemical solution deposition (CSD) were employed for the development of different thin film based electronic devices [149]. The recent drive for miniaturisation led to research efforts in developing FE thin film based devices that can meet or outperform the functionality of their bulk counterparts, owing to their smaller volume compared to bulk ceramics [151]. Thin film technology for EESSs is mainly fuelled by the advantages of small volume, large A, high $\rho$, low annealing temperature, large BDS, and excellent control of the microstructure [63]. In this chapter, some of the important considerations to be taken for designing thin film based EESSs will be discussed. Thin films differ from bulk ceramics and thick film multilayers for the presence of a substrate on which the films are deposited or grown. The desired FE material composition is deposited in a process that starts with random nucleation on the substrate surface followed by crystallisation and this is repeated until the desired thickness is reached. The temperature at which the film is crystallised is called annealing temperature [152].

a) Choice of substrate

The substrates for thin film deposition are selected based on the annealing temperatures and atmosphere used in the deposition process, and the lattice mismatch between the substrate and thin film. All these experimental parameters will have strong influence on microstructural properties such as defect formation, residual stress, grain size, etc. that will decide the ED performance of the thin film. Most widely available substrates can be divided in to two categories: metal-coated silicon and metal oxide [153].

*Metal coated Si* substrates pose several challenges in depositing thin film layers. First of all, most deposition methods require an oxidising atmosphere and annealing temperatures up to 700 °C [154]. Metals that does not oxidize and are thermally stable at high temperatures are required to endure such deposition procedures and Platinum (Pt) is one among them [155]. Some of the disadvantages of Pt coated silicon substrates are the high cost, the poor adhesion of Pt with Si substrate and the large difference of thermal expansion compared to the FE based thin films, which can introduce residual stresses in the multilayer structure [153]. To improve the adhesion between Pt and Si, an adhesion layer is often employed. Considering the deposition method and parameters, a stable adhesion material has to be chosen to avoid the appearance of defects [155]. Fox et al. thoroughly studied the effect of annealing temperatures on stability of Pt/Ti/$SiO_2$/Si substrates and demonstrated that, at temperatures higher than 600°C, the Pt layer undergoes chemical and microstructural changes, including changes in the grain size and formation of defects due to the poor stability of the Ti adhesion layer [156]. The appearance of these defects could affect the electrical properties of the film, including ED performance. Therefore, substrates have to be carefully selected depending on the choice of deposition methods and annealing temperatures.

An alternative to Pt as bottom electrode is *metal oxide materials*. Some promising candidates are $LaNiO_3$ (LNO), $SrRuO_3$ (SRO), (La,Sr)$MnO_3$ (LSMO), and $SrTiO_3$ (STO), due to their high electrical conductivity at room temperature [153], stability at higher annealing temperatures [157] and similar structural properties to BTO materials, since they possess the same perovskite structure. These substrates also allow depositing epitaxial films with a definite growing direction with anisotropic electrical properties. Zhang et al. deposited (100)-, (110)-, and (111)-oriented BTO films on (100), (110), and (111) SRO/STO substrates, respectively, by sputtering at 700 °C. Orientation of the film is found to have strong dependence on the $\varepsilon_r$ of obtained thin films in the following order: $\varepsilon_r$ (110) > $\varepsilon_r$ (111) > $\varepsilon_r$ (100) [158]. Nguyen et al. proved that also the BDS and $J_r$ of relaxor compositions can be influenced by the choice of the substrate. In a work on a lead-based system (lead zirconate titanate, PZT), thin films were deposited on Pt/Ti/$SiO_2$/Si and SRO/STO/Si using the pulsed laser deposition (PLD) method. Films deposited on Pt/Ti/$SiO_2$/Si substrate had a BDS of 1750 kV/cm and a $J_r$ of 17.6 J/cm$^3$ at 1500 kV/cm, whereas the films deposited on SRO/STO/Si presented an enhanced $J_r$ of 23.2 J/cm$^3$ and a BDS of 2500 kV/cm [159]. The difference in ED properties has been attributed to the dense and epitaxial microstructure of the films deposited on SRO compared to columnar and polycrystalline film on Pt substrate. Additionally, Pt as base electrode was not suitable to achieve high BDS in this particular film, lowering the $J_r$.

Recently, Zhu et al. showed that not only the type of bottom electrode but also the thickness will have a remarkable impact on the ED properties of thin films. Here, $Ba_{0.53}Sr_{0.47}TiO_3$ (BST) films were deposited on LSMO bottom electrodes of different thicknesses on a (001) $SrTiO_3$ (STO). The BDS and $J_r$ increased from 3075 kV/cm to 4822 kV/cm and 31 J/cm$^3$ to 51 J/cm$^3$, respectively, for a LSMO thickness ranging from 30 nm to 140 nm [160]. Increase in LSMO thickness ensured a smooth interface and better stress relaxation that profited the ED properties.

*b) Choice of deposition techniques*

Metal oxide thin film deposition techniques can be broadly classified into physical methods and chemical methods. Physical techniques include pulsed laser deposition (PLD), molecular beam epitaxy (MBE) and sputtering. Chemical techniques include the widely used chemical solution deposition (CSD) and Metal-organic Deposition (MOD) and metallo-organic chemical vapour deposition (MOCVD) [152, 161]. One major difference is that the physical techniques require high vacuum deposition environments while the latter can be carried out in ambient conditions, thus they can also be categorized as vacuum and non-vacuum techniques [149]. When selecting a deposition method, factors such as the desired properties, ease of use, eco-friendliness, the cost of the equipment will have to be considered. Table 1 summarizes the major advantages and disadvantages of each listed techniques. All techniques have been extensively described and compared by Wang et al. [161] and Heartling [149] in two comprehensive reviews.

Although some techniques were initially not suitable to produce high-quality thin films, improved methods were constantly developed in both physical and chemical deposition to ensure superior films (good uniformity and tight control of stoichiometry) while making the method more versatile. It is important to understand the scope of all the available deposition methods so that appropriate choice is made as required: To fabricate thin film MLCCs for EESSs, techniques such as molecular beam epitaxy (MBE) that facilitate the possibility to have complete control on the atomic-scale layer-by-layer film growth are too time-consuming and costly, which limits their applicability on an industrial scale [162].

- Sputtering and PLD allow having tight control over the stoichiometry and microstructure of the deposited film. This is especially advantageous to deposit films of complex chemistry (like relaxors) with desired microstructure and orientation [161]. The biggest drawback of physical methods like PLD is the high infrastructure cost [149], together with a non-trivial control on the microstructural properties of the films (unless in-situ characterization methods are used).
- The sol-gel technique has high versatility, suitability to deposit almost all the perovskite compounds, simplicity, low infrastructure cost and possibility of large-scale deposition. Also, control on microstructural properties of the films, including grain size and orientation, and film thickness, is possible by tuning the concentration of the precursor solutions and the thermal cycles. However, films produced by sol-gel usually contain more microstructural defects than by PLD, and this can strongly influence the ED of the deposited material [151, 152, 163]

**Table 1.** Most common thin films deposition methods and their respective advantages and disadvantages [149, 151, 152, 161, 162].

| . | Method | Advantages | Disadvantages |
|---|---|---|---|
| Physical Methods | PLD | Excellent stoichiometry control<br>Good versatility<br>Low substrate temperature | Defects formation<br>Poor scalability<br>High cost |
| | MBE | Possibility of epitaxial growth<br>Excellent film thickness control<br>Good stoichiometry control | Sophisticated apparatus<br>Precursor unavailability<br>Expensive |
| | Sputtering | High deposition rates<br>Uniform film<br>Low impurities<br>Ease in scalability | Poor stoichiometric control<br>High substrate temperature |
| Chemical Methods | CSD | Excellent stoichiometric control<br>Good uniformity<br>Affordable<br>Easiness<br>Precursors availability<br>Ease in scalability | Crack formation<br>Delamination<br>Defects formation<br>Low deposition rates<br>Poor thickness control |
| | MOCVD | Good Stoichiometry control<br>Excellent uniformity<br>Texture versatility<br>High deposition rates | Poor precursors availability<br>Bad reproducibility<br>High substrate temperatures |
| | MOD | Low temperature<br>High density<br>Good uniformity<br>Good stoichiometry control | Bad thickness control<br>Large volume shrinkage<br>Poor precursors availability<br>High annealing temperatures |

**c)** Choice of processing parameters

Regardless of the deposition techniques used, thin film layer thickness, microstructural defects [164] and microstructural properties (*G*, $\rho$, residual stresses) [165] influence the electrical properties such as $\varepsilon_r$, *Pr*, *Ps* and BDS [164, 166, 167]. The dependence of $\varepsilon_r$ on the grain size is already well established in the previous sections and by Buscaglia and Randall [21]. Song et al. investigated the dependence between *d* and *G* for barium stannate (BTS) thin films deposited with sol-gel on platinised silicon substrates. Varying the *d* from 80 nm to 600 nm, the grain size increased from 13 nm to 24 nm with an increase in $\varepsilon_r$ from 252 to 430 [164]. Moreover, a shift in $T_c/T_m$ was observed from -49.6 °C for a thickness of 160 nm to -27.3 °C for the 600 nm film. These effects are linked not only with the G but also the residual stresses in the films. For a thin film with constant thickness *d*, the annealing temperature has also demonstrated similar effects on $\varepsilon_r$ and $T_c/T_m$. Xu et al. investigated the influence of post-deposition annealing temperatures on $\varepsilon_r$ and $T_c/T_m$ of $BaZr_{0.3}Ti_{0.7}O_3$ (BZT30) thin films deposited via sputtering method at 650°C. The dielectric constant and $T_c/T_m$ increased from 600 to 3300 and -100°C to -40°C, respectively, for annealing temperatures increasing from 650°C to 1100°C. A higher annealing temperature not only resulted in increased *G* but also ensured low residual stresses in the films, which is beneficial for their ED properties [168]. In addition, Udayakumar et al. presented *d* dependence on BDS of PZT films deposited on platinised silicon with sol-gel technique [166]. This is in accordance with the discussion related to Eqn. 12 in the previous section.

d) Review on high ED thin-film systems

Although thin film technology has been around for many decades now, literature on lead-free thin-films for EESSs is very minimal. Karan et al. in an early work on BTO based lead-free thin-films for EESSs demonstrated a $J_s$ and BDS of 34 J/cm$^3$ and 3000 kV/cm on $Ba[(Ni_{1/2},W_{1/2})_{0.1}Ti_{0.9}]O_3$ thin films [169]. One requirement to assure high ED in addition to BDS in thin-films is to have low leakage currents and chemical methods such as CSD are very beneficial to realize those properties because of the advantages stated in Table 1. Also, for instance, a BDS of 2000 kV/cm and a $J_s$ of 37 J/cm$^3$ at 1900 kV/cm was realised in a $0.88BaTiO_3$-$0.12Bi(Mg,Ti)O_3$ thin film by Kwon and Lee. The CSD assisted thin film deposition used in this work ensured fine microstructure and low leakage currents, and exhibited great temperature stable ED performance [170]. In this work, space charge related conduction was proven to be dominating and possibly influencing the electrical properties, especially the thinning of PE hysteresis resulting in high $J_r$. Later, Zhu et al. showed that a combination of space charge and interlayer coupling can be positively reinforced in $BaTiO_3$-$BiFeO_3$ (BTO/BFO) thin film heterostructures to achieve a $J_r$ of 51 J/cm$^3$ compared to ~28 J/cm$^3$ for individual thin film structures [171]. Finally, Yang et al. showed the effect of Bi-based compounds (here: $Bi_{3.25}La_{0.75}Ti_3O_{12}$) in BTO (BTO-BLT) based thin-films on ED properties. Bi-based compounds limit defect concentration and obtain uniform microstructure that favours low leakage current and high BDS. As expected, for 0.6BT-0.4BLT thin films, $J_r$ and $\eta$ of 61.1 J/cm$^3$ and 84.2 % and a $J_r$ and $\eta$ of 58.4 J/cm$^3$ and 85 %, for 0.4BT-0.6BLT thin-films were reported, which is substantially better than BTO thin films [172].

On the other hand, using PLD, Instan et al. fabricated BZT30 thin films, with a ultrahigh $J_r$ of 156 J/cm$^3$ at 3000 kV/cm and a $\eta$ of ~73 %, which is by far one of the highest achieved ED values in relaxor based thin film structures [173]. Improving this work, Cheng et al. published a record high $J_s$ and $\eta$ of up to 166 J/cm$^3$ and 96% on BZT based films using RF magnetron sputtering. This was realized by introducing compressive stresses using a lattice mismatch between the substrate and the film. These stresses are expected to decay along the thickness direction, in addition to ensuring polydomain structure that contributes to superior ED properties [174]. Compositionally gradient multilayer structures were also tried on thin-film multilayer structures to control the electric field distribution and block the electric field path that causes electrical breakdown. Here BZT15 and BZT35 thin film stacks were alternated to benefit from their unique dielectric properties. The ED properties were strongly dependent on the number of alternating layers with a highest $J_r$ and BDS of 69 J/cm$^3$ and 8300 kV/cm for an optimum of 6 alternating layers [175].

e) Thin film based multilayer ceramic capacitors:

Despite thick film based MLCCs technology has revolutionized the capacitor industry for more than two decades, the use of powder-based approaches for producing thin films (below 0.5 μm) is not feasible, imposing a limit to further miniaturisation of the electronic components in a cost-effective fashion [176, 177]. Many thin film systems fabricated using alternative thin film fabrication technologies have demonstrated giant $J_r$ values. Nevertheless, due to the small volume, a single-layer capacitor will not be able to store a usable amount of energy. Hence, realizing thin film MLCCs seems the necessary route to increase the overall volume of the dielectric and consequently the amount of energy that can be stored, while maintaining the giant $J_r$ and BDS.

There is, however, a scarcity of literature reports on thin film MLCCs, which is likely due to the following reasons:

- MLCCs comprise numerous dielectric layers alternated with metallic electrodes, where the interface in-between plays a major role in final $J_r$ values, both due to effects on BDS and on leakage currents. This becomes especially critical for thin film MLCCs, since the material to electrode ratio can be substantially large compared to conventional thick film MLCCs. Therefore, interfacial reactions such as interdiffusion and oxidation, the appearance of defects at the interface and adhesion issues are all not trivial [178].
- It is difficult to guarantee the integrity of the multilayer structure upon several deposition cycles involving dielectric layer and electrode. The electrodes must survive several annealing cycles without any substantial changes [176]. Pt electrode is generally a good choice to improve the electrical properties, but the high cost could be a limiting factor for industrialization. For this reason, base metal electrodes such as nickel [179] or copper [178] have been investigated. Deposition in an oxidising atmosphere is not suitable for such metals, imposing constrains on the deposition and annealing conditions.
- Increasing the number of layers and electrodes, even if processing is well-tuned to avoid interfacial defects, increases the probability of finding a critical defect in the dielectric layers, with consequent increased probability of electrical failure.
- Using photolithography routes to prepare a thin film MLCC increases the risk of a non-perfect adhesion between layers, of uneven deposition of the top layers, difficulties in alignment for each layer, use of unrealistically wide margins (leading to poor area efficiency), and the extended times for stack buildup, leading to high costs [180].

Nevertheless, some promising thin film MLCCs based on lead-free materials have been lately produced using alternative stacking approaches. Nagata et al. used a PDMS stamp to realise stacked BTO-$LaNiO_3$ capacitors with up to five layers with good alignment and resulting electrical properties [180]. Wang et al. used CSD and sputtering techniques to sequentially deposit BTO dielectric layers and Ag electrodes, respectively, on a $SiO_2$ glass substrate. They used laser ablation to create the desired electrode pattern needed for the capacitor structure, and to remove the stack from the substrate once the desired thickness was reached. Using this cost-effective method, they were able to produce a thin film MLCC with monolayers of less than 200 nm [181]. These reports show that thin film MLCC realization is feasible, but there is no information on $J_r$ values that can be reached with these stacking techniques, which underlines the current need for investigation of these aspects in connection with ED studies.

*5.4. Final remarks*

The intention of this section is not to advocate any of the above stated material forms or techniques but to summarize some of the key parameters that influence ED properties in all forms. So far, we have learned some of the most important processing strategies that are commonly utilized to improve the ED of bulk ceramics, thick film and thin film based MLCCs, especially relaxors. Some conclusive remarks can be drawn out of these discussions. One is that a smaller grain size is always favourable for ED performance of the bulk ceramics [182, 183]. Grain size (*G*) and BDS are related by the following equation,

$$BDS \propto G^{-k} \tag{13}$$

where *K* is a constant. But, this does not ensure an everlasting enhancement of ED properties with the decreasing *G* because of size effects in ferroelectric (and relaxor) ceramics [184]. As previously discussed, with the decrease in grain size, the grain boundary density increases resulting in lowering of $\varepsilon_r$. In addition, stress in fine-grained ceramics results in strong electrostrictive coupling between the lattice strain and polarization, also lowering the $\varepsilon_r$. Because of the distribution of grain size, strong strain modulation is expected in the lattice and hence inhomogeneity in the polarization values.

To summarize, a tight control on grain size distribution and a smaller grain size can positively affect ED properties. In addition to the grain size, density has a profound effect on the BDS and hence ED properties. It was also discussed previously on how two-step sintering (TSS) processes exploit the difference in the kinetics of grain growth and densification mechanisms to attain favourable ED properties. This is similar to annealing temperature for thin film MLCCs. Pressure is proven to be a catalyst to effective particle rearrangement and also promoting the densifying mechanisms to attain high density at lower temperatures and time, thereby keeping the grain size low in bulk ceramics. Chemical modification is one of the key strategies when it comes to tuning ED for thick film MLCCs. And finally, the choice of techniques is very critical for thin film based MLCCs. In such systems, since the substrate has to also function as the bottom electrode, it has to be conductive in addition to ensuring nucleation and growth of the film. Therefore, the substrate has to be carefully chosen accordingly to the deposition temperature, growth direction and lattice parameter to limit stress (or sometimes benefit from the residual stresses) arising from the mismatch and to overall improve the dielectric properties. While most deposition techniques can be employed to produce good quality FE thin films, initial cost, scalability and easiness of use are the main factors to be evaluated. High quality films with favourable microstructural properties are essential to improve $J_r$ and BDS; especially, a small-grained microstructure is favourable, being cautious that very small *G* can dilute the polar characteristics of the film [21]. A very important point is the minimisation of the leakage currents, which has to be performed by tuning the electrode/dielectric interface not only by reducing defects, but also choosing the proper interface combination that allows establishing a high Schottky barrier at the interface (for example, by choosing high work function electrodes or depositing a buffer layer) [185, 186]. These aspects require a thorough, separate treatment, and thus were not addressed in this review. Finally, the stacking and patterning method for thin film MLCCs has to be selected to ensure good alignment and integrity of the device. This is not trivial and alternative cost-effective approaches to photolithography should be preferred.

**6. Guidelines for selection of materials and processing route**

Based on the currently available technology, it is clear that the achievable energy density on different forms of ceramics scales as follows: bulk ceramics < thick film multilayers < thin film multilayers, also shown in Figure 10. This is mainly attributed to the peculiarities in their respective processing routes. With the available techniques, it is not possible for bulk ceramics to demonstrate ED values comparable to thick films. On the other hand, it is possible for thick films to match ED properties of thin films when the appropriate method and chemistry is chosen. It is important to know the limitations of each available material forms and its related processing techniques to choose the material needs wisely.

Bulk ceramic processing is relatively well understood and is easily implemented, therefore can be the method of choice to experiment new compositions in order to screen promising candidates for high ED. Screening is viable only if the ED performances are compared between different systems in its bulk ceramic form since the BDS can be widely different for other thick and thin film forms as a result of reduced *d*.

When it comes to thick film processing, it is a complex methodology with several processing variables as discussed earlier. In additional to the processing parameters related to thick film fabrication, additional stacking, thermocompression, screen-printing and debonding steps are essential to functionalize it as MLCCs. This is however a widely investigated process and already industrialized. Apart from revolutionary innovations in MLCC fabrication like Cold Sintering [187] (which however are not industrially applied yet), the focus here should be accommodating the current facilities to new promising compositions that showed high ED performance as bulks, in order to achieve much higher recoverable ED through the MLCC structure.

As mentioned in this contribution, thin film processing allows to attain by far the highest ED and efficiency, compared to processing the same composition as bulk or thick film MLCC. The problem is that thin film processing is by far the most difficult and slow process, and building multilayers out of thin films may encounter integrity problems such as accumulation of micro-mechanical stresses, delamination, etc., as discussed above. Hence, realization of large thin film-based multilayers (with >> 10 layers) has not been demonstrated yet and is far from being industrially implemented.

All the compositions treated in this review are summarised in Table 2, and some selected compositions are reported visually in Figure 7, Figure 8, Figure 10 and Figure 11. It can clearly be seen that if we consider devices with the same volume, the thin films have the highest ED (in J/cm$^3$) and can potentially store the highest amount of energy (in J). The problem is that thin film-based devices with a volume (or a number of layers) comparable to a thick film multilayer cannot be produced yet.

Hence, with the choice of available material forms, we believe that,

- For high-energy and high-power applications, the thick film multilayer technology is currently the most attractive, because it can realise ED levels much higher than bulk ceramics (and not much lower than thin films), and the processing technology is well established, so that devices storing high amounts of energy (although much less than a battery) can be easily produced.
- For low-energy applications, thin film devices are still very attractive because of their small size and footprint. They can be used in applications were the amount of energy that has to be stored is low (mJ or below), for instance: autonomous sensors for the Internet of Things or small portable microelectronic devices. These applications may also require high-power as well as high-voltage, and need the flexibility in both high power and energy density offered by thin film relaxor capacitors. If the BDS allows, the higher the voltage the larger the amount of energy that can be stored, and so such thin film devices can be a suitable energy storage device for these low-energy applications.

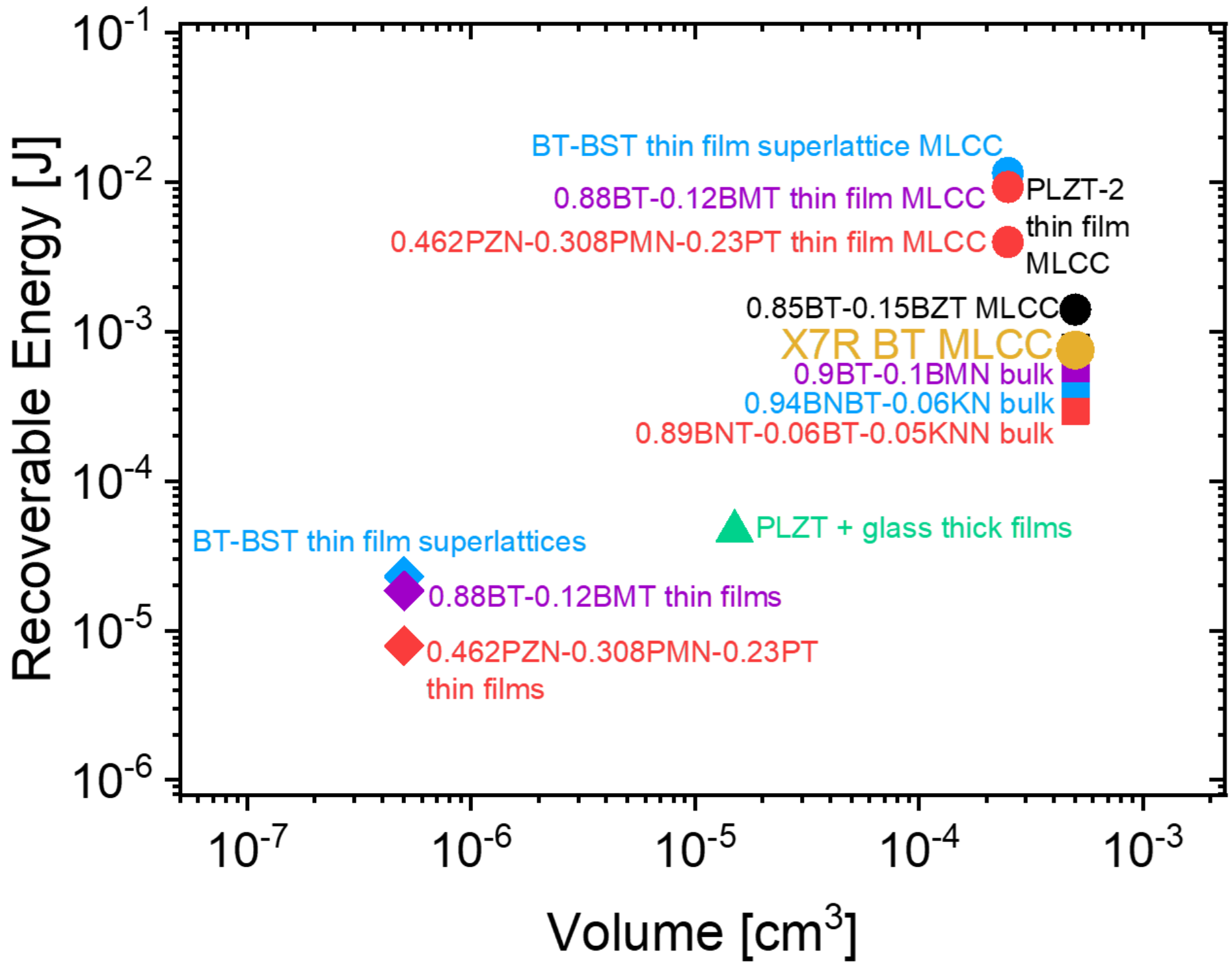

**Figure 10.** Recoverable energy density at the maximum applied electric field for some selected relaxor compositions in bulk, thick film multilayers and thin film multilayers Values for thin films MLCCs were extrapolated from the respective $J_r$ values of single-layer thin films assuming an MLCC consisting of 500 layers with 1 μm thickness and 0.5 x 1 $mm^2$ area [114, 178, 196, 197, 188–195].

*Table 2.* Energy density properties of selected perovskite lead-free relaxor ferroelectrics.

| S.No | Composition | Processing method | $J_s$ *[J/cm³]* | $J_r$ *[J/cm³]* | $\eta$ *[%]* | BDS *[kV/cm]* | Reference |
|---|---|---|---|---|---|---|---|
| | | | **BULK CERAMICS** | | | | |
| 1. | $0.9BaTiO_3$-$0.1Bi(Mg_{2/3}Nb_{1/3})O_3$-0.3 wt. % $MnCO_3$ | SSR | - | 1.7 | 90 | >210 | [48] |
| 2. | $0.60BiFeO_3$-$0.34BaTiO_3$-$0.06Ba(Zn_{1/3}Ta_{2/3})O_3$ | SSR | - | 2.56 | 80 | >160 | [70] |
| 3. | $BaTi_{0.7}Zr_{0.3}O_3$ | SPS | 0.51 | - | 70-80 | 170 | [71] |
| 4. | $Ba_{0.94}(Bi_{0.5}K_{0.5})_{0.06}Ti0_{.85}Zr_{0.15}O_3$ | TSS | - | 0.95 | 88 | 76 | [112] |
| 5. | $0.89Bi_{0.5}Na_{0.5}TiO_3$-0.06Ba $TiO_3$-$0.05K_{0.5}Na_{0.5}NbO_3$ | TSS | 0.9 | - | - | >99 | [113] |
| 6. | $0.89Bi_{0.5}Na_{0.5}TiO_3$-0.06Ba $TiO_3$-$0.05K_{0.5}Na_{0.5}NbO_3$ | SSR | 0.46 | - | - | >56 | [114] |
| 7. | $Bi_{0.5}K_{0.5}TiO_3$-$Ba(Mg_{1/3}Nb_{2/3})O_3$ | HP | - | 3.14 | 83.7 | >230 | [118] |
| 8. | $Bi_{1-x}Sm_xFe_{0.95}Sc_{0.05}O_3$ | HP | - | 2.21 | 76 | >230 | [119] |
| 9. | $0.55Bi_{0.5}Na_{0.5}TiO_3$-$0.45Ba_{0.85}Ca_{0.15}Ti_{0.85}Zr_{0.1}Sn_{0.05}O_3$ | MW | 1.21 | - | 72.08 | 130.2 | [125] |
| 10. | $BaTiO_3$-$Bi(Li_{0.5}Ta_{0.5})O_3$ | SSR | - | 2.2 | 89 | 280 | [129] |
| 11. | $Ba_{0.70}Ca_{0.30}TiO_3$-$Ba(Zr_{0.2}Ti_{0.8})O_3$ | SSR | 1.21 | 0.71 | | 150 | [132] |
| 12. | $0.61BiFeO_3$- $0.33(Ba_{0.8}Sr_{0.2})TiO_3$-$0.06La(Mg_{2/3}Nb_{1/3})O_3$ | SSR | - | 3.38 | 59 | 230 | [134] |
| 13. | $0.45SrTiO_3$-$0.2Na_{0.5}Bi_{0.5}TiO_3$-$0.35BaTiO_3$ | | - | 1.78 | - | >170 | [196] |
| 14. | $0.85BaTiO_3$-0.15Bi $(Mg_{2/3}Nb_{1/3})O_3$ | SSR | 1.18 | 1.13 | | >143.5 | [195] |
| 15. | $Ba_{0.997}Sm_{0.002}Zr_{0.15}Ti_{0.85}O_3$ | SSR | 1.15 | - | 92 | >10 | [198] |
| 16. | $0.67Bi_{0.9}Sm_{0.1}FeO_3$-$0.33BaTiO_3$ | | | 2.8 | 55.8 | 200 | [199] |

| S.No | Composition | Processing method | $J_s$ *[J/cm$^3$]* | $J_r$ *[J/cm$^3$]* | η *[%]* | BDS *[kV/cm]* | Reference |
|---|---|---|---|---|---|---|---|
| 17. | $BiScO_3$-$BaTiO_3$ + 20 wt% $(K_{1/2}Bi_{1/2})TiO_3$ | SSR | 1.28 | - | - | 100 | [200] |
| 18. | $0.89Bi_{0.5}Na_{0.5}TiO_3$-$0.06BaTiO_3$-$0.05K_{0.5}Na_{0.5}NbO_3$ | TSS | 0.9 | - | - | >100 | [113] |
| 19. | $1/3(Ba_{0.70}Sr_{0.30}TiO_3)$ + $1/3(Ba_{0.70}Ca_{0.30}TiO_3)$ + $1/3(BaZr_{0.20}Ti_{0.80}O_3)$ | SSR | 1.40 | 0.44 | - | >115 | [201] |
| 20. | $BaZr_{0.1}Ti_{0.9}O_3$ | SSR | 0.5 | - | - | 30 | [202] |
| 21. | $0.94Bi_{0.47}Na_{0.47}Ba_{0.06}TiO_3$-$0.06KNbO_3$ | SSR | 0.89 | - | - | 100 | [194] |
| 22. | $0.61BiFeO_3$-$0.33BaTiO_3$-$0.06Ba(Mg_{1/3}Nb_{2/3})O_3$ | SSR | 1.56 | - | 75 | >125 | [197] |
| 23. | $0.9\ Ba_{0.65}Sr_{0.35}TiO_3$-$0.1Bi(Mg_{2/3}Nb_{1/3})O_3$ | SSR | 3.9 | 3.34 | 85.71 | 400 | [203] |
| 24. | $0.94(Bi_{0.5}Na_{0.5})(Y_{0.5}Ta_{0.5})0.1Ti0.9O_3$-$0.06BaTiO_3$ | SSR | 1.215 | - | 68.7 | >98 | [204] |
| 25. | $0.94Bi_{0.5}Na_{0.5}TiO_3$-0.06BaTiO+0.03CaZrO3 | SSR | 0.7 | - | - | >70 | [205] |
| 26. | $0.95(0.93Bi_{0.5}Na_{0.5}TiO_3$-$0.07BaTiO_3)$-0.05 $KNbO_3$ | SG | 1.72 | - | - | >168 | [206] |
| 27. | $0.85[(0.94)Bi_{0.5}Na_{0.5}TiO_3$-$0.06BaTiO_3]$-$0.15Na_{0.73}Bi_{0.09}NbO_3$ | SSR | 1.4 | - | 66.3 | 142 | [207] |
| 28. | $0.86BaTiO_3$-$0.14Bi(Zn_{0.5}Ti_{0.5})O_3$ | SSR | 0.81 | - | 94 | 120 | [208] |
| 29. | $0.82[0.94Bi_{0.5}Na_{0.5}TiO_3$-$0.06BaTiO_3]$-$0.18K_{0.5}Na_{0.5}NbO$ | | 0.616 | - | 94 | >70 | [209] |
| 30. | $0.93Ba_{0.55}Sr_{0.45}TiO_3$-$0.07BiMg_{2/3}Nb_{1/3}O_3$ | SSR | - | 4.55 | 81.8 | 450 | [210] |
| 31. | $[(BaZr_{0.2}Ti_{0.80})O_3]_{0.85}$-$[(Ba_{0.70}Ca_{0.30})TiO_3]_{0.15}$ | SSR | 7.48 | - | - | 153 | [211] |
| 32. | $BaZr_{0.15}Ti_{0.85}O_3$ + 12 wt% $Bi_2O_3{\cdot}3TiO_2$ | SSR | - | - | - | 150.9 | [212] |
| 33. | $0.5(Ba_{0.7}Ca_{0.3})TiO_3$-$0.5Ba(Ti_{0.9}Zr_{0.1})O_3$ | SSR | - | 0.164 | 74 | - | [213] |

| S.No | Composition | Processing method | $J_s$ *[J/cm$^3$]* | $J_r$ *[J/cm$^3$]* | η *[%]* | BDS *[kV/cm]* | Reference |
|---|---|---|---|---|---|---|---|
| 34. | 0.6 Ba(Zr0.2Ti0.8)O3-0.4Na0.5Bi0.5TiO3 | SSR | - | 3.22 | 91.2 | 241 | [214] |
| 35. | $0.88BaTiO_3$-$0.12Bi(Mg_{1/2}Ti_{1/2})O_3$ | SSR | 1.81 | - | 88 | 535.5 | [215] |
| 36. | $0.85BaTiO_3$-$0.15Bi(Mg_{1/2}Zr_{1/2})O_3$ | SSR | 1.31 | 1.25 | 95 | 185 | [216] |
| 37. | $0.88BaTiO_3$-$0.12(Mg_{2/3}Ta_{1/3})O_3$ | SSR | - | 3.28 | 93 | 395 | [217] |
| 38. | $0.96(1-x)BaTiO_3$-$0.04KNbO_3$ | SSR | - | 2.03 | 94.5 | 300 | [218] |
| 39. | $0.6Bi(Mg_{1/2}Ti_{1/2})O_3$-$0.4BaTiO_3$ @ 120 °C | SSR | 0.7 | - | - | 60 | [219] |
| 40. | $0.9BaTiO_3$-$0.1Ba(Mg_{1/3}Nb_{2/3})O_3$ | SSR | 1.01 | - | - | 158 | [220] |
| 41. | $0.85BaTiO_3$-$0.15Bi(Zn_{2/3}Nb_{1/3})O_3$ | SSR | 0.79 | - | 93.5 | 131 | [221] |
| 42. | $0.90(Na_{1/2}Bi_{1/2})_{0.92}Ba_{0.08}TiO_3$-$0.10Bi(Mg_{1/2}Ti_{1/2})O_3$ | SSR | 2 | - | - | >135 | [222] |
| 43. | $0.084BiTi_{0.5}Zn_{0.5}O_3$-$0.916(0.935Bi_{0.5}Na_{0.5}TiO_3$-$0.065BaTiO_3)$ | SSR | 1.04 | - | 80 | >95 | [223] |
| 44. | $0.4(Na_{0.5}Bi_{0.5}TiO_3)$-$0.225BaTiO_3$-$0.375BiFeO_3$ | SPS | 1.4 | - | 90 | - | [224] |
| | *Ceramic + (Glass/organics) composites* | | | | | | |
| 45. | $0.88BaTiO_3$–$0.12Bi(Mg_{1/2}Ti_{1/2})O_3$ + 4% ($SiO_2$–$B_2O_3$) | SSR | 1.97 | - | 94.5 | >270 | [75] |
| 46. | $Ba_{0.9995}La_{0.0005}TiO_3$ + 20 wt% 65PbO-$20B_2O_3$-$15SiO_2$ | SSR | 0.56 | 0.31 | 54.2 | >300 | [76] |
| 47. | $BaTiO_3$ + 3 wt% $Al_2O_3$ + 1 wt% $SiO_2$ | SSR | 0.725 | | 80 | 190 | [77] |
| 48. | $Ba_{0.4}Sr_{0.6}TiO_3$ + 2 wt% (30.8 SrO-$58.9B_2O_3$-$10.3SiO_2$) | SSR | | 0.44 | 67.4 | | [78] |
| 49. | $Ba_{0.4}Sr_{0.6}TiO_3$ + 9 wt% ($65Bi_2O_3$-$20B_2O_3$-$15SiO_2$) | SSR | 2.18 | 1.98 | 90.57 | >279 | [83] |
| 50. | $Ba_{0.4}Sr_{0.6}TiO_3$ + 4 wt% BaO-$B_2O_3$-$SiO_2$-$Na_2CO_3$-$K_2CO_3$ | SSR | - | 0.72 | - | 280.5 | [84] |

| S.No | Composition | Processing method | $J_s$ *[J/cm^3]* | $J_r$ *[J/cm^3]* | η *[%]* | BDS *[kV/cm]* | Reference |
|---|---|---|---|---|---|---|---|
| 51. | $Ba_{0.85}Ca_{0.15}Zr_{0.1}Ti_{0.9}O_3$ + 5 wt% $B_2O_3$-$Al_2O_3$-$SiO_2$ | SSR | 1.153 | - | - | 200 | [85] |
| 52. | $BaTiO_3$ + 2.5 wt% BaO–$Bi_2O_3$–$P_2O_5$ | SSR | 0.0069 | - | 69.21 | >15 | [86] |
| 53. | $Ba_{0.4}Sr_{0.6}TiO_3$ + 55 wt% BaO-$B_2O_3$-$Al_2O_3$-$SiO_2$ | SSR | 3.1 | - | - | >405 | [88] |
| 54. | $BaTiO_3$ + 2wt% $SiO_2$ | SSR | 2.23 | 1.2 | 53.8 | 200 | [89] |
| 55. | 0.715$Bi_{0.5}Na_{0.5}TiO_3$-0.065$BaTiO_3$-0.22$SrTiO_3$ + 4 wt% 3BaO-3$TiO_2$-$B_2O_3$ | SSR | 0.203 | - | 60 | - | [225] |
| 56. | $Ba_{0.3}Sr_{0.7}TiO_3$ + 1.6 wt% ZnO | SSR | 3.9 | - | - | 400 | [90] |
| 57. | 0.95(0.76$Na_{1/2}Bi_{1/2}TiO_3$-0.24$SrTiO_3$)-0.05$AgNbO_3$:$SiO_2$ | SSR | - | 3.22 | - | 316 | [91] |
| 58. | $Ba_{0.4}Sr_{0.6}TiO_3$ + 0.5 wt% $SiO_2$ | SSR | - | 0.86 | 79 | 134 | [92] |
| 59. | $Ba(Zr_{0.2}Ti_{0.8})O_3$-0.15$(Ba_{0.7}Ca_{0.3})TiO_3$ + 11 wt% BaO-SrO-$TiO_2$-$Al_2O_3$-$SiO_2$-$BaF_2$ | SSR | 1.45 | - | - | 108 | [226] |
| 60. | $Ba_{0.4}Sr_{0.6}TiO_3$ + 5 wt% MgO | SPS | 1.7 | 1.5 | 88.5 | 300 | [95] |
| 62. | $BaTi_{0.85}Sn_{0.15}O_3$ +10 wt% MgO | SPS | 0.5107 | - | 92.11 | 190 | [96] |
| 63. | 0.55$Bi_{0.5}Na_{0.5}TiO_3$- 0.45$Ba_{0.85}Ca_{0.15}Ti_{0.85}Zr_{0.1}Sn_{0.05}O_3$ + 5 wt% MgO | MW | 2.09 | - | 79.51 | 189.7 | [125] |
| 64. | $Ba_{0.4}Sr_{0.6}(Ti_{0.996}Mn_{0.004})O_3$-2 wt% MgO | - | - | 2.014 | 88.6 | 300 | [227] |
| 65. | 0.9(0.94$Na_{0.5}Bi_{0.5}TiO_3$-0.06$BaTiO_3$)-0.1$NaNbO_3$ + 1 wt% ZnO | SSR | 1.27 | | 67 | >100 | [98] |

| S.No | Composition | Processing method | $J_s$ *[J/cm³]* | $J_r$ *[J/cm³]* | η *[%]* | BDS *[kV/cm]* | Reference |
|---|---|---|---|---|---|---|---|
| 66. | $Bi_{0.5}Na_{0.5}TiO_3$-$BaTiO_3$-$K_{0.5}Na_{0.5}NbO_3$ + 40 wt% ZnO | SSR | - | 1.03 | 72.7 | 140 | [99] |
| 67. | $0.85BaTiO_3$-$0.15Bi(Mg_{1/2}Zr_{1/2})O_3$ + 10 wt% $MnCO_3$ | | - | 1.61 | 94.3 | 230 | [228] |
| | | | *Ceramic + Polymer composites* | | | | |
| 68. | BTO-(P(VDF-HFP)-20 vol % BTO | SC | 8.13 | - | 57 | 3300 | [103] |
| 69. | Ba0.95Ca0.05Zr0.15Ti0.85O3 + 40% PVDF | SSR-SC | 2.0 | - | - | 600 | [100] |
| 70. | BaTiO3 + 80 wt% PVDF | SSR- SC | 3.54 | - | - | >2000 | [105] |
| | | | **THICK FILM MULTILAYERS** | | | | |
| 71. | $0.62BiFeO_3$-$0.3BaTiO_3$-$0.08NdZn_{0.5}Zr_{0.5}O_3$. | SSR-TC | - | 10.5 | 87 | >700 | [67] |
| 72. | $0.87BaTiO_3$-$0.13Bi(Zn_{2/3}(Nb_{0.85}Ta_{0.15})_{1/3})O_3$ | SSR-TC-TSS | - | 10.12 | 90 | >1047 | [74] |
| 73. | $0.87BaTiO_3$-$0.13Bi(Zn_{2/3}(Nb_{0.85}Ta_{0.15})_{1/3})O_3$ | SSR-TSS | - | 10.5 | 93.7 | 1000 | [128] |
| 74. | $BaTiO_3$-$Bi(Li_{0.5}Ta_{0.5})O_3$ | SSR-TC | - | 4.05 | 95.5 | 466 | [129] |
| 75. | $BaZr_{0.2}Ti_{0.8}O_3$ | SSR-TC | - | 6.2 | 98% | 800 | [131] |
| 76. | $0.90BaTiO_3$– $0.10Bi(Li_{0.5}Nb_{0.5})O_3$ | SSR-TC | - | 4.5 | 91.5 | 450 | [133] |
| 77. | $0.61BiFeO_3$- $0.33(Ba_{0.8}Sr_{0.2})TiO_3$-$0.06La(Mg_{2/3}Nb_{1/3})O_3$ | SSR-TC | - | 10 | 72 | >730 | [134] |
| | | | **THIN FILM MULTILAYERS** | | | | |
| 78. | $Ba_{0.53}Sr_{0.47}TiO_3$ | PLD | | 51.2 | 67.3 | 4800 | [160] |
| 79. | $Ba[(Ni_{1/2},W_{1/2})_{0.1}Ti_{0.9}]O_3$ | CSD | 34 | - | - | 3000 | [169] |
| 80. | $0.88BaTiO_3$–$0.12Bi(Mg,Ti)O_3$ | CSD | 37 | - | - | 1900 | [170] |
| 81. | $BaTiO_3$/$BiFeO_3$ (bilayer) | RFMS | - | 51 | 73 | >2700 | [171] |

| S.No | Composition | Processing method | $J_s$ *[J/cm^3]* | $J_r$ *[J/cm^3]* | η *[%]* | BDS *[kV/cm]* | Reference |
|---|---|---|---|---|---|---|---|
| 82. | $0.6BaTiO_3$-$0.4Bi_{0.25}La_{0.75}Ti_3O$ | CSD | 61.1 | | 84.2 | 3230 | [172] |
| 83. | $BaZr_{0.3}Ti_{0.7}O_3$ | PLD | 214 | 156 | 72.8 | 3000 | [173] |
| 84. | $Ba(Zr_{0.15}Ti_{0.85})O_3/Ba(Zr_{0.35}Ti_{0.65})O_3$ | RFMS | - | 83.9 | 78.4 | 1470 | [175] |
| 85. | $Ba_2Bi_4Ti_5O_{18}$ | CSD | - | 37.1 | 91.5 | 2340 | [229] |
| 86. | $Ba_{0.664}Y_{0.006}Sr_{0.33}Ti_{0.995}Mn_{0.005}O_3$ | RFMS | - | 9.75 | 77.7 | 450 | [230] |
| 87. | $Ba_{0.65}Sr_{0.35}TiO_3$ | CSD | | 0.128 | | >100 | [231] |

*SSR-solid state reaction; SPS-Spark Plasma Sintering; SG-Sol-Gel; SC-Slip casting; TC-Tape casting; TSS-Two Step sintering; HP-Hot press sintering; MW-Microwave sintering; PLD-Pulsed Laser Deposition; CSD- Chemical solution deposition; RFMS- RF Magnetron sputtering

## 7. Future directions

Achieving high energy density properties in ceramics is – as testified by this review – a very complex problem that involves several length scales. Not only the chemical composition, but also density, grain size and shape, and microstructural features like microcracks, pores or aggregates play a role. The design and control of energy density is possible only if the structural complexity of hierarchical materials – like relaxors are [232] – is understood. As mentioned in the previous sections, high $J_r$ can be achieved if the polarization hysteresis is reduced at the same time maintaining an acceptable $P_s$ – or $\varepsilon_r$ – value. This relies on a competitive mechanism, namely the disruption of long-range ferroelectric correlation and the availability of reorientable electrical dipoles spanning several unit cells, respectively. Permittivity is also influenced by porosity, defects, and microstructure, and so are the losses – which additionally depend on electronic or ionic conduction, too. Energy density design must thus embrace all these aspects, and simple trial-and-error procedures seem currently too simplistic (or give only partial results) given the complexity of the problem. In spite of the complexity of the problem, dedicated research has kept pushing the ED performance of bulk ceramics up for almost a decade, as shown in Figure 11. Considering the complexity, future approaches must take advantage of ever-improving computational methods and computational resources in order to channel, generate and accelerate knowledge on structure-property relationships to achieve high energy density. There are in principle three directions that are currently gaining importance:

### *7.1. Model-aided characterisation*

Advanced characterisation methods for material's atomic structure, micro/mesostructure and/or chemistry often require the implementation of models or databases for precise identification of phases, defects, or to obtain quantitative information from the investigated material. Of paramount importance is (i) the availability of extremely precise and reliable characterisation equipment and (ii) the development of complex models and the needed expansion of computational facilities. Both are made increasingly available nowadays. Examples of this approach are the determination of the local structure in relaxor materials by Reverse Monte Carlo modelling of neutron diffuse scattering [233], the detection of defects in $BaTiO_3$ solid solutions aided by ab-initio phonon calculations [38, 234], the use of Machine Learning to quantify phases and atomic structures in ferroelectrics by correlative spectroscopy/microscopy [235–237], the model-aided calculation of local, atomic-scale polarization in relaxors from drift-free STEM analyses [238] and the quantification of porosity and tortuosity in ceramics from tomography scans aided by image correlation based on convolutional neural networks [239].

### *7.2. Model-aided structural simulation*

In order to understand how hierarchical materials need to be modified in order to attain specific properties, structural simulation methods need to be employed on several length scales. Methods based on a higher scale generally require inputs from lower-scale methods (for example: Density Functional Theory, DFT, constructs potentials that can be used for Molecular Dynamics, MD, simulations), and all methods necessitate inputs from model-aided characterisation in order to construct realistic simulation supercells. Examples are in this sense the use of effective Hamiltonians to calculate macroscopic properties (like phase transitions) in substituted BTO systems [240], the use of MD calculations to determine dynamic disorder in cubic BTO [241], and the use of phase field modelling to capture ferroelectric domain structures and their dynamics [242, 243]. With increased computational resources it can be expected that these approaches will thrive in the next five to ten years, allowing researchers to bridge several length scales with integrated DFT-MD and phenomenological modelling approaches.

### *7.3. Accelerated materials search*

Given the vastity of the parameter space for compositional tuning in lead-free perovskites, a trial-and-error materials design involving synthesis and property evaluation is, practically, an impossible task. This is even more valid when processing parameters and their influence on microstructure must be included in the picture. Hence, accelerated materials search using optimisation routines based on Artificial Intelligence (AI) attracted much interest recently. These methods start with defining a parameter space constituted by possible chemical substituents, possible processing parameters, and desired properties (for example, energy density). The parameter space may be populated by literature data or by dedicated experiments. Further, a target function is defined and is made "walk through" the parameter space efficiently using optimisation routines. The global minimum corresponds to a proposed new composition and a set of processing parameters that needs then to be tested (effectively synthesising the material with that composition and measuring its structure and properties). The results of this test are then included in the parameter space and are used to further tune the target function as well. This way, an iterative procedure is set up, which ideally should converge – after a few rounds – into the best-possible composition fulfilling the goals of the materials search. Several attempts were made already and can be found in the literature [59, 244–246]. In particular, accelerated search allowed to identify that crossover compositions are the most advantageous to achieve high energy density and efficiency in $BaTiO_3$ HoV solid solutions [59]. The drawback of these methods is the degrees of freedom in the definition of the parameter space and of the target function, where human intervention is needed, and thus may lead to the accelerated search being ill-defined. Notably, this procedure may end up suggesting compositions that either cannot be synthesised or possess additional properties that are negative for the application and were not considered in the optimisation loop (such as high conductivity, for instance) [245]. These methods are certainly very powerful and need to be further developed in the future, but need also to embrace expert inputs from advanced model-based characterisation and structural simulation tools.

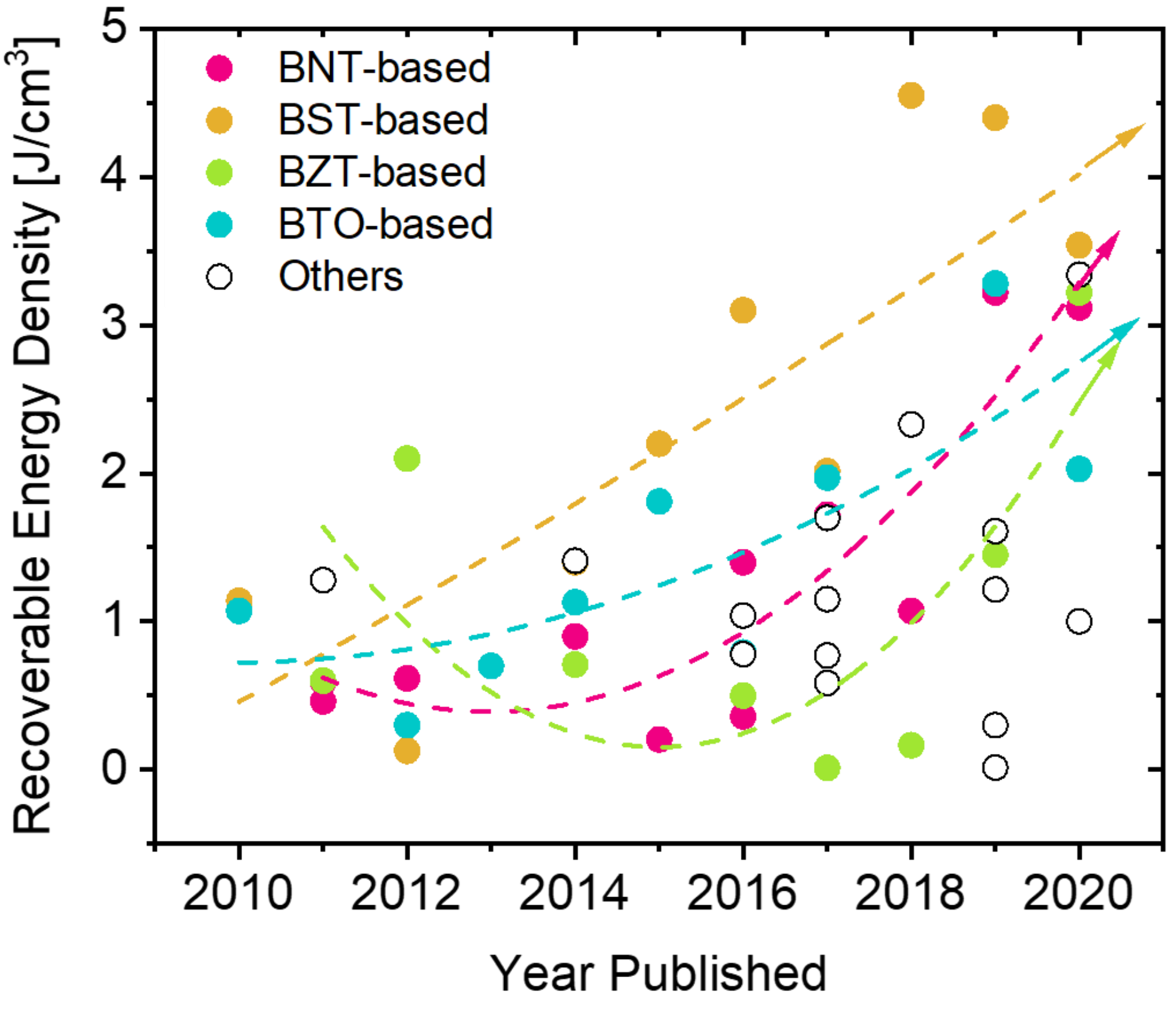

**Figure 11.** Recoverable energy density of selected lead-free relaxor systems from 2010 to 2020 [88, 91, 200–209, 94, 210–219, 113, 220–228, 230, 114, 231, 247–254, 132, 195, 196, 198, 199].

Note: BNT-Bismuth Sodium Titanate, BST-Barium Strontium Titanate and BZT-Barium Zirconium Titanate. Dashed arrows are merely a guide to the eye.

## 8. Concluding remarks

Relaxor ferroelectrics are very interesting materials that have shown extraordinary potentials in several applications. Relaxors for energy storage based on perovskite lead-free BTO, in particular, are of great interest. In this review, some of the common strategies used to tune ED properties in such materials were presented. The aspects covered here were referred to lead-free BTO based relaxors, but are largely valid in other ceramic systems for energy storage. The main macroscopic properties that have to be targeted to achieve superior ED properties were summarized for each available material forms (bulk ceramics, thick film multilayers and thin film multilayers). It was made clear that BDS enhancement through microstructure is decisive to achieve high ED properties, which however depends primarily on chemistry. While the microstructure is tuned mostly by choosing the processing route and controlling the processing parameters, chemical tuning is performed traditionally by a 'trial and error' approach combined with current scientific understanding (as discussed in Chapter 3) or with more recently developed computational procedures (as discussed in Chapter 7). The usefulness of each available material forms and ways to bridge the research developments between them were presented. More importantly, the implications of each available form and the real performance considerations between them was highlighted. This review does not cover aspects related to electrode/dielectric interfaces that might have influence on leakage currents or BDS. This is in fact a topic that would require a dedicated review. Also, the information we present on thin film multilayers is limited due to the scarcity of scientific literature on the topic. However, it is worth mentioning that more patents are filed within thin film MLCC technology than scientific papers, which highlights the increasing industrial interest on this type of EESSs. Overall, perovskite lead-free relaxors for EESSs is a very promising research area and there are great possibilities to introduce new material and methodological innovations also implementing the considerations we presented in this review.

**Author Contributions: Author Contributions:** Conceptualization: M. D., V. V..; Supervision: M. D.; Research design: V. V., M. D., F. B., T. G.; Review analysis: V. V., M. D.; Writing: V. V., F. B., T. G., M. D.; Editing and Reviewing: M. D., V. V.; Reviewing and Support: M. D.. All authors have read and agreed to the published version of the manuscript. All authors have read and agreed to the published version of the manuscript.

**Funding: Funding:** This project has received funding from the European research Council (ERC) under the European Union's Horizon 2020 research and innovation programme (grant agreement No 817190). V. V. acknowledges funding from the Austrian Science Fund (FWF): Project I4581-N.

**Conflicts of Interest:** The authors declare no conflict of interest.

## *References:*

[1] Whittingham, M. S. History, Evolution, and Future Status of Energy Storage. *Proc. IEEE*, **2012**, *100* (SPL CONTENT), 1518–1534. https://doi.org/10.1109/JPROC.2012.2190170.

[2] Sherrill, S. A.; Banerjee, P.; Rubloff, G. W.; Lee, S. B. High to Ultra-High Power

Electrical Energy Storage. *Phys. Chem. Chem. Phys.*, **2011**, *13* (46), 20714–20723. https://doi.org/10.1039/c1cp22659b.

[3] Christen, T.; Carlen, M. W. Theory of Ragone Plots. *J. Power Sources*, **2000**, *91* (2), 210–216. https://doi.org/10.1016/S0378-7753(00)00474-2.

[4] Fletcher, N. H.; Hilton, A. D.; Ricketts, B. W. Optimization of Energy Storage Density in Ceramic Capacitors. *J. Phys. D. Appl. Phys.*, **1996**, *29* (1), 253–258. https://doi.org/10.1088/0022-3727/29/1/037.

[5] Moulson, A. J.; Herbert, J. M. Dielectrics and Insulators. In *Electroceramics*; John Wiley & Sons, Ltd: Chichester, UK, 2003; pp 243–335. https://doi.org/10.1002/0470867965.

[6] Love, G. R. Energy Storage in Ceramic Dielectrics. *J. Am. Ceram. Soc.*, **1990**, *73* (2), 323–328. https://doi.org/10.1111/j.1151-2916.1990.tb06513.x.

[7] Kaliyaperumal Veerapandiyan, V. Inducing Diffuse Phase Transitions in Barium Titanate Using $Ga^{3+}$-$Ta^{5+}$ Dipole Pair Substituents, Alfred University, 2017.

[8] Haertling, G. Ferroelectric Ceramics: History and Technology. *J. Am. Ceram. Soc.*, **1999**, *82* (4), 718–818. https://doi.org/10.1111/j.1151-2916.1999.tb01840.x.

[9] Tilley, R. J. D. Insulating Solids. In *Understanding Solids: The Science of Materials*; John Wiley & Sons, Ltd: Chichester, UK, 2004. https://doi.org/10.1002/0470020849.

[10] Zhu, X. L.; Zhuang, K. Y.; Wu, S. Y.; Chen, X. M. Energy Storage Properties in $Ba_5LaTi_3Ta_7O_{30}$ Tungsten Bronze Ceramics. *J. Am. Ceram. Soc.*, **2019**, *102* (6), 3438–3447. https://doi.org/10.1111/jace.16181.

[11] Cao, L.; Yuan, Y.; Tang, B.; Li, E.; Zhang, S. Excellent Thermal Stability, High Efficiency and High Power Density of $(Sr_{0.7}Ba_{0.3})_5LaNb_7Ti_3O_{30}$–Based Tungsten Bronze Ceramics. *J. Eur. Ceram. Soc.*, **2020**, *40* (6), 2366–2374. https://doi.org/10.1016/j.jeurceramsoc.2020.01.022.

[12] Bhalla, A. S.; Guo, R.; Roy, R. The Perovskite Structure - a Review of Its Role in Ceramic Science and Technology. *Mater. Res. Innov.*, **2000**, *4* (1), 3–26. https://doi.org/10.1007/s100190000062.

[13] Jonker, G. H.; Havinga, E. E. The Influence of Foreign Ions on the Crystal Lattice of

Barium Titanate. *Mater. Res. Bull.*, **1982**, *17* (3), 345–350. https://doi.org/10.1016/0025-5408(82)90083-6.

[14] Lemanov, V. V. Barium Titanate-Based Solid Solutions. *Ferroelectrics*, **2007**, *354* (1), 69–76. https://doi.org/10.1080/00150190701454545.

[15] Goldschmidt, V. M. Die Gesetze Der Krystallochemie. *Naturwissenschaften*, **1926**, *14* (21), 477–485. https://doi.org/10.1007/BF01507527.

[16] Schütz, D.; Deluca, M.; Krauss, W.; Feteira, A.; Jackson, T.; Reichmann, K. Lone-Pair-Induced Covalency as the Cause of Temperature- and Field-Induced Instabilities in Bismuth Sodium Titanate. *Adv. Funct. Mater.*, **2012**, *22* (11), 2285–2294. https://doi.org/10.1002/adfm.201102758.

[17] Cohen, R. E. Origin of Ferroelectricity in Perovskite Oxides. *Nature*, **1992**, *358* (6382), 136–138. https://doi.org/10.1038/358136a0.

[18] Bell, A. J.; Deubzer, O. Lead-Free Piezoelectrics-The Environmental and Regulatory Issues. *MRS Bull.*, **2018**, *43* (8), 581–587. https://doi.org/10.1557/mrs.2018.154.

[19] Megaw, H. D. Origin of Ferroelectricity in Barium Titanate and Other Perovskite-Type Crystals. *Acta Crystallogr.*, **1952**, *5* (6), 739–749. https://doi.org/10.1107/S0365110X52002069.

[20] von Hippel, A. Piezoelectricity, Ferroelectricity, and Crystal Structure. *Zeitschrift für Phys. A Hadron. Nucl.*, **1952**, *133* (1–2), 158–173. https://doi.org/10.1007/BF01948692.

[21] Buscaglia, V.; Randall, C. A. Size and Scaling Effects in Barium Titanate. An Overview. *J. Eur. Ceram. Soc.*, **2020**, *40* (November 2019), 0–1. https://doi.org/10.1016/j.jeurceramsoc.2020.01.021.

[22] Acosta, M.; Novak, N.; Rojas, V.; Patel, S.; Vaish, R.; Koruza, J.; Rossetti, G. A.; Rödel, J. BaTiO 3 -Based Piezoelectrics: Fundamentals, Current Status, and Perspectives. *Appl. Phys. Rev.*, **2017**, *4* (4), 041305. https://doi.org/10.1063/1.4990046.

[23] Burns, G.; Burstein, E. Index of Refraction in “Dirty” Displacive Ferroelectrics. *Ferroelectrics*, **1974**, *7* (1), 297–299. https://doi.org/10.1080/00150197408238026.

[24] Smolensky, G. Ferroelectrics with Diffuse Phase Transition. *Ferroelectrics*, **1984**, *53* (1), 129–135. https://doi.org/10.1080/00150198408245041.

[25] Cross, L. E. Relaxor Ferroelectrics. *J. Ceram. Soc. Japan*, **1987**, *76*, 241–267. https://doi.org/10.2109/jcersj.99.829.

[26] Cross, L. E. Relaxorferroelectrics: An Overview. *Ferroelectrics*, **1994**, *151* (1), 305–320. https://doi.org/10.1080/00150199408244755.

[27] Bokov, A. A.; Ye, Z.-G. Z.-G. Recent Progress in Relaxor Ferroelectrics with Perovskite Structure. *J. Mater. Sci.*, **2006**, *41* (1), 31–52. https://doi.org/10.1007/s10853-005-5915-7.

[28] Cowley, R. A.; Gvasaliya, S. N.; Lushnikov, S. G.; Roessli, B.; Rotaru, G. M. Relaxing with Relaxors: A Review of Relaxor Ferroelectrics. *Adv. Phys.*, **2011**, *60* (2), 229–327. https://doi.org/10.1080/00018732.2011.555385.

[29] Hao, X. A Review on the Dielectric Materials for High Energy-Storage Application. *J. Adv. Dielectr.*, **2013**, *03* (01), 1330001. https://doi.org/10.1142/S2010135X13300016.

[30] Setter, N. What Is a Ferroelectric–a Materials Designer Perspective. *Ferroelectrics*, **2016**, *500* (1), 164–182. https://doi.org/10.1080/00150193.2016.1232104.

[31] Ye, Z. G. Relaxor Ferroelectric Complex Perovskites: Structure, Properties and Phase Transitions. *Key Eng. Mater.*, **1998**, *155–156* (155–156), 81–122. https://doi.org/10.4028/www.scientific.net/KEM.155-156.81.

[32] Ahn, C. W.; Hong, C. H.; Choi, B. Y.; Kim, H. P.; Han, H. S.; Hwang, Y.; Jo, W.; Wang, K.; Li, J. F.; Lee, J. S.; et al. A Brief Review on Relaxor Ferroelectrics and Selected Issues in Lead-Free Relaxors. *J. Korean Phys. Soc.*, **2016**, *68* (12), 1481–1494. https://doi.org/10.3938/jkps.68.1481.

[33] Hlinka, J. Do We Need the Ether of Polar Nanoregions? *J. Adv. Dielectr.*, **2012**, *02* (02), 1241006. https://doi.org/10.1142/s2010135x12410068.

[34] KLEEMANN, W. RANDOM FIELDS IN RELAXOR FERROELECTRICS — A JUBILEE REVIEW. *J. Adv. Dielectr.*, **2012**, *02* (02), 1241001. https://doi.org/10.1142/S2010135X12410019.

[35] Takenaka, H.; Grinberg, I.; Liu, S.; Rappe, A. M. Slush-like Polar Structures in Single-Crystal Relaxors. *Nature*, **2017**, *546* (7658), 391–395. https://doi.org/10.1038/nature22068.

[36] Fontana, M. P.; Lambert, M. Linear Disorder and Temperature Dependence of Raman Scattering in $BaTiO_3$. *Solid State Commun.*, **1972**, *10* (1), 1–4. https://doi.org/10.1016/0038-1098(72)90334-1.

[37] Bencan, A.; Oveisi, E.; Hashemizadeh, S.; Veerapandiyan, V. K.; Hoshina, T.; Rojac, T.; Deluca, M.; Drazic, G.; Damjanovic, D. Atomic Scale Symmetry and Polar Nanoclusters in the Paraelectric Phase of Ferroelectric Materials. **2020**.

[38] Veerapandiyan, V. K.; Khosravi H, S.; Canu, G.; Feteira, A.; Buscaglia, V.; Reichmann, K.; Deluca, M. B-Site Vacancy Induced Raman Scattering in $BaTiO_3$-Based Ferroelectric Ceramics. *J. Eur. Ceram. Soc.*, **2020**, *40* (13), 4684–4688. https://doi.org/10.1016/j.jeurceramsoc.2020.05.051.

[39] Spear, W. E.; Le Comber, P. G. Substitutional Doping of Amorphous Silicon. *Solid State Commun.*, **1993**, *88* (11–12), 1015–1018. https://doi.org/10.1016/0038-1098(93)90286-V.

[40] Farhi, R.; El Marssi, M.; Simon, A.; Ravez, J. Relaxor-like and Spectroscopic Properties of Niobium Modified Barium Titanate. *Eur. Phys. J. B*, **2000**, *18* (4), 605–610. https://doi.org/10.1007/s100510070008.

[41] Maiti, T.; Guo, R.; Bhalla, A. S. Structure-Property Phase Diagram of $BaZr_xTi_{1-x}O_3$ System. *J. Am. Ceram. Soc.*, **2008**, *91* (6), 1769–1780. https://doi.org/10.1111/j.1551-2916.2008.02442.x.

[42] Buscaglia, V.; Tripathi, S.; Petkov, V.; Dapiaggi, M.; Deluca, M.; Gajović, A.; Ren, Y. Average and Local Atomic-Scale Structure in $BaZr_xTi_{1-x}O_3$ (x = 0.10, 0.20, 0.40) Ceramics by High-Energy x-Ray Diffraction and Raman Spectroscopy. *J. Phys. Condens. Matter*, **2014**, *26* (6). https://doi.org/10.1088/0953-8984/26/6/065901.

[43] Pramanick, A.; Dmowski, W.; Egami, T.; Budisuharto, A. S.; Weyland, F.; Novak, N.; Christianson, A. D.; Borreguero, J. M.; Abernathy, D. L.; Jørgensen, M. R. V. Stabilization of Polar Nanoregions in Pb-Free Ferroelectrics. *Phys. Rev. Lett.*, **2018**,

*120* (20), 207603. https://doi.org/10.1103/PhysRevLett.120.207603.

[44] Morrison, F. D.; Coats, A. M.; Sinclair, D. C.; West, A. R. Charge Compensation Mechanisms in La-Doped $BaTiO_3$. *J. Electroceramics*, **2001**, *6* (3), 219–232. https://doi.org/https://doi.org/10.1023/A:1011400630449.

[45] Shvartsman, V. V.; Lupascu, D. C. Lead-Free Relaxor Ferroelectrics. *J. Am. Ceram. Soc.*, **2012**, *95* (1), 1–26. https://doi.org/10.1111/j.1551-2916.2011.04952.x.

[46] Veerapandiyan, V. K.; Deluca, M.; Misture, S. T.; Schulze, W. A.; Pilgrim, S. M.; Tidrow, S. C. Dielectric and Structural Studies of Ferroelectric Phase Evolution in Dipole-pair Substituted Barium Titanate Ceramics. *J. Am. Ceram. Soc.*, **2020**, *103* (1), 287–296. https://doi.org/10.1111/jace.16713.

[47] Ravez, J.; Simon, A. Some Solid State Chemistry Aspects of Lead-Free Relaxor Ferroelectrics. *J. Solid State Chem.*, **2001**, *162* (2), 260–265. https://doi.org/10.1006/jssc.2001.9285.

[48] Li, W.-B.; Zhou, D.; Pang, L.-X. Enhanced Energy Storage Density by Inducing Defect Dipoles in Lead Free Relaxor Ferroelectric BaTiO 3 -Based Ceramics. *Appl. Phys. Lett.*, **2017**, *110* (13), 132902. https://doi.org/10.1063/1.4979467.

[49] Ravez, J. Ferroelectricity in Solid State Chemistry. *Comptes Rendus l'Academie des Sci. - Ser. IIc Chem.*, **2000**, *3* (4), 267–283. https://doi.org/10.1016/S1387-1609(00)00127-4.

[50] Lu, D. Y.; Sun, X. Y.; Liu, B.; Zhang, J. L.; Ogata, T. Structural and Dielectric Properties, Electron Paramagnetic Resonance, and Defect Chemistry of Pr-Doped $BaTiO_3$ Ceramics. *J. Alloys Compd.*, **2014**, *615*, 25–34. https://doi.org/10.1016/j.jallcom.2014.06.067.

[51] Morrison, F. D.; Sinclair, D. C.; West, A. R. Electrical and Structural Characteristics of Lanthanum-Doped Barium Titanate Ceramics. *J. Appl. Phys.*, **1999**, *86* (11), 6355–6366. https://doi.org/10.1063/1.371698.

[52] Anwar, S.; Sagdeo, P. R.; Lalla, N. P. Crossover from Classical to Relaxor Ferroelectrics in $BaTi_{1-x}Hf_xO_3$ Ceramics. *J. Phys. Condens. Matter*, **2006**, *18* (13), 3455–3468. https://doi.org/10.1088/0953-8984/18/13/013.

[53] Canu, G.; Confalonieri, G.; Deluca, M.; Curecheriu, L.; Buscaglia, M. T.; Asandulesa, M.; Horchidan, N.; Dapiaggi, M.; Mitoseriu, L.; Buscaglia, V. Structure-Property Correlations and Origin of Relaxor Behaviour in $BaCe_xTi_{1-x}O_3$. *Acta Mater.*, **2018**, *152*, 258–268. https://doi.org/10.1016/j.actamat.2018.04.038.

[54] Farhi, R.; El Marssi, M.; Simon, A.; Ravez, J. A Raman and Dielectric Study of Ferroelectric Ceramics. *Eur. Phys. J. B*, **1999**, *9* (4), 599–604. https://doi.org/10.1007/s100510050803.

[55] Lin, J. N.; Wu, T. B. Effects of Isovalent Substitutions on Lattice Softening and Transition Character of $BaTiO_3$ Solid Solutions. *J. Appl. Phys.*, **1990**, *68* (3), 985–993. https://doi.org/10.1063/1.346665.

[56] Wei, X.; Yao, X. Preparation, Structure and Dielectric Property of Barium Stannate Titanate Ceramics. *Mater. Sci. Eng. B Solid-State Mater. Adv. Technol.*, **2007**, *137* (1–3), 184–188. https://doi.org/10.1016/j.mseb.2006.11.012.

[57] Zhou, L.; Vilarinho, P. M.; Baptista, J. L. Dependence of the Structural and Dielectric Properties of $Ba_{1-x}Sr_xTiO_3$ Ceramic Solid Solutions on Raw Material Processing. *J. Eur. Ceram. Soc.*, **1999**, *19* (11), 2015–2020. https://doi.org/10.1016/s0955-2219(99)00010-2.

[58] Shannon, R. D. Revised Effective Ionic Radii and Systematic Studies of Interatomic Distances in Halides and Chalcogenides. *Acta Crystallogr. Sect. A*, **1976**, *32* (5), 751–767. https://doi.org/10.1107/S0567739476001551.

[59] Yuan, R.; Tian, Y.; Xue, D.; Xue, D.; Zhou, Y.; Ding, X.; Sun, J.; Lookman, T. Accelerated Search for BaTiO3-Based Ceramics with Large Energy Storage at Low Fields Using Machine Learning and Experimental Design. *Adv. Sci.*, **2019**, *6* (21). https://doi.org/10.1002/advs.201901395.

[60] Pilgrim, S. M.; Sutherland, A. E.; Winzer, S. R. Diffuseness as a Useful Parameter for Relaxor Ceramics. *J. Am. Ceram. Soc.*, **1990**, *73* (10), 3122–3125. https://doi.org/10.1111/j.1151-2916.1990.tb06733.x.

[61] Uchino, K.; Nomura, S. Critical Exponents of the Dielectric Constants in Diffused-Phase-Transition Crystals. *Ferroelectrics*, **1982**, *44* (1), 55–61.

https://doi.org/10.1080/00150198208260644.

[62] Hsiang, H.-I.; Yen, F.-S.; Huang, C.-Y. Effects of Porosity on Dielectric Properties of $BaTiO_3$ Ceramics. *Jpn. J. Appl. Phys.*, **1995**, *34* (Part 1, No. 4A), 1922–1925. https://doi.org/10.1143/JJAP.34.1922.

[63] Yang, L.; Kong, X.; Li, F.; Hao, H.; Cheng, Z.; Liu, H.; Li, J. F.; Zhang, S. Perovskite Lead-Free Dielectrics for Energy Storage Applications. *Prog. Mater. Sci.*, **2019**, *102* (December 2018), 72–108. https://doi.org/10.1016/j.pmatsci.2018.12.005.

[64] Rahaman, M. N. Grain Growth and Microstructure Control. In *Ceramic Processing and Sintering, Second Edition*; CRC Press, 2017; pp 540–619. https://doi.org/10.1201/9781315274126.

[65] Tian, H. Y.; Wang, Y.; Miao, J.; Chan, H. L. W.; Choy, C. L. Preparation and Characterization of Hafnium Doped Barium Titanate Ceramics. *J. Alloys Compd.*, **2007**, *431* (1–2), 197–202. https://doi.org/10.1016/j.jallcom.2006.05.037.

[66] Dechakupt, T.; Tangsritrakul, J.; Ketsuwan, P.; Yimnirun, R. Microstructure and Electrical Properties of Niobium Doped Barium Titanate Ceramics. *Ferroelectrics*, **2011**, *415* (1), 141–148. https://doi.org/10.1080/00150193.2011.577386.

[67] Wang, G.; Li, J.; Zhang, X.; Fan, Z.; Yang, F.; Feteira, A.; Zhou, D.; Sinclair, D. C.; Ma, T.; Tan, X.; et al. Ultrahigh Energy Storage Density Lead-Free Multilayers by Controlled Electrical Homogeneity. *Energy Environ. Sci.*, **2019**, *12* (2), 582–588. https://doi.org/10.1039/C8EE03287D.

[68] Liu, G.; Zhang, S.; Jiang, W.; Cao, W. Losses in Ferroelectric Materials. *Mater. Sci. Eng. R Reports*, **2015**, *89*, 1–48. https://doi.org/10.1016/j.mser.2015.01.002.

[69] Guyonnet, J. *Ferroelectric Domain Walls*; 2014. https://doi.org/10.1007/978-3-319-05750-7.

[70] Liu, N.; Liang, R.; Zhou, Z.; Dong, X. Designing Lead-Free Bismuth Ferrite-Based Ceramics Learning from Relaxor Ferroelectric Behavior for Simultaneous High Energy Density and Efficiency under Low Electric Field. *J. Mater. Chem. C*, **2018**, *6* (38), 10211–10217. https://doi.org/10.1039/C8TC03855D.

[71] Liu, B.; Wu, Y.; Huang, Y. H.; Song, K. X.; Wu, Y. J. Enhanced Dielectric Strength

and Energy Storage Density in $BaTi_{0.7}Zr_{0.3}O_3$ Ceramics via Spark Plasma Sintering. *J. Mater. Sci.*, **2019**, *54* (6), 4511–4517. https://doi.org/10.1007/s10853-018-3170-y.

[72] Seitz, F. On the Theory of Electron Multiplication in Crystals. *Phys. Rev.*, **1949**, *76* (9), 1376–1393. https://doi.org/10.1103/PhysRev.76.1376.

[73] Frohlich, H.; A, P. R. S. L. On the Theory of Dielectric Breakdown in Solids. *Proc. R. Soc. London. Ser. A. Math. Phys. Sci.*, **1947**, *188* (1015), 521–532. https://doi.org/10.1098/rspa.1947.0023.

[74] Zhao, P.; Wang, H.; Wu, L.; Chen, L.; Cai, Z.; Li, L.; Wang, X. High-Performance Relaxor Ferroelectric Materials for Energy Storage Applications. *Adv. Energy Mater.*, **2019**, *9* (17), 1–7. https://doi.org/10.1002/aenm.201803048.

[75] Hu, Q.; Wang, T.; Zhao, L.; Jin, L.; Xu, Z.; Wei, X. Dielectric and Energy Storage Properties of $BaTiO_3$–$Bi(Mg_{1/2}Ti_{1/2})O_3$ Ceramic: Influence of Glass Addition and Biasing Electric Field. *Ceram. Int.*, **2017**, *43* (1), 35–39. https://doi.org/10.1016/j.ceramint.2016.08.005.

[76] Puli, V. S.; Pradhan, D. K.; Adireddy, S.; Kothakonda, M.; Katiyar, R. S.; Chrisey, D. B. Effect of Lead Borosilicate Glass Addition on the Crystallization, Ferroelectric and Dielectric Energy Storage Properties of Ba0.9995La0.0005TiO3ceramics. *J. Alloys Compd.*, **2016**, *688*, 721–728. https://doi.org/10.1016/j.jallcom.2016.07.025.

[77] Liu, B.; Wang, X.; Zhao, Q.; Li, L. Improved Energy Storage Properties of Fine-Crystalline $BaTiO_3$ Ceramics by Coating Powders with $Al_2O_3$ and $SiO_2$. *J. Am. Ceram. Soc.*, **2015**, *98* (8), 2641–2646. https://doi.org/10.1111/jace.13614.

[78] Chen, K.; Pu, Y.; Xu, N.; Luo, X. Effects of SrO-$B_2O_3$-$SiO_2$ Glass Additive on Densification and Energy Storage Properties of $Ba_{0.4}Sr_{0.6}TiO_3$ Ceramics. *J. Mater. Sci. Mater. Electron.*, **2012**, *23* (8), 1599–1603. https://doi.org/10.1007/s10854-012-0635-7.

[79] Kingery, W. D. Densification during Sintering in the Presence of a Liquid Phase. I. Theory. *J. Appl. Phys.*, **1959**, *30* (3), 301–306. https://doi.org/10.1063/1.1735155.

[80] Hsiang, H. I.; Hsi, C. S.; Huang, C. C.; Fu, S. L. Sintering Behavior and Dielectric Properties of $BaTiO_3$ Ceramics with Glass Addition for Internal Capacitor of LTCC.

*J. Alloys Compd.*, **2008**, *459* (1–2), 307–310. https://doi.org/10.1016/j.jallcom.2007.04.218.

[81] Jeon, H. P.; Lee, S. K.; Kim, S. W.; Choi, D. K. Effects of BaO-$B_2O_3$-$SiO_2$ Glass Additive on Densification and Dielectric Properties of $BaTiO_3$ Ceramics. *Mater. Chem. Phys.*, **2005**, *94* (2–3), 185–189. https://doi.org/10.1016/j.matchemphys.2005.04.049.

[82] Lin, J. C. C.; Wei, W. C. J. Low-Temperature Sintering of $BaTiO_3$ with Mn-Si-O Glass. *J. Electroceramics*, **2010**, *25* (2–4), 179–187. https://doi.org/10.1007/s10832-010-9613-8.

[83] Yang, H.; Yan, F.; Lin, Y.; Wang, T. Enhanced Energy Storage Properties of $Ba_{0.4}Sr_{0.6}TiO_3$ Lead-Free Ceramics with $Bi_2O_3$-$B_2O_3$-$SiO_2$ Glass Addition. *J. Eur. Ceram. Soc.*, **2018**, *38* (4), 1367–1373. https://doi.org/10.1016/j.jeurceramsoc.2017.11.058.

[84] Wang, T.; Jin, L.; Shu, L.; Hu, Q.; Wei, X. Energy Storage Properties in Ba0.4Sr0.6TiO 3 Ceramics with Addition of Semi-Conductive BaO-B2O 3-SiO2-Na2CO3-K2CO 3 Glass. *J. Alloys Compd.*, **2014**, *617*, 399–403. https://doi.org/10.1016/j.jallcom.2014.08.038.

[85] Yang, H.; Yan, F.; Zhang, G.; Lin, Y.; Wang, F. Dielectric Behavior and Impedance Spectroscopy of Lead-Free Ba0.85Ca0.15Zr0.1Ti0.9O3 Ceramics with B2O3-Al2O3-SiO2 Glass-Ceramics Addition for Enhanced Energy Storage. *J. Alloys Compd.*, **2017**, *720*, 116–125. https://doi.org/10.1016/j.jallcom.2017.05.158.

[86] Haily, E.; Bih, L.; El bouari, A.; Lahmar, A.; Elmarssi, M.; Manoun, B. Effect of BaO–Bi2O3–P2O5 Glass Additive on Structural, Dielectric and Energy Storage Properties of BaTiO3 Ceramics. *Mater. Chem. Phys.*, **2020**, *241* (November 2019), 122434. https://doi.org/10.1016/j.matchemphys.2019.122434.

[87] Wang, S. F.; Yang, T. C. K.; Wang, Y. R.; Kuromitsu, Y. Effect of Glass Composition on the Densification and Dielectric Properties of BaTiO3 Ceramics. *Ceram. Int.*, **2001**, *27* (2), 157–162. https://doi.org/10.1016/S0272-8842(00)00055-9.

[88] Xiao, S.; Xiu, S.; Zhang, W.; Shen, B.; Zhai, J.; Zhang, Y. Effects of BaxSr1−xTiO3

Ceramics Additives on Structure and Energy Storage Properties of Ba0.4Sr0.6TiO3–BaO–B2O3–Al2O3–SiO2 Glass-Ceramic. *J. Alloys Compd.*, **2016**, *675*, 15–21. https://doi.org/10.1016/j.jallcom.2016.03.105.

[89] Zhang, Y.; Cao, M.; Yao, Z.; Wang, Z.; Song, Z.; Ullah, A.; Hao, H.; Liu, H. Effects of Silica Coating on the Microstructures and Energy Storage Properties of BaTiO3 Ceramics. *Mater. Res. Bull.*, **2015**, *67*, 70–76. https://doi.org/10.1016/j.materresbull.2015.01.056.

[90] Dong, G.; Ma, S.; Du, J.; Cui, J. Dielectric Properties and Energy Storage Density in ZnO-Doped Ba0.3Sr0.7TiO3 Ceramics. *Ceram. Int.*, **2009**, *35* (5), 2069–2075. https://doi.org/10.1016/j.ceramint.2008.11.003.

[91] Ma, W.; Fan, P.; Salamon, D.; Kongparakul, S.; Samart, C.; Zhang, T.; Zhang, G.; Jiang, S.; Chang, J. J.; Zhang, H. Fine-Grained BNT-Based Lead-Free Composite Ceramics with High Energy-Storage Density. *Ceram. Int.*, **2019**, *45* (16), 19895–19901. https://doi.org/10.1016/j.ceramint.2019.06.245.

[92] Diao, C.; Liu, H.; Hao, H.; Cao, M.; Yao, Z. Effect of SiO2 Additive on Dielectric Response and Energy Storage Performance of Ba0.4Sr0.6TiO3 Ceramics. *Ceram. Int.*, **2016**, *42* (11), 12639–12643. https://doi.org/10.1016/j.ceramint.2016.04.169.

[93] Huang, Y. H.; Wu, Y. J.; Liu, B.; Yang, T. N.; Wang, J. J.; Li, J.; Chen, L. Q.; Chen, X. M. From Core-Shell Ba0.4Sr0.6TiO3@SiO2 Particles to Dense Ceramics with High Energy Storage Performance by Spark Plasma Sintering. *J. Mater. Chem. A*, **2018**, *6* (10), 4477–4484. https://doi.org/10.1039/c7ta10821d.

[94] Zhang, Q.; Wang, L.; Luo, J.; Tang, Q.; Du, J. Ba0.4Sr0.6TiO3/MgO Composites with Enhanced Energy Storage Density and Low Dielectric Loss for Solid-State Pulse-Forming Line. *Int. J. Appl. Ceram. Technol.*, **2010**, *7* (SUPPL. 1), 124–128. https://doi.org/10.1111/j.1744-7402.2009.02456.x.

[95] Huang, Y. H.; Wu, Y. J.; Qiu, W. J.; Li, J.; Chen, X. M. Enhanced Energy Storage Density of Ba0.4Sr0.6TiO3-MgO Composite Prepared by Spark Plasma Sintering. *J. Eur. Ceram. Soc.*, **2015**, *35* (5), 1469–1476. https://doi.org/10.1016/j.jeurceramsoc.2014.11.022.

[96] Ren, P.; Wang, Q.; Li, S.; Zhao, G. Energy Storage Density and Tunable Dielectric Properties of BaTi0.85Sn0.15O3/MgO Composite Ceramics Prepared by SPS. *J. Eur. Ceram. Soc.*, **2017**, *37* (4), 1501–1507. https://doi.org/10.1016/j.jeurceramsoc.2016.12.016.

[97] Chou, C. S.; Wu, C. Y.; Yang, R. Y.; Ho, C. Y. Preparation and Characterization of the Bismuth Sodium Titanate (Na 0.5Bi 0.5TiO 3) Ceramic Doped with ZnO. *Adv. Powder Technol.*, **2012**, *23* (3), 358–365. https://doi.org/10.1016/j.apt.2011.04.015.

[98] Yao, Y.; Li, Y.; Sun, N.; Du, J.; Li, X.; Zhang, L.; Zhang, Q.; Hao, X. Enhanced Dielectric and Energy-Storage Properties in ZnO-Doped 0.9(0.94Na0.5Bi0.5TiO3−0.06BaTiO3)−0.1NaNbO3 Ceramics. *Ceram. Int.*, **2018**, *44* (6), 5961–5966. https://doi.org/10.1016/j.ceramint.2017.12.174.

[99] Tao, C. W.; Geng, X. Y.; Zhang, J.; Wang, R. X.; Gu, Z. Bin; Zhang, S. T. Bi0.5Na0.5TiO3-BaTiO3-K0.5Na0.5NbO3:ZnO Relaxor Ferroelectric Composites with High Breakdown Electric Field and Large Energy Storage Properties. *J. Eur. Ceram. Soc.*, **2018**, *38* (15), 4946–4952. https://doi.org/10.1016/j.jeurceramsoc.2018.07.006.

[100] Luo, B.; Wang, X.; Wang, Y.; Li, L. Fabrication, Characterization, Properties and Theoretical Analysis of Ceramic/PVDF Composite Flexible Films with High Dielectric Constant and Low Dielectric Loss. *J. Mater. Chem. A*, **2014**, *2* (2), 510–519. https://doi.org/10.1039/c3ta14107a.

[101] Li, H.; Liu, F.; Fan, B.; Ai, D.; Peng, Z.; Wang, Q. Nanostructured Ferroelectric-Polymer Composites for Capacitive Energy Storage. *Small Methods*, **2018**, *2* (6), 1700399. https://doi.org/10.1002/smtd.201700399.

[102] Singh, P.; Borkar, H.; Singh, B. P.; Singh, V. N.; Kumar, A. Ferroelectric Polymer-Ceramic Composite Thick Films for Energy Storage Applications. *AIP Adv.*, **2014**, *4* (8). https://doi.org/10.1063/1.4892961.

[103] Luo, H.; Zhang, D.; Jiang, C.; Yuan, X.; Chen, C.; Zhou, K. Improved Dielectric Properties and Energy Storage Density of Poly(Vinylidene Fluoride- Co -Hexafluoropropylene) Nanocomposite with Hydantoin Epoxy Resin Coated $BaTiO_3$.

*ACS Appl. Mater. Interfaces*, **2015**, *7* (15), 8061–8069. https://doi.org/10.1021/acsami.5b00555.

[104] Song, Y.; Shen, Y.; Hu, P.; Lin, Y.; Li, M.; Nan, C. W. Significant Enhancement in Energy Density of Polymer Composites Induced by Dopamine-Modified Ba 0.6Sr 0.4TiO 3 Nanofibers. *Appl. Phys. Lett.*, **2012**, *101* (15), 1–5. https://doi.org/10.1063/1.4760228.

[105] Yu, K.; Wang, H.; Zhou, Y.; Bai, Y.; Niu, Y. Enhanced Dielectric Properties of BaTiO3/Poly(Vinylidene Fluoride) Nanocomposites for Energy Storage Applications. *J. Appl. Phys.*, **2013**, *113* (3). https://doi.org/10.1063/1.4776740.

[106] Rahaman, M. N. Sintering of Ceramic: Fundamentals. In *Ceramic Processing and Sintering, Second Edition*; CRC Press, 2017; pp 425–468. https://doi.org/10.1201/9781315274126.

[107] Schomann, K. D. Electric Breakdown of Barium Titanate: A Model. *Appl. Phys.*, **1975**, *6* (1), 89–92. https://doi.org/10.1007/BF00883554.

[108] Guillon, O.; Gonzalez-Julian, J.; Dargatz, B.; Kessel, T.; Schierning, G.; Räthel, J.; Herrmann, M. Field-Assisted Sintering Technology/Spark Plasma Sintering: Mechanisms, Materials, and Technology Developments. *Adv. Eng. Mater.*, **2014**, *16* (7), 830–849. https://doi.org/10.1002/adem.201300409.

[109] Herring, C. Effect of Change of Scale on Sintering Phenomena. *J. Appl. Phys.*, **1950**, *21* (4), 301–303. https://doi.org/10.1063/1.1699658.

[110] Chen, I.-W.; Wang, X.-H. Sintering Dense Nanocrystalline Ceramics without Final-Stage Grain Growth. *Nature*, **2000**, *404* (6774), 168–171. https://doi.org/10.1038/35004548.

[111] Lóh, N. J.; Simão, L.; Faller, C. A.; De Noni, A.; Montedo, O. R. K. A Review of Two-Step Sintering for Ceramics. *Ceram. Int.*, **2016**, *42* (11), 12556–12572. https://doi.org/10.1016/j.ceramint.2016.05.065.

[112] Wang, X. W.; Zhang, B. H.; Feng, G.; Sun, L. Y.; Hu, Y. C.; Shang, S. Y.; Yin, S. Q.; Shang, J.; Wang, X. E. Enhanced Energy Storage Performance of Ba 0.94 (Bi 0.5 K 0.5 ) 0.06 Ti 0.85 Zr 0.15 O 3 Relaxor Ceramics by Two-Step Sintering Method.

*Mater. Res. Bull.*, **2019**, *114* (February), 74–79. https://doi.org/10.1016/j.materresbull.2019.02.004.

[113] Ding, J.; Liu, Y.; Lu, Y.; Qian, H.; Gao, H.; Chen, H.; Ma, C. Enhanced Energy-Storage Properties of 0.89Bi 0.5 Na 0.5 TiO 3 –0.06BaTiO 3 –0.05K 0.5 Na 0.5 NbO 3 Lead-Free Anti-Ferroelectric Ceramics by Two-Step Sintering Method. *Mater. Lett.*, **2014**, *114*, 107–110. https://doi.org/10.1016/j.matlet.2013.09.103.

[114] Gao, F.; Dong, X.; Mao, C.; Liu, W.; Zhang, H.; Yang, L.; Cao, F.; Wang, G. Energy-Storage Properties of 0.89Bi0.5Na0.5TiO 3-0.06BaTiO3-0.05K0.5Na0.5NbO 3 Lead-Free Anti-Ferroelectric Ceramics. *J. Am. Ceram. Soc.*, **2011**, *94* (12), 4382–4386. https://doi.org/10.1111/j.1551-2916.2011.04731.x.

[115] VASILOS, T.; SPRIGGS, R. M. Pressure Sintering: Mechanisms and Microstructures for Alumina and Magnesia. *J. Am. Ceram. Soc.*, **1963**, *46* (10), 493–496. https://doi.org/10.1111/j.1151-2916.1963.tb13781.x.

[116] Xue, L. A.; Chen, Y.; Gilbart, E.; Brook, R. J. The Kinetics of Hot-Pressing for Undoped and Donor-Doped BaTiO3 Ceramics. *J. Mater. Sci.*, **1990**, *25* (2), 1423–1428. https://doi.org/10.1007/BF00585460.

[117] Li, F.; Jiang, T.; Zhai, J.; Shen, B.; Zeng, H. Exploring Novel Bismuth-Based Materials for Energy Storage Applications. *J. Mater. Chem. C*, **2018**, *6* (30), 7976–7981. https://doi.org/10.1039/C8TC02801J.

[118] Li, F.; Hou, X.; Li, T.; Si, R.; Wang, C.; Zhai, J. Fine-Grain Induced Outstanding Energy Storage Performance in Novel Bi0.5K0.5TiO3-Ba(Mg1/3Nb2/3)O3 Ceramics: Via a Hot-Pressing Strategy. *J. Mater. Chem. C*, **2019**, *7* (39), 12127–12138. https://doi.org/10.1039/c9tc04320a.

[119] Gao, X.; Li, Y.; Chen, J.; Yuan, C.; Zeng, M.; Zhang, A.; Gao, X.; Lu, X.; Li, Q.; Liu, J. M. High Energy Storage Performances of Bi 1−x Sm x Fe 0.95 Sc 0.05 O 3 Lead-Free Ceramics Synthesized by Rapid Hot Press Sintering. *J. Eur. Ceram. Soc.*, **2019**, *39* (7), 2331–2338. https://doi.org/10.1016/j.jeurceramsoc.2019.02.009.

[120] Becker, M. Z. ev; Shomrat, N.; Tsur, Y. Recent Advances in Mechanism Research and Methods for Electric-Field-Assisted Sintering of Ceramics. *Adv. Mater.*, **2018**, *30*

(41), 1–8. https://doi.org/10.1002/adma.201706369.

[121] Langer, J.; Hoffmann, M. J.; Guillon, O. Electric Field-Assisted Sintering and Hot Pressing of Semiconductive Zinc Oxide: A Comparative Study. *J. Am. Ceram. Soc.*, **2011**, *94* (8), 2344–2353. https://doi.org/10.1111/j.1551-2916.2011.04396.x.

[122] Munir, Z. A.; Anselmi-Tamburini, U.; Ohyanagi, M. The Effect of Electric Field and Pressure on the Synthesis and Consolidation of Materials: A Review of the Spark Plasma Sintering Method. *J. Mater. Sci.*, **2006**, *41* (3), 763–777. https://doi.org/10.1007/s10853-006-6555-2.

[123] Huang, Y. H.; Wu, Y. J.; Li, J.; Liu, B.; Chen, X. M. Enhanced Energy Storage Properties of Barium Strontium Titanate Ceramics Prepared by Sol-Gel Method and Spark Plasma Sintering. *J. Alloys Compd.*, **2017**, *701*, 439–446. https://doi.org/10.1016/j.jallcom.2017.01.150.

[124] Qu, B.; Du, H.; Yang, Z. Lead-Free Relaxor Ferroelectric Ceramics with High Optical Transparency and Energy Storage Ability. *J. Mater. Chem. C*, **2016**, *4* (9), 1795–1803. https://doi.org/10.1039/C5TC04005A.

[125] Pu, Y.; Zhang, L.; Guo, X.; Yao, M. Improved Energy Storage Properties of 0.55Bi0.5Na0.5TiO3-0.45Ba0.85Ca0.15Ti0.85Zr0.1Sn0.05O3 Ceramics by Microwave Sintering. *Ceram. Int.*, **2018**, *44* (August), S242–S245. https://doi.org/10.1016/j.ceramint.2018.08.105.

[126] RANDALL, C. A. Scientific and Engineering Issues of the State-of-the-Art and Future Multilayer Capacitors. *J. Ceram. Soc. Japan*, **2001**, *109* (1265), S2–S6. https://doi.org/10.2109/jcersj.109.S2.

[127] Chen, G.; Zhao, J.; Li, S.; Zhong, L. Origin of Thickness Dependent Dc Electrical Breakdown in Dielectrics. *Appl. Phys. Lett.*, **2012**, *100* (22). https://doi.org/10.1063/1.4721809.

[128] Wang, H.; Zhao, P.; Chen, L.; Wang, X. Effects of Dielectric Thickness on Energy Storage Properties of 0.87BaTiO3-0.13Bi(Zn2/3(Nb0.85Ta0.15)1/3)O3 Multilayer Ceramic Capacitors. *J. Eur. Ceram. Soc.*, **2020**, *40* (5), 1902–1908. https://doi.org/10.1016/j.jeurceramsoc.2020.01.032.

[129] Li, W. B.; Zhou, D.; Xu, R.; Pang, L. X.; Reaney, I. M. BaTiO3-Bi(Li0.5Ta0.5)O3, Lead-Free Ceramics, and Multilayers with High Energy Storage Density and Efficiency. *ACS Appl. Energy Mater.*, **2018**, *1* (9), 5016–5023. https://doi.org/10.1021/acsaem.8b01001.

[130] Li, J.; Li, F.; Xu, Z.; Zhang, S. Multilayer Lead-Free Ceramic Capacitors with Ultrahigh Energy Density and Efficiency. *Adv. Mater.*, **2018**, *30* (32), 1–7. https://doi.org/10.1002/adma.201802155.

[131] Xia, W.; Zhang, N.; Yang, H.; Cao, C.; Li, J. Energy Storage BaZr0.2Ti0.8O3 Bilayer Relaxor Ferroelectric Ceramic Thick Films with High Discharging Efficiency and Fatigue Resistance. *J. Alloys Compd.*, **2019**, *788*, 978–983. https://doi.org/10.1016/j.jallcom.2019.02.332.

[132] Puli, V. S.; Pradhan, D. K.; Riggs, B. C.; Chrisey, D. B.; Katiyar, R. S. Structure, Ferroelectric, Dielectric and Energy Storage Studies of Ba 0.70Ca0.30TiO3, Ba(Zr0.2Ti 0.8)O3 Ceramic Capacitors. *Integr. Ferroelectr.*, **2014**, *157* (1), 139–146. https://doi.org/10.1080/10584587.2014.912939.

[133] Li, W. B.; Zhou, D.; Xu, R.; Wang, D. W.; Su, J. Z.; Pang, L. X.; Liu, W. F.; Chen, G. H. BaTiO3-Based Multilayers with Outstanding Energy Storage Performance for High Temperature Capacitor Applications. *ACS Appl. Energy Mater.*, **2019**, *2* (8), 5499–5506. https://doi.org/10.1021/acsaem.9b00664.

[134] Wang, G.; Lu, Z.; Li, J.; Ji, H.; Yang, H.; Li, L.; Sun, S.; Feteira, A.; Yang, H.; Zuo, R.; et al. Lead-Free (Ba,Sr)TiO3 – BiFeO3 Based Multilayer Ceramic Capacitors with High Energy Density. *J. Eur. Ceram. Soc.*, **2020**, *40* (4), 1779–1783. https://doi.org/10.1016/j.jeurceramsoc.2019.12.009.

[135] Lu, Z.; Wang, G.; Bao, W.; Li, J.; Li, L.; Mostaed, A.; Yang, H.; Ji, H.; Li, D.; Feteira, A.; et al. Superior Energy Density through Tailored Dopant Strategies in Multilayer Ceramic Capacitors. *Energy Environ. Sci.*, **2020**, *13* (9), 2938–2948. https://doi.org/10.1039/d0ee02104k.

[136] Kosec, M.; Kuscer, D.; Holc, J. Multifunctional Polycrystalline Ferroelectric Materials. In *Springer Series in Materials Science*; 2011; Vol. 140, pp 39–61.

https://doi.org/10.1007/978-90-481-2875-4_2.

[137] Pan, M.-J.; Randall, C. A. A Brief Introduction to Ceramic Capacitors. *IEEE Electr. Insul. Mag.*, **2010**, *26* (3), 44–50. https://doi.org/10.1109/MEI.2010.5482787.

[138] Goossens, D. J.; Weekes, C. J.; Avdeev, M.; Hutchison, W. D. Crystal and Magnetic Structure of (X=0.2, 0.3, 0.4 and 0.8). *J. Solid State Chem.*, **2013**, *207*, 111–116. https://doi.org/10.1016/j.jssc.2013.09.024.

[139] Liu, H.; Yang, X. Structural, Dielectric, and Magnetic Properties of BiFeO 3 -SrTiO 3 Solid Solution Ceramics. *Ferroelectrics*, **2016**, *500* (1), 310–317. https://doi.org/10.1080/00150193.2016.1230445.

[140] Cai, Z.; Zhu, C.; Wang, H.; Zhao, P.; Chen, L.; Li, L.; Wang, X. High-Temperature Lead-Free Multilayer Ceramic Capacitors with Ultrahigh Energy Density and Efficiency Fabricated: Via Two-Step Sintering. *J. Mater. Chem. A*, **2019**, *7* (24), 14575–14582. https://doi.org/10.1039/c9ta04317a.

[141] Li, J.; Shen, Z.; Chen, X.; Yang, S.; Zhou, W.; Wang, M.; Wang, L.; Kou, Q.; Liu, Y.; Li, Q.; et al. Grain-Orientation-Engineered Multilayer Ceramic Capacitors for Energy Storage Applications. *Nat. Mater.*, **2020**, *19* (9), 999–1005. https://doi.org/10.1038/s41563-020-0704-x.

[142] Yao, Z.; Song, Z.; Hao, H.; Yu, Z.; Cao, M.; Zhang, S.; Lanagan, M. T.; Liu, H. Homogeneous/Inhomogeneous-Structured Dielectrics and Their Energy-Storage Performances. *Adv. Mater.*, **2017**, *29* (20). https://doi.org/10.1002/adma.201601727.

[143] Kishi, H.; Mizuno, Y.; Chazono, H. Base-Metal Electrode-Multilayer Ceramic Capacitors: Past, Present and Future Perspectives. *Japanese J. Appl. Physics, Part 1 Regul. Pap. Short Notes Rev. Pap.*, **2003**, *42* (1), 1–5. https://doi.org/10.1143/jjap.42.1.

[144] Yoon, J.-R.; Moon, B. H.; Lee, H. Y.; Jeong, D. Y.; Rhie, D. H. Design and Analysis of Electrical Properties of a Multilayer Ceramic Capacitor Module for DC-Link of Hybrid Electric Vehicles. *J. Electr. Eng. Technol.*, **2013**, *8* (4), 808–812. https://doi.org/10.5370/JEET.2013.8.4.808.

[145] Cai, Z.; Wang, X.; Luo, B.; Hong, W.; Wu, L.; Li, L. Multiscale Design of High-

Voltage Multilayer Energy-Storage Ceramic Capacitors. *J. Am. Ceram. Soc.*, **2018**, *101* (4), 1607–1615. https://doi.org/10.1111/jace.15322.

[146] Cai, Z.; Wang, H.; Zhao, P.; Chen, L.; Zhu, C.; Hui, K.; Li, L.; Wang, X. Significantly Enhanced Dielectric Breakdown Strength and Energy Density of Multilayer Ceramic Capacitors with High Efficiency by Electrodes Structure Design. *Appl. Phys. Lett.*, **2019**, *115* (2). https://doi.org/10.1063/1.5110527.

[147] Yoon, D. H.; Lee, B. I. Processing of Barium Titanate Tapes with Different Binders for MLCC Applications-Part I: Optimization Using Design of Experiments. *J. Eur. Ceram. Soc.*, **2004**, *24* (5), 739–752. https://doi.org/10.1016/S0955-2219(03)00333-9.

[148] Yoon, D. H.; Lee, B. I. Processing of Barium Titanate Tapes with Different Binders for MLCC Applications-Part II: Comparison of the Properties. *J. Eur. Ceram. Soc.*, **2004**, *24* (5), 753–761. https://doi.org/10.1016/S0955-2219(03)00334-0.

[149] Haertling, G. H. Ferroelectric Thin Films for Electronic Applications. *J. Vac. Sci. Technol. A Vacuum, Surfaces, Film.*, **1991**, *9* (3), 414–420. https://doi.org/10.1116/1.577424.

[150] Setter, N.; Damjanovic, D.; Eng, L.; Fox, G.; Gevorgian, S.; Hong, S.; Kingon, A.; Kohlstedt, H.; Park, N. Y.; Stephenson, G. B.; et al. Ferroelectric Thin Films: Review of Materials, Properties, and Applications. *J. Appl. Phys.*, **2006**, *100* (5). https://doi.org/10.1063/1.2336999.

[151] Xu, Y.; Mackenzie, J. D. Ferroelectric Thin Films Prepared by Sol-Gel Processing. *Integr. Ferroelectr.*, **1992**, *1* (1), 17–42. https://doi.org/10.1080/10584589208215563.

[152] Wasa, K. *Thin Film Technologies for Manufacturing Piezoelectric Materials*; Woodhead Publishing Limited, 2010. https://doi.org/10.1533/9781845699758.2.441.

[153] *Chemical Solution Deposition of Functional Oxide Thin Films*; Schneller, T., Waser, R., Kosec, M., Payne, D., Eds.; Springer Vienna: Vienna, 2013. https://doi.org/10.1007/978-3-211-99311-8.

[154] ULRICH, R.; SCHAPER, L.; NELMS, D.; LEFTWICH, M. Comparison of Paraelectric and Ferroelectric Materials for Applications as Dielectrics in Thin Film Integrated Capacitors. *Int. J. microcircuits Electron. Packag.*, **2000**, *23* (2), 172–181.

[155] Jeong, Y. S.; Lee, H. U.; Lee, S. A.; Kim, J. P.; Kim, H. G.; Jeong, S. Y.; Cho, C. R. Annealing Effect of Platinum-Based Electrodes on Physical Properties of PZT Thin Films. *Curr. Appl. Phys.*, **2009**, *9* (1), 115–119. https://doi.org/10.1016/j.cap.2007.12.006.

[156] Fox, G. R.; Trolier-McKinstry, S.; Krupanidhi, S. B.; Casas, L. M. Pt/Ti/SiO 2 /Si Substrates. *J. Mater. Res.*, **1995**, *10* (6), 1508–1515. https://doi.org/10.1557/JMR.1995.1508.

[157] Wang, D. Y.; Wang, J.; Chan, H. L. W.; Choy, C. L. Structural and Electro-Optic Properties of Ba0.7Sr 0.3TiO3 Thin Films Grown on Various Substrates Using Pulsed Laser Deposition. *J. Appl. Phys.*, **2007**, *101* (4), 0–6. https://doi.org/10.1063/1.2646014.

[158] Zhang, W.; Cheng, H.; Yang, Q.; Hu, F.; Ouyang, J. Crystallographic Orientation Dependent Dielectric Properties of Epitaxial BaTiO3 Thin Films. *Ceram. Int.*, **2016**, *42* (3), 4400–4405. https://doi.org/10.1016/j.ceramint.2015.11.122.

[159] Nguyen, M. D.; Nguyen, C. T. Q.; Vu, H. N.; Rijnders, G. Controlling Microstructure and Film Growth of Relaxor-Ferroelectric Thin Films for High Break-down Strength and Energy-Storage Performance. *J. Eur. Ceram. Soc.*, **2018**, *38* (1), 95–103. https://doi.org/10.1016/j.jeurceramsoc.2017.08.027.

[160] Zhu, X.; Shi, P.; Lou, X.; Gao, Y.; Guo, X.; Sun, H.; Liu, Q.; Ren, Z. Remarkably Enhanced Energy Storage Properties of Lead-Free Ba0.53Sr0.47TiO3 Thin Films Capacitors by Optimizing Bottom Electrode Thickness. *J. Eur. Ceram. Soc.*, **2020**, *40* (15), 5475–5482. https://doi.org/10.1016/j.jeurceramsoc.2020.06.038.

[161] Wang, Y.; Chen, W.; Wang, B.; Zheng, Y. Ultrathin Ferroelectric Films: Growth, Characterization, Physics and Applications. *Materials (Basel).*, **2014**, *6* (9), 6377–6485. https://doi.org/10.3390/ma7096377.

[162] Jayadevan, K. P.; Tseng, T. Y. Review Composite and Multilayer Ferroelectric Thin Films: Processing, Properties and Applications. *J. Mater. Sci. Mater. Electron.*, **2002**, *13* (8), 439–459. https://doi.org/10.1023/A:1016129318548.

[163] *Oxide Thin Films, Multilayers, and Nanocomposites*; Mele, P., Endo, T., Arisawa, S.,

Li, C., Tsuchiya, T., Eds.; Springer International Publishing: Cham, 2015. https://doi.org/10.1007/978-3-319-14478-8.

[164] Song, S.; Zhai, J.; Gao, L.; Yao, X.; Lu, S.; Xu, Z. Thickness-Dependent Dielectric and Tunable Properties of Barium Stannate Titanate Thin Films. *J. Appl. Phys.*, **2009**, *106* (2), 20–25. https://doi.org/10.1063/1.3181060.

[165] Ohno, T.; Matsuda, T.; Ishikawa, K.; Suzuki, H. Thickness Dependence of Residual Stress in Alkoxide-Derived Pb(Zr 0.3Ti0.7)O3 Thin Film by Chemical Solution Deposition. *Japanese J. Appl. Physics, Part 1 Regul. Pap. Short Notes Rev. Pap.*, **2006**, *45* (9 B), 7265–7269. https://doi.org/10.1143/JJAP.45.7265.

[166] Udayakumar, K. R.; Schuele, P. J.; Chen, J.; Krupanidhi, S. B.; Cross, L. E. Thickness-Dependent Electrical Characteristics of Lead Zirconate Titanate Thin Films. *J. Appl. Phys.*, **1995**, *77* (8), 3981–3986. https://doi.org/10.1063/1.359508.

[167] Canedy, C. L.; Li, H.; Alpay, S. P.; Salamanca-Riba, L.; Roytburd, A. L.; Ramesh, R. Dielectric Properties in Heteroepitaxial Ba0.6Sr0.4TiO3 Thin Films: Effect of Internal Stresses and Dislocation-Type Defects. *Appl. Phys. Lett.*, **2000**, *77* (11), 1695–1697. https://doi.org/10.1063/1.1308531.

[168] Xu, J.; Menesklou, W.; Ivers-Tiffée, E. Annealing Effects on Structural and Dielectric Properties of Tunable BZT Thin Films. *J. Electroceramics*, **2004**, *13* (1–3), 229–233. https://doi.org/10.1007/s10832-004-5103-1.

[169] Karan, N. K.; Saavedra-Arias, J. J.; Perez, M.; Thomas, R.; Katiyar, R. S. High Energy Density Metal-Insulator-Metal Capacitors with Ba [(Ni12, W12)0.1 Ti0.9] O3 Thin Films. *Appl. Phys. Lett.*, **2008**, *92* (1), 1–4. https://doi.org/10.1063/1.2828700.

[170] Do-Kyun Kwon; Min Hyuk Lee. Temperature-Stable High-Energy-Density Capacitors Using Complex Perovskite Thin Films. *IEEE Trans. Ultrason. Ferroelectr. Freq. Control*, **2012**, *59* (9), 1894–1899. https://doi.org/10.1109/TUFFC.2012.2403.

[171] Zhu, H.; Liu, M.; Zhang, Y.; Yu, Z.; Ouyang, J.; Pan, W. Increasing Energy Storage Capabilities of Space-Charge Dominated Ferroelectric Thin Films Using Interlayer Coupling. *Acta Mater.*, **2017**, *122*, 252–258. https://doi.org/10.1016/j.actamat.2016.09.051.

[172] Yang, B. B.; Guo, M. Y.; Song, D. P.; Tang, X. W.; Wei, R. H.; Hu, L.; Yang, J.; Song, W. H.; Dai, J. M.; Lou, X. J.; et al. Energy Storage Properties in BaTiO3-Bi3.25La0.75Ti3O12 Thin Films. *Appl. Phys. Lett.*, **2018**, *113* (18), 0–5. https://doi.org/10.1063/1.5053446.

[173] Instan, A. A.; Pavunny, S. P.; Bhattarai, M. K.; Katiyar, R. S. Ultrahigh Capacitive Energy Storage in Highly Oriented Ba(ZrxTi1-x)O3 Thin Films Prepared by Pulsed Laser Deposition. *Appl. Phys. Lett.*, **2017**, *111* (14), 0–4. https://doi.org/10.1063/1.4986238.

[174] Cheng, H.; Ouyang, J.; Zhang, Y.-X.; Ascienzo, D.; Li, Y.; Zhao, Y.-Y.; Ren, Y. Demonstration of Ultra-High Recyclable Energy Densities in Domain-Engineered Ferroelectric Films. *Nat. Commun.*, **2017**, *8* (1), 1999. https://doi.org/10.1038/s41467-017-02040-y.

[175] Fan, Q.; Liu, M.; Ma, C.; Wang, L.; Ren, S.; Lu, L.; Lou, X.; Jia, C.-L. Significantly Enhanced Energy Storage Density with Superior Thermal Stability by Optimizing Ba(Zr0.15Ti0.85)O3/Ba(Zr0.35Ti0.65)O3 Multilayer Structure. *Nano Energy*, **2018**, *51*, 539–545. https://doi.org/10.1016/j.nanoen.2018.07.007.

[176] Pan, H.; Kursumovic, A.; Lin, Y. H.; Nan, C. W.; MacManus-Driscoll, J. L. Dielectric Films for High Performance Capacitive Energy Storage: Multiscale Engineering. *Nanoscale*, **2020**, *12* (38), 19582–19591. https://doi.org/10.1039/d0nr05709f.

[177] Brennecka, G. L.; Parish, C. M.; Tuttle, B. A.; Brewer, L. N. Multilayer Thin and Ultrathin Film Capacitors Fabricated by Chemical Solution Deposition. *J. Mater. Res.*, **2008**, *23* (1), 176–181. https://doi.org/10.1557/jmr.2008.0010.

[178] Brennecka, G. L.; Ihlefeld, J. F.; Maria, J.-P.; Tuttle, B. A.; Clem, P. G. Processing Technologies for High-Permittivity Thin Films in Capacitor Applications. *J. Am. Ceram. Soc.*, **2010**, *93* (12), 3935–3954. https://doi.org/10.1111/j.1551-2916.2010.04211.x.

[179] Yoshida, K.; Saita, H.; Kariya, T. Ultra Low Profile Thin Film Capacitor for High Performance Electronic Packages. *Proc. - Electron. Components Technol. Conf.*, **2020**, *2020-June*, 414–418. https://doi.org/10.1109/ECTC32862.2020.00073.

[180] Nagata, H.; Ko, S. W.; Hong, E.; Randall, C. A.; Trolier-McKinstry, S.; Pinceloup, P.; Skamser, D.; Randall, M.; Tajuddin, A. Microcontact Printed BaTiO3 and LaNiO3 Thin Films for Capacitors. *J. Am. Ceram. Soc.*, **2006**, *89* (9), 060612075903001-??? https://doi.org/10.1111/j.1551-2916.2006.01137.x.

[181] Wang, P.; Xu, H.; Zhu, G.; Zhao, Y.; Li, J.; Yu, A. An Efficient Method to Achieve MLCC Miniaturization and Ensure Its Reliability. *J. Mater. Sci. Mater. Electron.*, **2017**, *28* (5), 4102–4106. https://doi.org/10.1007/s10854-016-6029-5.

[182] BEAUCHAMP, E. K. Effect of Microstructure on Pulse Electrical Strength of MgO. *J. Am. Ceram. Soc.*, **1971**, *54* (10), 484–487. https://doi.org/10.1111/j.1151-2916.1971.tb12184.x.

[183] Song, Z.; Liu, H.; Zhang, S.; Wang, Z.; Shi, Y.; Hao, H.; Cao, M.; Yao, Z.; Yu, Z. Effect of Grain Size on the Energy Storage Properties of (Ba0.4Sr0.6)TiO3 Paraelectric Ceramics. *J. Eur. Ceram. Soc.*, **2014**, *34* (5), 1209–1217. https://doi.org/10.1016/j.jeurceramsoc.2013.11.039.

[184] Buscaglia, M. T.; Buscaglia, V.; Viviani, M.; Petzelt, J.; Savinov, M.; Mitoseriu, L.; Testino, A.; Nanni, P.; Harnagea, C.; Zhao, Z.; et al. Ferroelectric Properties of Dense Nanocrystalline BaTiO3 Ceramics. *Nanotechnology*, **2004**, *15* (9), 1113–1117. https://doi.org/10.1088/0957-4484/15/9/001.

[185] Pintilie, L.; Vrejoiu, I.; Hesse, D.; Alexe, M. The Influence of the Top-Contact Metal on the Ferroelectric Properties of Epitaxial Ferroelectric Pb(Zr0.2Ti0.8)O3 Thin Films. *J. Appl. Phys.*, **2008**, *104* (11), 114101. https://doi.org/10.1063/1.3021293.

[186] Klein, A. Interface Properties of Dielectric Oxides. *J. Am. Ceram. Soc.*, **2016**, *99* (2), 369–387. https://doi.org/10.1111/jace.14074.

[187] Wang, D.; Zhou, D.; Song, K.; Feteira, A.; Randall, C. A.; Reaney, I. M. Cold-Sintered C0G Multilayer Ceramic Capacitors. *Adv. Electron. Mater.*, **2019**, *5* (7), 1900025. https://doi.org/10.1002/aelm.201900025.

[188] Posadas, A.-B.; Lippmaa, M.; Walker, F. J.; Dawber, M.; Ahn, C. H.; Triscone, J.-M. Growth and Novel Applications of Epitaxial Oxide Thin Films. In *Physics of Ferroelectrics*; Springer Berlin Heidelberg: Berlin, Heidelberg; pp 219–304.

https://doi.org/10.1007/978-3-540-34591-6_6.

[189] Muralt, P. Recent Progress in Materials Issues for Piezoelectric MEMS. *J. Am. Ceram. Soc.*, **2008**, *91* (5), 1385–1396. https://doi.org/10.1111/j.1551-2916.2008.02421.x.

[190] Kui Yao; Shuting Chen; Rahimabady, M.; Mirshekarloo, M. S.; Shuhui Yu; Tay, F. E. H.; Sritharan, T.; Li Lu. Nonlinear Dielectric Thin Films for High-Power Electric Storage with Energy Density Comparable with Electrochemical Supercapacitors. *IEEE Trans. Ultrason. Ferroelectr. Freq. Control*, **2011**, *58* (9), 1968–1974. https://doi.org/10.1109/TUFFC.2011.2039.

[191] Hao, X.; Wang, P.; Zhang, X.; Xu, J. Microstructure and Energy-Storage Performance of PbO–B2O3–SiO2–ZnO Glass Added (Pb0.97La0.02)(Zr0.97Ti0.03)O3 Antiferroelectric Thick Films. *Mater. Res. Bull.*, **2013**, *48* (1), 84–88. https://doi.org/10.1016/j.materresbull.2012.10.005.

[192] Birnie, D. P. Spin Coating: Art and Science. In *Chemical Solution Deposition of Functional Oxide Thin Films*; Springer Vienna: Vienna, 2013; pp 263–274. https://doi.org/10.1007/978-3-211-99311-8_11.

[193] Bassiri-Gharb, N.; Bastani, Y.; Bernal, A. Chemical Solution Growth of Ferroelectric Oxide Thin Films and Nanostructures. *Chem. Soc. Rev.*, **2014**, *43* (7), 2125–2140. https://doi.org/10.1039/C3CS60250H.

[194] Wang, B.; Luo, L.; Jiang, X.; Li, W.; Chen, H. Energy-Storage Properties of (1−x)Bi0.47Na0.47Ba0.06TiO3–XKNbO3 Lead-Free Ceramics. *J. Alloys Compd.*, **2014**, *585*, 14–18. https://doi.org/10.1016/j.jallcom.2013.09.052.

[195] Wang, T.; Jin, L.; Li, C.; Hu, Q.; Wei, X. Relaxor Ferroelectric BaTiO3-Bi(Mg2/3Nb1/3)O3 Ceramics for Energy Storage Application. *J. Am. Ceram. Soc.*, **2014**, *98* (2), 559–566. https://doi.org/10.1111/jace.13325.

[196] Cui, C.; Pu, Y.; Shi, R. High-Energy Storage Performance in Lead-Free (0.8-x)SrTiO3-0.2Na0.5Bi0.5TiO3-XBaTiO3 Relaxor Ferroelectric Ceramics. *J. Alloys Compd.*, **2018**, *740*, 1180–1187. https://doi.org/10.1016/j.jallcom.2018.01.106.

[197] Zheng, D.; Zuo, R.; Zhang, D.; Li, Y. Novel BiFeO3-BaTiO3-Ba(Mg1/3Nb2/3)O3 Lead-Free Relaxor Ferroelectric Ceramics for Energy-Storage Capacitors. *J. Am.*

*Ceram. Soc.*, **2015**, *98* (9), 2692–2695. https://doi.org/10.1111/jace.13737.

[198] Sun, Z.; Li, L.; Yu, S.; Kang, X.; Chen, S. Energy Storage Properties and Relaxor Behavior of Lead-Free Ba1-:XSm2 x/3Zr0.15Ti0.85O3 Ceramics. *Dalt. Trans.*, **2017**, *46* (41), 14341–14347. https://doi.org/10.1039/c7dt03140h.

[199] Chen, Z.; Bai, X.; Wang, H.; Du, J.; Bai, W.; Li, L.; Wen, F.; Zheng, P.; Wu, W.; Zheng, L.; et al. Achieving High-Energy Storage Performance in 0.67Bi1-XSmxFeO3-0.33BaTiO3 Lead-Free Relaxor Ferroelectric Ceramics. *Ceram. Int.*, **2020**, *46* (8), 11549–11555. https://doi.org/10.1016/j.ceramint.2020.01.181.

[200] Lim, J. B.; Zhang, S.; Shrout, T. R. High Temperature Capacitors Using a BiScO3-BaTiO3-(K1/2Bi1/2)TiO3 Ternary System. *Electron. Mater. Lett.*, **2011**, *7* (1), 71–75. https://doi.org/10.1007/s13391-011-0311-8.

[201] Puli, V. S.; Pradhan, D. K.; Riggs, B. C.; Adireddy, S.; Katiyar, R. S.; Chrisey, D. B. Synthesis and Characterization of Lead-Free Ternary Component BST–BCT–BZT Ceramic Capacitors. *J. Adv. Dielectr.*, **2014**, *04* (02), 1450014. https://doi.org/10.1142/s2010135x14500143.

[202] Zhang, Y.; Li, Y.; Zhu, H.; Fu, Z.; Zhang, Q. Sintering Temperature Dependence of Dielectric Properties and Energy-Storage Properties in (Ba,Zr)TiO3 Ceramics. *J. Mater. Sci. Mater. Electron.*, **2017**, *28* (1), 514–518. https://doi.org/10.1007/s10854-016-5552-8.

[203] Dai, Z.; Xie, J.; Liu, W.; Wang, X.; Zhang, L.; Zhou, Z.; Li, J.; Ren, X. Effective Strategy to Achieve Excellent Energy Storage Properties in Lead-Free BaTiO3-Based Bulk Ceramics. *ACS Appl. Mater. Interfaces*, **2020**, *12* (27), 30289–30296. https://doi.org/10.1021/acsami.0c02832.

[204] Yan, B.; Fan, H.; Wang, C.; Zhang, M.; Yadav, A. K.; Zheng, X.; Wang, H.; Du, Z. Giant Electro-Strain and Enhanced Energy Storage Performance of (Y0.5Ta0.5)4+ Co-Doped 0.94(Bi0.5Na0.5)TiO3-0.06BaTiO3 Lead-Free Ceramics. *Ceram. Int.*, **2020**, *46* (1), 281–288. https://doi.org/10.1016/j.ceramint.2019.08.261.

[205] Li, Q.; Wang, J.; Ma, Y.; Ma, L.; Dong, G.; Fan, H. Enhanced Energy-Storage Performance and Dielectric Characterization of 0.94Bi0.5Na0.5TiO3–0.06BaTiO3

Modified by CaZrO3. *J. Alloys Compd.*, **2016**, *663*, 701–707. https://doi.org/10.1016/j.jallcom.2015.12.194.

[206] Xu, Q.; Xie, J.; He, Z.; Zhang, L.; Cao, M.; Huang, X.; Lanagan, M. T.; Hao, H.; Yao, Z.; Liu, H. Energy-Storage Properties of Bi0.5Na0.5TiO3-BaTiO3-KNbO3 Ceramics Fabricated by Wet-Chemical Method. *J. Eur. Ceram. Soc.*, **2017**, *37* (1), 99–106. https://doi.org/10.1016/j.jeurceramsoc.2016.07.011.

[207] Xu, Q.; Lanagan, M. T.; Huang, X.; Xie, J.; Zhang, L.; Hao, H.; Liu, H. Dielectric Behavior and Impedance Spectroscopy in Lead-Free BNT–BT–NBN Perovskite Ceramics for Energy Storage. *Ceram. Int.*, **2016**, *42* (8), 9728–9736. https://doi.org/https://doi.org/10.1016/j.ceramint.2016.03.062.

[208] Zhao, X.; Bai, W.; Ding, Y.; Wang, L.; Wu, S.; Zheng, P.; Li, P.; Zhai, J. Tailoring High Energy Density with Superior Stability under Low Electric Field in Novel (Bi0.5Na0.5)TiO3-Based Relaxor Ferroelectric Ceramics. *J. Eur. Ceram. Soc.*, **2020**, *40* (13), 4475–4486. https://doi.org/https://doi.org/10.1016/j.jeurceramsoc.2020.05.078.

[209] LIU, Y.; XU, Z.; FENG, Y. TEMPERATURE-INDEPENDENT DIELECTRIC PROPERTIES OF 0.82[0.94Bi0.5Na0.5TiO3–0.06BaTiO3]–0.18K0.5Na0.5NbO3 CERAMICS. *J. Adv. Dielectr.*, **2012**, *02* (01), 1250006. https://doi.org/10.1142/S2010135X12500063.

[210] Huang, W.; Chen, Y.; Li, X.; Wang, G.; Liu, N.; Li, S.; Zhou, M.; Dong, X. Ultrahigh Recoverable Energy Storage Density and Efficiency in Barium Strontium Titanate-Based Lead-Free Relaxor Ferroelectric Ceramics. *Appl. Phys. Lett.*, **2018**, *113* (20), 203902. https://doi.org/10.1063/1.5054000.

[211] Puli, V. S.; Kumar, A.; Katiyar, R. S.; Su, X.; Busta, C. M.; Chrisey, D. B.; M.Tomozawa. Dielectric Breakdown of BaO–B2O3–ZnO–[(BaZr0.2Ti0.80)O3]0.85 [(Ba0.70Ca0.30)TiO3]0.15 Glass-Ceramic Composites. *J. Non. Cryst. Solids*, **2012**, *358* (24), 3510–3516. https://doi.org/https://doi.org/10.1016/j.jnoncrysol.2012.05.018.

[212] Zhang, Y.; Li, Y.; Zhu, H.; Fu, Z.; Zhang, Q. Low Dielectric Loss of Bi-Doped

BaZr0.15Ti0.85O3 Ceramics for High-Voltage Capacitor Applications. *Ceram. Int.*, **2017**, *43* (15), 12186–12190. https://doi.org/https://doi.org/10.1016/j.ceramint.2017.06.077.

[213] Swain, A. B.; Subramanian, V.; Murugavel, P. The Role of Precursors on Piezoelectric and Ferroelectric Characteristics of 0.5BCT-0.5BZT Ceramic. *Ceram. Int.*, **2018**, *44* (6), 6861–6865. https://doi.org/https://doi.org/10.1016/j.ceramint.2018.01.110.

[214] Wang, Y.; Gao, S.; Wang, T.; Liu, J.; Li, D.; Yang, H.; Hu, G.; Kong, L.; Wang, F.; Liu, G. Structure, Dielectric Properties of Novel Ba(Zr,Ti)O3 Based Ceramics for Energy Storage Application. *Ceram. Int.*, **2020**, *46* (8), 12080–12087. https://doi.org/10.1016/j.ceramint.2020.01.251.

[215] Hu, Q.; Jin, L.; Wang, T.; Li, C.; Xing, Z.; Wei, X. Dielectric and Temperature Stable Energy Storage Properties of 0.88BaTiO3–0.12Bi(Mg1/2Ti1/2)O3 Bulk Ceramics. *J. Alloys Compd.*, **2015**, *640*, 416–420. https://doi.org/10.1016/j.jallcom.2015.02.225.

[216] Jiang, X.; Hao, H.; Zhang, S.; Lv, J.; Cao, M.; Yao, Z.; Liu, H. Enhanced Energy Storage and Fast Discharge Properties of BaTiO3 Based Ceramics Modified by Bi(Mg1/2Zr1/2)O3. *J. Eur. Ceram. Soc.*, **2019**, *39* (4), 1103–1109. https://doi.org/10.1016/j.jeurceramsoc.2018.11.025.

[217] Chen, X.; Li, X.; Sun, J.; Sun, C.; Shi, J.; Pang, F.; Zhou, H. Simultaneously Achieving Ultrahigh Energy Storage Density and Energy Efficiency in Barium Titanate Based Ceramics. *Ceram. Int.*, **2020**, *46* (3), 2764–2771. https://doi.org/10.1016/j.ceramint.2019.09.265.

[218] Huang, Y.; Zhao, C.; Wu, B.; Wu, J. Multifunctional BaTiO3-Based Relaxor Ferroelectrics toward Excellent Energy Storage Performance and Electrostrictive Strain Benefiting from Crossover Region. *ACS Appl. Mater. Interfaces*, **2020**, *12* (21), 23885–23895. https://doi.org/10.1021/acsami.0c03677.

[219] Zhang, Q.; Li, Z. Weakly Coupled Relaxor Behavior of BaTiO 3 -Bi(Mg 1/2 Ti 1/2 )O 3 Lead-Free Ceramics. *J. Adv. Dielectr.*, **2013**, *03* (01), 1320001. https://doi.org/10.1142/S2010135X13200014.

[220] Li, Y. Q.; Liu, H. X.; Yao, Z. H.; Xu, J.; Cui, Y. J.; Hao, H.; Cao, M. H.; Yu, Z. Y.

Characterization and Energy Storage Density of $BaTiO_3$ - $Ba(Mg_{1/3}Nb_{2/3})O_3$ Ceramics. *Mater. Sci. Forum*, **2010**, *654–656*, 2045–2048. https://doi.org/10.4028/www.scientific.net/MSF.654-656.2045.

[221] Wu, L.; Wang, X.; Li, L. Lead-Free BaTiO3–Bi(Zn2/3Nb1/3)O3 Weakly Coupled Relaxor Ferroelectric Materials for Energy Storage. *RSC Adv.*, **2016**, *6* (17), 14273–14282. https://doi.org/10.1039/C5RA21261H.

[222] Chen, P.; Chu, B. Improvement of Dielectric and Energy Storage Properties in Bi(Mg1/2Ti1/2)O3-Modified (Na1/2Bi1/2)0.92Ba0.08TiO3 Ceramics. *J. Eur. Ceram. Soc.*, **2016**, *36* (1), 81–88. https://doi.org/10.1016/j.jeurceramsoc.2015.09.029.

[223] Liu, X. X. X.; Du, H.; Liu, X. X. X.; Shi, J.; Fan, H. Energy Storage Properties of BiTi0.5Zn0.5O3-Bi0.5Na0.5TiO3-BaTiO3 Relaxor Ferroelectrics. *Ceram. Int.*, **2016**, *42* (15), 17876–17879. https://doi.org/10.1016/j.ceramint.2016.08.087.

[224] Mishra, A.; Majumdar, B.; Ranjan, R. A Complex Lead-Free (Na, Bi, Ba)(Ti, Fe)O 3 Single Phase Perovskite Ceramic with a High Energy-Density and High Discharge-Efficiency for Solid State Capacitor Applications. *J. Eur. Ceram. Soc.*, **2017**, *37* (6), 2379–2384. https://doi.org/10.1016/j.jeurceramsoc.2017.01.036.

[225] Patel, S.; Chauhan, A.; Vaish, R.; Thomas, P. Enhanced Energy Storage Performance of Glass Added 0.715Bi 0.5 Na 0.5 TiO 3 -0.065BaTiO 3 -0.22SrTiO 3 Ferroelectric Ceramics. *J. Asian Ceram. Soc.*, **2015**, *3* (4), 383–389. https://doi.org/10.1016/j.jascer.2015.07.004.

[226] Ping, W.; Liu, W.; Li, S. Enhanced Energy Storage Property in Glass-Added Ba(Zr0.2Ti0.8)O3-0.15(Ba0.7Ca0.3)TiO3 Ceramics and the Charge Relaxation. *Ceram. Int.*, **2019**, *45* (9), 11388–11394. https://doi.org/https://doi.org/10.1016/j.ceramint.2019.03.003.

[227] Li, W. B.; Zhou, D.; Pang, L. X. Structure and Energy Storage Properties of Mn-Doped (Ba,Sr)TiO3–MgO Composite Ceramics. *J. Mater. Sci. Mater. Electron.*, **2017**, *28* (12), 8749–8754. https://doi.org/10.1007/s10854-017-6600-8.

[228] Zhu, C.; Cai, Z.; Li, L.; Wang, X. High Energy Density, High Efficiency and Excellent Temperature Stability of Lead Free Mn–Doped BaTiO3–Bi(Mg1/2Zr1/2)O3

Ceramics Sintered in a Reducing Atmosphere. *J. Alloys Compd.*, **2020**, *816*, 152498. https://doi.org/10.1016/j.jallcom.2019.152498.

[229] Yang, B.; Guo, M.; Tang, X.; Wei, R.; Hu, L.; Yang, J.; Song, W.; Dai, J.; Lou, X.; Zhu, X.; et al. Lead-Free A 2 Bi 4 Ti 5 O 18 Thin Film Capacitors (A = Ba and Sr) with Large Energy Storage Density, High Efficiency, and Excellent Thermal Stability. *J. Mater. Chem. C*, **2019**, *7* (7), 1888–1895. https://doi.org/10.1039/c8tc05558k.

[230] Xu, Z.; Qiang, H.; Chen, Y. Improved Energy Storage Properties of Mn and Y Co-Doped BST Films. *Mater. Lett.*, **2020**, *259*, 126894. https://doi.org/https://doi.org/10.1016/j.matlet.2019.126894.

[231] Huang, K.; Wang, J. B.; Zhong, X. L.; Liu, B. L.; Chen, T.; Zhou, Y. C. Significant Polarization Variation near Room Temperature of Ba0.65Sr0.35TiO3 Thin Films for Pyroelectric Energy Harvesting. *Sensors Actuators B Chem.*, **2012**, *169*, 208–212. https://doi.org/https://doi.org/10.1016/j.snb.2012.04.068.

[232] Goodwin, A. L. Opportunities and Challenges in Understanding Complex Functional Materials. *Nat. Commun.*, **2019**, *10* (1), 10–13. https://doi.org/10.1038/s41467-019-12422-z.

[233] Paściak, M.; Welberry, T. R.; Kulda, J.; Kempa, M.; Hlinka, J. Polar Nanoregions and Diffuse Scattering in the Relaxor Ferroelectric PbMg1/3Nb2/3O3. *Phys. Rev. B*, **2012**, *85* (22), 224109. https://doi.org/10.1103/PhysRevB.85.224109.

[234] Popov, M. N.; Spitaler, J.; Veerapandiyan, V. K.; Bousquet, E.; Hlinka, J.; Deluca, M. Raman Spectra of Fine-Grained Materials from First Principles. *npj Comput. Mater.*, **2020**, *6* (1), 121. https://doi.org/10.1038/s41524-020-00395-3.

[235] Li, L.; Yang, Y.; Zhang, D.; Ye, Z. G.; Jesse, S.; Kalinin, S. V.; Vasudevan, R. K. Machine Learning-Enabled Identification of Material Phase Transitions Based on Experimental Data: Exploring Collective Dynamics in Ferroelectric Relaxors. *Sci. Adv.*, **2018**, *4* (3). https://doi.org/10.1126/sciadv.aap8672.

[236] Cui, A.; Jiang, K.; Jiang, M.; Shang, L.; Zhu, L.; Hu, Z.; Xu, G.; Chu, J. Decoding Phases of Matter by Machine-Learning Raman Spectroscopy. *Phys. Rev. Appl.*, **2019**, *12* (5), 1. https://doi.org/10.1103/PhysRevApplied.12.054049.

[237] Ziatdinov, M.; Nelson, C.; Vasudevan, R. K.; Chen, D. Y.; Kalinin, S. V. Building Ferroelectric from the Bottom up: The Machine Learning Analysis of the Atomic-Scale Ferroelectric Distortions. *Appl. Phys. Lett.*, **2019**, *115* (5). https://doi.org/10.1063/1.5109520.

[238] Kumar, A.; Baker, J. N.; Bowes, P. C.; Cabral, M. J.; Zhang, S.; Dickey, E. C.; Irving, D. L.; LeBeau, J. M. Atomic-Resolution Electron Microscopy of Nanoscale Local Structure in Lead-Based Relaxor Ferroelectrics. *Nat. Mater.*, **2020**. https://doi.org/10.1038/s41563-020-0794-5.

[239] Vorauer, T.; Kumar, P.; Berhaut, C. L.; Chamasemani, F. F.; Jouneau, P. H.; Aradilla, D.; Tardif, S.; Pouget, S.; Fuchsbichler, B.; Helfen, L.; et al. Multi-Scale Quantification and Modeling of Aged Nanostructured Silicon-Based Composite Anodes. *Commun. Chem.*, **2020**, *3* (1), 1–11. https://doi.org/10.1038/s42004-020-00386-x.

[240] Mentzer, C.; Lisenkov, S.; Fthenakis, Z. G.; Ponomareva, I. Phase Evolution in the Ferroelectric Relaxor Ba(Ti1-x,Zrx) O3 from Atomistic Simulations. *Phys. Rev. B*, **2019**, *99* (6), 1–7. https://doi.org/10.1103/PhysRevB.99.064111.

[241] Paściak, M.; Welberry, T. R.; Kulda, J.; Leoni, S.; Hlinka, J. Dynamic Displacement Disorder of Cubic BaTiO3. *Phys. Rev. Lett.*, **2018**, *120* (16), 167601. https://doi.org/10.1103/PhysRevLett.120.167601.

[242] Wang, B.; Chen, H. N.; Wang, J. J.; Chen, L. Q. Ferroelectric Domain Structures and Temperature-Misfit Strain Phase Diagrams of K1-XNaxNbO3 Thin Films: A Phase-Field Study. *Appl. Phys. Lett.*, **2019**, *115* (9), 092902. https://doi.org/10.1063/1.5116910.

[243] Wang, J. J.; Wang, B.; Chen, L. Q. Understanding, Predicting, and Designing Ferroelectric Domain Structures and Switching Guided by the Phase-Field Method. *Annu. Rev. Mater. Res.*, **2019**, *49*, 127–152. https://doi.org/10.1146/annurev-matsci-070218-121843.

[244] Xue, D.; Balachandran, P. V.; Hogden, J.; Theiler, J.; Xue, D.; Lookman, T. Accelerated Search for Materials with Targeted Properties by Adaptive Design. *Nat.*

*Commun.*, **2016**, *7*, 1–9. https://doi.org/10.1038/ncomms11241.

[245] Balachandran, P. V.; Kowalski, B.; Sehirlioglu, A.; Lookman, T. Experimental Search for High-Temperature Ferroelectric Perovskites Guided by Two-Step Machine Learning. *Nat. Commun.*, **2018**, *9* (1). https://doi.org/10.1038/s41467-018-03821-9.

[246] Yuan, R.; Liu, Z.; Balachandran, P. V.; Xue, D.; Zhou, Y.; Ding, X.; Sun, J.; Xue, D.; Lookman, T. Accelerated Discovery of Large Electrostrains in BaTiO3-Based Piezoelectrics Using Active Learning. *Adv. Mater.*, **2018**, *30* (7), 1–8. https://doi.org/10.1002/adma.201702884.

[247] Vuong, L. D.; Gio, P. D. Enhancement in Dielectric, Ferroelectric, and Piezoelectric Properties of BaTiO3- Modified Bi0.5(Na0.4K0.1)TiO3 Lead-Free Ceramics. *J. Alloys Compd.*, **2020**, *817*, 152790. https://doi.org/10.1016/j.jallcom.2019.152790.

[248] Hanani, Z.; Mezzane, D.; Amjoud, M.; Razumnaya, A. G.; Fourcade, S.; Gagou, Y.; Hoummada, K.; El Marssi, M.; Gouné, M. Phase Transitions, Energy Storage Performances and Electrocaloric Effect of the Lead-Free Ba 0.85 Ca 0.15 Zr 0.10 Ti 0.90 O 3 Ceramic Relaxor. *J. Mater. Sci. Mater. Electron.*, **2019**, *30* (7), 6430–6438. https://doi.org/10.1007/s10854-019-00946-5.

[249] Li, F.; Zhai, J.; Shen, B.; Liu, X.; Zeng, H. Simultaneously High-Energy Storage Density and Responsivity in Quasi-Hysteresis-Free Mn-Doped Bi0.5Na0.5TiO3-BaTiO3-(Sr0.7Bi0.2□0.1)TiO3 Ergodic Relaxor Ceramics. *Mater. Res. Lett.*, **2018**, *6* (7), 345–352. https://doi.org/10.1080/21663831.2018.1457095.

[250] Wang, S.; Huang, X.; Wang, G.; Wang, Y.; He, J.; Jiang, P. Increasing the Energy Efficiency and Breakdown Strength of High-Energy-Density Polymer Nanocomposites by Engineering the Ba0.7Sr0.3TiO3 Nanowire Surface via Reversible Addition–Fragmentation Chain Transfer Polymerization. *J. Phys. Chem. C*, **2015**, *119* (45), 25307–25318. https://doi.org/10.1021/acs.jpcc.5b09066.

[251] Puli, V. S.; Kumar, A.; Chrisey, D. B.; Tomozawa, M.; Scott, J. F.; Katiyar, R. S. Barium Zirconate-Titanate/Barium Calcium-Titanate Ceramics via Sol–Gel Process: Novel High-Energy-Density Capacitors. *J. Phys. D. Appl. Phys.*, **2011**, *44* (39), 395403. https://doi.org/10.1088/0022-3727/44/39/395403.

[252] Zhang, Q.; Zhang, Y.; Wang, X.; Ma, T.; Yuan, Z. Influence of Sintering Temperature on Energy Storage Properties of BaTiO3–(Sr1−1.5xBix) TiO3 Ceramics. *Ceram. Int.*, **2012**, *38* (6), 4765–4770. https://doi.org/https://doi.org/10.1016/j.ceramint.2012.02.064.

[253] Yang, H.; Yan, F.; Lin, Y.; Wang, T.; Wang, F.; Wang, Y.; Guo, L.; Tai, W.; Wei, H. Lead-Free BaTiO3-Bi0.5Na0.5TiO3-Na0.73Bi0.09NbO3 Relaxor Ferroelectric Ceramics for High Energy Storage. *J. Eur. Ceram. Soc.*, **2017**, *37* (10), 3303–3311. https://doi.org/10.1016/j.jeurceramsoc.2017.03.071.

[254] Ghosh, S. K.; Chauhan, V.; Hussain, A.; Rout, S. K. Phase Transition and Energy Storage Properties of BaTiO3-Modified Bi0.5(Na0.8K0.2)0.5TiO3 Ceramics. *Ferroelectrics*, **2017**, *517* (1), 97–103. https://doi.org/10.1080/00150193.2017.1370266.